\documentclass[12pt]{article}
\usepackage[utf8]{inputenc}
\usepackage[DIV=13]{typearea}\usepackage{ragged2e}
\usepackage{calligra,amsmath,amsfonts,bbm,mathrsfs,amssymb}
\usepackage{mathrsfs}
\usepackage{slashed,cancel}
\usepackage{cite}
\usepackage{hyperref}
\hypersetup{colorlinks=true,urlcolor=magenta,anchorcolor=blue,citecolor=blue,filecolor=blue,linkcolor=magenta,menucolor=blue,linktocpage=true,pdfproducer=medialab,pdfauthor={Carla Biggio, Marta Fuentes Zamoro, Xu Li, Luca Merlo, Luca Ottonello},pdftitle={How to Identify a Majoron: Effective Field Theories of Spontaneous Lepton Number Breaking}}
\usepackage[english]{babel}
\usepackage{indentfirst}
\usepackage{graphicx}
\usepackage{tikz}
\usetikzlibrary{positioning,arrows.meta,fit,backgrounds}
\usepackage{float}
\usepackage{enumerate}
\usepackage{caption}
\usepackage{subcaption}
\usepackage{multirow}
\usepackage{xcolor}
\usepackage{booktabs}
\usepackage{physics}
\usepackage{url}
\usepackage{soul}
\usepackage{verbatim}
\usepackage{multicol}
\usepackage{bbold}
\usepackage{fontawesome}
\usepackage{tablefootnote}
\usepackage[normalem]{ulem}
\newcommand{\email}[1]{\href{mailto:#1}{\tt #1}}

\numberwithin{equation}{section}

\newcommand{\blue}[1]{\color{blue} #1 \color{black}}
\newcommand{\magenta}[1]{\color{magenta} #1 \color{black}}
\newcommand{\be}{\begin{equation}}
\newcommand{\ee}{\end{equation}}
\newcommand{\ba} {\begin{equation}\begin{aligned}}
\newcommand{\ea} {\end{aligned}\end{equation}}
\newcommand{\bea}{\begin{eqnarray}}
\newcommand{\eea}{\end{eqnarray}}

\newcommand{\sL}{\mathscr{L}}

\newcommand{\cC}{\mathcal{C}}

\newcommand{\cJ}{\mathcal{J}}

\newcommand{\cO}{\mathcal{O}}
\newcommand{\cQ}{\mathcal{Q}}

\newcommand{\derp}{\partial}
\newcommand{\hc}{\text{h.c.}}
\renewcommand{\sL}{\mathscr{L}}
\newcommand{\nn}{\nonumber}
\renewcommand{\vev}[1]{\langle #1\rangle}
\newcommand{\ov}[1]{\overline{#1}}

\newcommand\subsetsim{\mathrel{
  \ooalign{\raise0.2ex\hbox{$\subset$}\cr\hidewidth\raise-0.8ex\hbox{\scalebox{0.9}{$\sim$}}\hidewidth\cr}}}

\newcommand{\TeV}{\ \text{TeV}}
\newcommand{\GeV}{\ \text{GeV}}
\newcommand{\MeV}{\ \text{MeV}}

\usepackage{color}
\usepackage{colortbl}
\definecolor{rossoc}{cmyk}{0,1,1,0.2}
\definecolor{dgreen}{rgb}{0.0, 0.5, 0.0}

\begin{document} 
\renewcommand*{\thefootnote}{\fnsymbol{footnote}}

\begin{titlepage}

\vspace*{-1cm}
\flushleft{\magenta{IFT-UAM/CSIC-26-92}} 
\\[1cm]

\begin{center}
\blue{\bf \Large How to Identify a Majoron:}\\
\vskip .3cm
\blue{\bf \Large Effective Field Theories of}\\
\blue{\bf \Large Spontaneous Lepton Number Breaking}
\centering
\vskip .3cm
\end{center}
\vskip 0.5  cm
\begin{center}
{\large\bf Carla Biggio}$^{a,b}$~\footnote{\email{carla.biggio@ge.infn.it}},
{\large\bf Marta Fuentes Zamoro}$^{c}$~\footnote{\email{marta.zamoro@uam.es}},
{\large\bf Xu Li}$^{b}$~\footnote{\email{xu.li@ge.infn.it}},
\vskip 0.5cm
{\large\bf Luca Merlo}$^{c}$~\footnote{\email{luca.merlo@uam.es}}, and {\large\bf Luca Ottonello}$^{a}$~\footnote{\email{S4872814@studenti.unige.it}},
\vskip .7cm
{\footnotesize
$^a$~Dipartimento di Fisica, Universit\`a di Genova,
I-16146 Genova, Italy{\par\centering \vskip 0.25 cm\par}
$^b$~INFN, Sezione di Genova,
Via Dodecaneso 33, I-16146 Genova, Italy{\par\centering \vskip 0.25 cm\par}
$^c$~Departamento de F\'isica Te\'orica and Instituto de F\'isica Te\'orica UAM/CSIC,\\
Universidad Aut\'onoma de Madrid, Cantoblanco, 28049, Madrid, Spain
}

\end{center}
\vskip 1cm

\begin{abstract}
\justify

We revisit the traditional type~I, II, and III seesaw mechanisms in the presence of a complex scalar field charged under a global $U(1)$ symmetry that can be identified with lepton number and the Peccei-Quinn symmetry. After symmetry breaking, the radial mode becomes heavy while the angular mode appears as an axionlike particle, traditionally dubbed the Majoron. We construct the effective field theory obtained after integrating out the heavy states and analyze two matching orders: first removing the radial mode and then the seesaw fields, and vice versa. Both procedures yield the same low-energy Lagrangian containing only Standard Model fields and the Majoron. Because a single vacuum expectation value fixes the mediator masses, the radial mode, and every Majoron coupling, these models predict relations among observables rather than their individual size, and it is these relations that are testable. Indeed, the invisible Higgs width is locked to the universal suppression of the Higgs couplings, while the Majoron-lepton coupling is fixed by the measured nonunitarity of the leptonic mixing matrix, and the two independently give comparable lower bounds on the same lepton-number-breaking scale, of the order of $1$--$10$~TeV. Neutrinoless double beta decay with Majoron emission, by contrast, has no sensitivity in this class of models. The framework is, thus, falsifiable even when the new states lie far beyond experimental reach.
\end{abstract}
\end{titlepage}
\setcounter{footnote}{0}

\pdfbookmark[1]{Table of Contents}{tableofcontents}
\tableofcontents
\renewcommand*{\thefootnote}{\arabic{footnote}}

\bigskip
\section{Introduction}
\label{sec:intro}

The discovery of neutrino oscillations provided the first unambiguous evidence for physics beyond the Standard Model (SM). The observation that at least two neutrinos possess nonzero masses and mix with large angles calls for an extension of the theory that explains both the smallness of neutrino masses and the structure of the leptonic mixing matrix. Among the most compelling and economical possibilities are the type~I~\cite{Minkowski:1977sc,Gell-Mann:1979vob,Yanagida:1979as,Mohapatra:1979ia}, II~\cite{Magg:1980ut,Schechter:1980gr,Cheng:1980qt,Lazarides:1980nt,Mohapatra:1980yp,Wetterich:1981bx}, and III~\cite{Foot:1988aq} seesaw mechanisms, which generate Majorana masses for the active neutrinos through the exchange of fermionic singlets, scalar triplets, or fermionic triplets, respectively. These mechanisms are characterized by the explicit breaking of the lepton number (LN), an accidental symmetry of the SM Lagrangian. They provide a description for the active neutrino masses that can be generically written as $m_\nu \sim g_\nu v^2/M_\text{SS}$, where $M_\text{SS}$ denotes the mass scale of the new degrees of freedom (d.o.f.) and $g_\nu$ a combination of Lagrangian parameters. The active neutrino mass scale can be naturally suppressed either with large $M_\text{SS}$ and $g_\nu\sim\cO(1)$, with relatively small $M_\text{SS}$ and tiny $g_\nu$, or with interpolating cases. For example, deviating from the traditional constructions, a well-known subclass of type~I seesaw scenarios, dubbed low-scale seesaw~\cite{Wyler:1982dd,Mohapatra:1986bd,Bernabeu:1987gr,Malinsky:2005bi,Kersten:2007vk,Abada:2007ux}, exists, where the lepton number invariance is an approximate symmetry of the relevant Lagrangian and active neutrino masses can be described with TeV right-handed (RH) neutrinos and a small explicit lepton-number-breaking parameter.
In addition, seesaw frameworks can be embedded in grand unified theories, can accommodate leptogenesis, and often arise as low-energy limits of more fundamental ultraviolet (UV) completions.

Parallel to the neutrino sector, the strong CP problem remains one of the most persistent puzzles in particle physics. The Peccei-Quinn (PQ) mechanism~\cite{Peccei:1977hh} offers a dynamical solution by promoting the QCD $\theta$ parameter to a dynamical field, whose relaxation to zero is ensured by the presence of a pseudo-Nambu-Goldstone boson: the axion~\cite{Peccei:1977hh,Weinberg:1977ma,
Wilczek:1977pj,Zhitnitsky:1980tq,Dine:1981rt,Kim:1979if,Shifman:1979if}. While the QCD axion is tightly constrained by astrophysical and cosmological observations, a broader class of pseudoscalar particles known as axionlike particles (ALPs) arises generically in theories with spontaneously broken approximate global symmetries. 

One of the main conceptual differences between the QCD axion and an ALP is the existence or not of a relation between its mass $m_a$ and its characteristic scale $f_a$. For the QCD axion, a strict inverse proportionality holds:
\be
m_a f_a\approx m_\pi f_\pi\,,
\label{AxionMassRelation}
\ee
where $m_\pi$ and $f_\pi$ are the pion mass and its decay constant, respectively. For a long time, the community considered that any sizable deviation from this relation would nullify the explanation of the strong CP problem, but recent results~\cite{DiLuzio:2016sbl,DiLuzio:2017pfr,Gaillard:2018xgk,Hook:2019qoh,DiLuzio:2020wdo,DiLuzio:2020oah,DiLuzio:2021pxd,DiLuzio:2021gos,Gavela:2023tzu,Cox:2023dou,deGiorgi:2024elx,FernandezNavarro:2026cyu} demonstrated that the parameter space accommodating a consistent solution to this problem exceeds that of the very constrained QCD axion band. A paradigm change followed, and nowadays an ALP, for which no explicit relation between $m_a$ and $f_a$ is assumed, may represent a low-energy description of a nontraditional QCD axion that nevertheless solves the strong CP problem. Even so, the ALP mass is almost always taken below its characteristic scale, $m_a<f_a$, a habit inherited from the QCD axion rather than a requirement of the effective description. What the derivative expansion asks is that $m_a$ stay below the cutoff $4\pi f_a$, so that the window $f_a<m_a<4\pi f_a$ remains open, as recently explored for leptophilic ALPs in Ref.~\cite{Zamoro:2026ily}.

ALPs appear in a wide range of UV scenarios, including string compactifications~\cite{Witten:1984dg,Choi:2006qj,
Svrcek:2006yi,Arvanitaki:2009fg,Cicoli:2012sz}, supersymmetric extensions~\cite{Bellazzini:2017neg}, and composite Higgs models~\cite{Merlo:2017sun,Brivio:2017sdm,
Alonso-Gonzalez:2018vpc,Alonso-Gonzalez:2020wst}. They have been considered as possible dark matter candidates~\cite{Gelmini:1984pe,Berezinsky:1993fm,Lattanzi:2007ux,
Bazzocchi:2008fh,Lattanzi:2013uza,Queiroz:2014yna,Akita:2026gzk,deGiorgi:2026jqn,Batell:2026avi}, and, more generally, their possible impact on cosmological observables has been investigated~\cite{Ferreira:2018vjj,DEramo:2018vss,Escudero:2019gvw,Arias-Aragon:2020qtn,Arias-Aragon:2020qip, Arias-Aragon:2020shv,Ferreira:2020bpb,Escudero:2021rfi,Araki:2021xdk,DEramo:2021psx,DEramo:2021lgb, DEramo:2022nvb}. Last but not least, ALPs remain an evergreen topic in flavor model building~\cite{Davidson:1981zd,Wilczek:1982rv,Ema:2016ops, Calibbi:2016hwq,Arias-Aragon:2017eww,Arias-Aragon:2022ats,DiLuzio:2023ndz,Greljo:2024evt}. A recent and complete review that includes models with ALPs is Ref.~\cite{Albertus:2026fbe}. Still from the theoretical point of view, their interactions with SM fields are well captured by an effective field theory (EFT) expansion in inverse powers of the ALP scale $f_a$, and the community has extensively worked to improve and complete this description~\cite{Choi:1986zw,Salvio:2013iaa,
Brivio:2017ije,Alonso-Alvarez:2018irt,Gavela:2019wzg,Chala:2020wvs,DiLuzio:2020oah,Bonilla:2021ufe,Arias-Aragon:2022byr,
Arias-Aragon:2022iwl,DiLuzio:2023cuk,DiLuzio:2023lmd}.

Additionally, ALPs are currently the subject of an extensive experimental program spanning astrophysics, cosmology, and laboratory searches at both colliders~\cite{Jaeckel:2012yz,Mimasu:2014nea,
Jaeckel:2015jla,Alves:2016koo,Knapen:2016moh,Brivio:2017ije,Bauer:2017nlg,Mariotti:2017vtv,Bauer:2017ris,
Baldenegro:2018hng,Craig:2018kne,Bauer:2018uxu,Gavela:2019cmq,Haghighat:2020nuh,Wang:2021uyb,deGiorgi:2022oks,
Bonilla:2022pxu,Ghebretinsaea:2022djg,Vileta:2022jou,Calibbi:2022izs,Marcos:2024yfm,Arias-Aragon:2024gpm,Biekotter:2025fll,Ema:2025bww,Bhattacharya:2025hme} and low-energy
facilities~\cite{Izaguirre:2016dfi,Marciano:2016yhf,Merlo:2019anv,Aloni:2019ruo,Bauer:2019gfk,Cornella:2019uxs,Bauer:2020jbp,Calibbi:2020jvd,Bauer:2021mvw,Carmona:2021seb,Guerrera:2021yss,Gallo:2021ame,Bertholet:2021hjl,Cheng:2021kjg,Bonilla:2022qgm,Bonilla:2022vtn,deGiorgi:2022vup,Guerrera:2022ykl,Bonilla:2023dtf,Arias-Aragon:2023ehh,DiLuzio:2024jip,deGiorgi:2024str,Alda:2024cxn,Alda:2024xxa,Arias-Aragon:2024qji,Arias-Aragon:2024gdz,Bisht:2024hbs,Calibbi:2024rcm,MartinCamalich:2025srw,Alda:2025uwo,Alda:2025nsz,Ardu:2026vsr}. 
A very recent review on ALPs can be found in Ref.~\cite{Arza:2026rsl}.

The coexistence of neutrino mass generation mechanisms and spontaneously broken global symmetries naturally raises the question of whether these two sectors may be structurally connected. A particularly appealing possibility is that the global symmetry responsible for the ALP is (or is aligned with) the lepton number. In such scenarios, the breaking of the symmetry that controls neutrino mass generation simultaneously gives rise to an ALP, thereby linking the neutrino and axion sectors in a unified framework. This idea has appeared in various guises, including Majoron models~\cite{Chikashige:1980qk,Chikashige:1980ui,Gelmini:1980re,Biggio:2023gtm,Debnath:2024vpf} and, more in general, theories in which the PQ and seesaw scales are related or even identified~\cite{deGiorgi:2023tvn,Liang:2024vnd,Greljo:2025suh,Debnath:2026ebg}.
However, a systematic and model-independent EFT treatment of the interplay between the heavy seesaw d.o.f., the radial mode of the PQ-breaking scalar, and the resulting ALP remains incomplete (though recently addressed for Kim-Shifman-Vainshtein-Zakharov (KSVZ)-type completions~\cite{Palavric:2026vej}). In particular, the order in which heavy fields are integrated out can obscure the physical origin of the effective operators and complicate the interpretation of ALP couplings in terms of the underlying neutrino mass mechanism.

In this paper, we revisit the three traditional seesaw mechanisms in the presence of an additional complex scalar field $\phi$ charged under a global $U(1)$ symmetry. We take this symmetry to be identifiable with the LN, although, due to its chiral nature, it stands for a PQ symmetry. After spontaneous symmetry breaking, the radial mode of the scalar, denoted as $\rho$, acquires a large mass proportional to the scale of the breaking, while the angular degree of freedom becomes an ALP. In this particular case in which the global symmetry coincides with the LN, it will be dubbed Majoron and labeled by $J$. We perform a systematic matching of the UV theory onto the low-energy EFT in two complementary orders:
\begin{enumerate}
    \item first integrating out the radial mode of the LN-breaking scalar and subsequently the heavy seesaw d.o.f.;
    \item first integrating out the heavy Seesaw d.o.f.\ and subsequently the radial mode.
\end{enumerate}
Both procedures lead to the same low-energy Lagrangian containing only SM fields and the Majoron, as required by EFT consistency, but the intermediate EFTs differ in structure and provide complementary physical insight. Our analysis yields the complete set of effective operators up to dimension 7 involving the Majoron and SM fields that arise from each seesaw type, including the contributions to the Weinberg operator, Majoron-neutrino couplings, Majoron-Higgs interactions, and Majoron-gauge couplings induced at tree level or through anomaly matching. The resulting EFT provides a unified and model-independent description of Majoron interactions in seesaw frameworks, valid for arbitrary hierarchies between the LN scale and the seesaw scale.

The effective interactions derived in this work have direct implications for ongoing and future experimental efforts targeting Majoron or ALPs and neutrino physics. The structure of Majoron couplings to leptons, gauge bosons, and the Higgs sector depends sensitively on the underlying mechanism of neutrino mass generation. This dependence opens the possibility of using Majoron phenomenology as a diagnostic tool to distinguish between different seesaw realizations, even when the heavy d.o.f.\ themselves lie far above the reach of colliders. Our EFT results can, in principle, be mapped onto observables as diverse as rare meson and lepton decays, neutrino experiments, astrophysical probes such as stellar cooling and supernova energy loss, and cosmological signatures. Which of these actually retain sensitivity, however, depends strongly on the scale at which the lepton number is broken and turns out to be a rather selective question: we address it in Sec.~\ref{sec:pheno}. In particular, the interplay between neutrino mass generation and Majoron couplings leads to characteristic patterns of correlations among Majoron interactions that are experimentally testable.

The framework developed here, thus, establishes the theoretical foundation for exploiting Majoron searches as an indirect probe of the origin of neutrino masses. By connecting UV seesaw physics with low-energy Majoron signatures in a consistent EFT language, our results enhance the physics potential of current and future Majoron experiments and motivate dedicated strategies to identify the mechanism of neutrino mass generation through the Majoron portal.

It may be useful to anticipate the outcome of the phenomenological discussion, since it is the part of the analysis with the most direct experimental relevance. Because a single vacuum expectation value (VEV) $v_\phi$ fixes the mediator mass, the radial mode, and every Majoron coupling, the models do not predict the magnitude of individual observables, which always depend on unknown couplings, but relations among them, and it is these relations that are testable. Two are particularly sharp. In the Higgs sector, the invisible decay into a Majoron pair is tied to the universal suppression of the Higgs couplings $\kappa_V$, and a measured enhancement $\kappa_V>1$ would exclude the whole class of models. In the leptonic sector, the Majoron-lepton coupling is fixed by the measured nonunitarity of the leptonic mixing matrix. Remarkably, these two independent lower bounds on $v_\phi$---one from the invisible Higgs branching ratio and the other from $\mu\to eJ$---currently sit in the same window, $v_\phi\gtrsim\mathcal{O}(1\text{--}10)\TeV$. We also find that neutrinoless double beta decay with Majoron emission, usually regarded as the flagship Majoron signature, has no sensitivity whatsoever in this class of models. Indeed, the derivative nature of the Goldstone couplings implies that this process is suppressed by the active neutrino mass. On the contrary, the informative observables lie instead in Higgs precision physics and in charged-lepton flavor violation.

The paper is organized as follows. Section~\ref{sec:HEEFTs} introduces the three Majoron seesaw models, fixes conventions and spectra, and carries out the first integration step for each of them: since the radial mode and the seesaw mediator may be ordered in either way, this produces seven intermediate effective theories, summarized in Table~\ref{tab:EFTsummary}, in which the Majoron coexists with one heavy state. Section~\ref{sec:JSMEFT} completes the reduction and gives the three low-energy Lagrangians, dubbed the JSMEFT, i.e., the Standard Model effective field theory (SMEFT) extended by the Majoron $J$, obtained by removing the remaining heavy field, verifies that the two matching orders agree operator by operator, reduces the redundant structures to the Warsaw basis, and collects the outcome in Table~\ref{tab:JSMEFTsummary}. Section~\ref{sec:pheno} is devoted to the phenomenology and to the question of how the Majoron sector could actually be identified. Our conclusions are summarized in Sec.~\ref{sec:Conclusions}, while Appendix~\ref{app:EOM} collects the equations of motion and their perturbative solutions for each model.

\section{High-Energy EFTs}
\label{sec:HEEFTs}
In this section, we consider the three Majoron seesaw models and discuss the possible hierarchies of scales. For each case, we perform the integration out of the heaviest degree of freedom, leading to seven different intermediate-energy EFTs, where the SM, the Majoron, and the intermediate scale d.o.f.\ are present. A summary of the different cases can be found in Table~\ref{tab:EFTsummary}.

\begin{table}[ht!]
\centering
\begin{tabular}{@{}ccccc@{}}
\toprule
Seesaw & Mass hierarchy & Field integrated out first & Intermediate EFT & Section \\
\midrule
\multirow{2}{*}{Type~I}
 & $m_\rho > M_N$                                            & $\rho$      & NJSMEFT          & \ref{sec:NJSMEFT} \\
 & $M_N > m_\rho$                                             & $N$         & $\rho$JSMEFT-I   & \ref{sec:rhoJSMEFT-I} \\
\midrule
\multirow{2}{*}{Type~III}
 & $m_\rho > M_\Sigma$                                        & $\rho$      & $\Sigma$JSMEFT   & \ref{sec:SigmaJSMEFT} \\
 & $M_\Sigma > m_\rho$                                        & $\Sigma$    & $\rho$JSMEFT-III & \ref{sec:rhoJSMEFT-III} \\
\midrule
\multirow{3}{*}{Type~II}
 & $m_\rho > M_\Delta$                                        & $\rho$      & $\Delta$JSMEFT   & \ref{sec:DeltaJSMEFT} \\
 & $M_\Delta^2=\mu_3^2+\frac12\beta_2v_\phi^2 > m_\rho^2$          & $\Delta$    & $\rho$JSMEFT-IIa & \ref{sec:rhoJSMEFT-IIa} \\
 & $M_\Delta=\mu_3 > m_\rho$                                  & $\Delta$    & $\rho$JSMEFT-IIb & \ref{sec:rhoJSMEFT-IIb} \\
\bottomrule
\end{tabular}
\caption{\em The seven intermediate high-energy EFTs obtained by integrating out the heaviest d.o.f.\ in each seesaw realization, for the two (type~I and III) or three (type~II) possible mass hierarchies. In every case, the resulting Lagrangian additionally contains the SM fields and the Majoron~$J$.}
\label{tab:EFTsummary}
\end{table}

\subsection{The type~I Majoron model}
\label{sec:typeI}

The type~I Majoron model consists in the generalization of the traditional type I seesaw scenario with the addition of a scalar singlet~\cite{Chikashige:1980qk,Chikashige:1980ui}. The complete spectrum is, thus, provided by supplementing the SM one with RH neutrinos $N_R$ and a complex scalar boson $\phi$, invariant under the SM gauge transformations, while transforming under the global lepton number $U(1)_L$, as shown in Table~\ref{tab:typeI-spectrum}. For simplicity, we will explicitly consider only the case with three RH neutrinos, although the discussion can be generalized to any number of RH neutrinos larger than 1. Throughout the paper, the SM spectrum comes with its $n_g=3$ generations, and, unless otherwise stated, all flavor indices are understood to be summed over; correspondingly, we take $n_N=n_g=3$ RH neutrinos here and, in Sec.~\ref{sec:typeIII}, $n_\Sigma=n_g=3$ fermionic triplets. This is immaterial for all the tree-level Wilson coefficients, whose flavor contractions are left implicit, but it does fix the overall normalization of the anomalous couplings of Eq.~\eqref{Jops-typeI}, which are generated at one loop and, therefore, receive an explicit sum over the fields running in the triangle.

\begin{table}[ht!]
\centering
\begin{tabular}{@{}lcccc@{}}
\toprule
Field & Spin & $SU(3)_c\times SU(2)_L\times U(1)_Y$ & $U(1)_L$ \\
\midrule
$L_L$   & $1/2$     & $(\mathbf{1},\mathbf{2},-1/2)$ & $+1$ \\
$e_R$   & $1/2$     & $(\mathbf{1},\mathbf{1},-1)$ & $+1$ \\
$H$     & $0$       & $(\mathbf{1},\mathbf{2},+1/2)$ & $0$ \\
\midrule
$N_R$   & $1/2$     & $(\mathbf{1},\mathbf{1},0)$ & $+1$ \\
$\phi$  & $0$       & $(\mathbf{1},\mathbf{1},0)$ & $-2$ \\
\bottomrule
\end{tabular}
\caption{\em Field content of the SM spectrum and of the type~I Majoron model.}
\label{tab:typeI-spectrum}
\end{table}

The corresponding Lagrangian is given by
\begin{equation}
    \sL^\text{type~I}=
    \sL_{\textrm{SM}} 
    + \overline{N}_R\, i\slashed{\partial} N_R
    -\left(\ov{L_L} \widetilde{H}Y_N N_R + \dfrac12\phi\overline{N_R^c}Y_{NN} N_R
    +\hc\right)+\derp_\mu\phi^\ast\derp^\mu\phi
    -V(\phi,H)\,,
\end{equation}
where, throughout the paper, $\psi^c\equiv C\overline\psi^T$ denotes the charge conjugate of a fermion field $\psi$, with $C$ the charge-conjugation matrix satisfying $C\gamma^{\mu T}C^{-1}=-\gamma^\mu$, and $\sL_{\text{SM}}$ is the SM Lagrangian without its Higgs scalar potential, which is instead subsumed, together with all the terms involving the additional scalar field $\phi$, into the complete scalar potential
$V(\phi,H)$:
\begin{equation}    
V\left(\phi,H\right)=-\mu_1^2\phi^\ast\phi+\lambda_1\left(\phi^\ast\phi\right)^2-\mu_2^2H^\dag H+\lambda_2\left(H^\dag H\right)^2+\beta_1 H^\dag H\phi^\ast\phi \, .
\label{CompleteScalarPotential}
\end{equation}
While all the parameters entering $V(\phi,H)$ are real by construction, we can work in the basis where $Y_{NN}$ is taken to be real and diagonal. Likewise, we also assume, without any loss of generality, that the Yukawa coupling matrix of the charged leptons is also real and diagonal. As a consequence, $Y_N$ is necessarily a generic $3\times 3$ complex matrix. 

In principle, one should study the complete potential $V(\phi,H)$, identify its true minima, and expand the theory around them. The only term entangling the Higgs and $\phi$ sectors is the mixed quartic coupling $\beta_1 H^\dag H\,\phi^\ast\phi$: once $\phi$ acquires a vacuum expectation value (VEV), this operator feeds directly into the effective Higgs mass parameter through a shift of the order of $\beta_1 v_\phi^2$. Since we take $v_\phi\gg v_{\rm EW}$, treating this mixing exactly would reintroduce, and generically worsen, the electroweak hierarchy problem, for a generic $\beta_1$. To avoid this, we take $\beta_1$ to be a small parameter, so that its effect on the minimization of $V(\phi,H)$ can be neglected at leading order. Under this assumption, motivated also by the absence, so far, of any collider hint of new physics up to the TeV scale, the $\phi$- and $H$-dependent parts of the potential effectively decouple, and the minimization can be performed step by step, first along $\phi$ and then along $H$. Working at leading order in $\beta_1$, we can, therefore, set $v_{\rm EW}=0$ when extremizing along $\phi$, which then acquires a VEV 
\be
v_\phi^2=\dfrac{\mu_1^2}{\lambda_1}\,.
\label{eq:vphiDEF}
\ee
Once $\phi$ acquires this VEV, it can be written as
\be
\phi=\dfrac{(v_\phi+\rho)}{\sqrt2}e^{iJ/v_\phi}\,,
\label{eq:phidecomp}
\ee
where $J$ is the massless Majoron and $\rho$ is the heavy radial mode with mass 
\be
m_\rho^2=2\lambda_1 v_\phi^2\,.
\label{mrhoDEF}
\ee 
Simultaneously, $N_R$ acquires a mass after LN spontaneous symmetry breaking (SSB): 
\be
M_N=Y_{NN}\,v_\phi/\sqrt2\,.
\ee
Notably, although $m_\rho$ and $M_N$ originate from the same breaking scale $v_\phi$, they need not be comparable in size, since they depend independently on $\lambda_1$ and $Y_{NN}$, respectively.

After the LN SSB, the terms of the original Lagrangian involving $\phi$ can be rewritten in terms of $\rho$ and $J$:
\begin{align}
\sL^\text{type~I}\supset&\dfrac{1}{2}\derp_\mu\rho\derp^\mu\rho+\dfrac{\left(\rho+v_\phi\right)^2}{2v_\phi^2}\derp_\mu J\derp^\mu J-\left(\dfrac12\frac{(v_\phi+\rho)}{\sqrt2}e^{iJ/v_\phi}\overline{N_R^c}Y_{NN}N_R+\hc\right)+\label{eq:SSBLagTypeI}\\
&-\dfrac{\lambda_1}{4}\rho^4-\lambda_1v_\phi\rho^3-\frac{m_\rho^2}{2}\rho^2-\frac{\beta_1}{2}v_\phi^2\left(H^\dag H\right)-\beta_1 v_\phi\rho\left(H^\dag H\right)-\frac{\beta_1}{2}\rho^2\left(H^\dag H\right)\,.
\nn
\end{align}
The $\rho$-independent piece $-\tfrac12\beta_1v_\phi^2(H^\dag H)$ is simply a shift of the Higgs mass parameter, $\mu_2^2\to\mu_2^2-\beta_1v_\phi^2/2$, usually reabsorbed into $\mu_2^2$; we keep it explicit here, and in every Lagrangian below where the same $(v_\phi+\rho)^2$ expansion appears, for consistency with the operator lists in Sec.~\ref{sec:NJSMEFT} and beyond.
The $J$ dependence in the Yukawa term proportional to $Y_{NN}$ can be rotated away by performing the following vectorial transformation on the lepton fields:
\be
\label{rotJ}
\left\{L_L,N_R,{e}_R\right\}\rightarrow\left\{L_L,N_R,{e}_R\right\}e^{-iJ/2v_\phi} \, .
\ee
While this is useful to simplify the calculation we are going to perform, this procedure reintroduces the $J$ dependence in chirality-preserving couplings with fermions. By applying that transformation to the lepton field kinetic terms, new operators containing the Majoron do appear and the relevant part of the Lagrangian looks like
\be
\begin{split}
\sL^\text{type~I}\supset&-\dfrac{(v_\phi+\rho)}{2v_\phi}\ov{N}{M}_NN+\frac{\partial_\mu J}{2v_\phi}\left(\ov{L_L}\gamma^\mu L_L+\ov{e_R}\gamma^\mu e_R+\frac12\ov{N}\gamma^\mu \gamma_5 N\right)+\\
&+\dfrac{J}{v_\phi}\left(\dfrac{g^{\prime2}\,c_B}{16\pi^2}B_{\mu\nu}\widetilde{B}^{\mu\nu}+\dfrac{g^2\,c_W}{16\pi^2}W^i_{\mu\nu}\widetilde{W}^{i\mu\nu}\right)\,,
\end{split}
\label{Jops-typeI}
\ee
where we adopted the Majorana notation for the heavy neutral leptons (HNLs), $N\equiv N_R+N_R^c$. The terms proportional to $J/v_\phi$ are the so-called anomalous terms and arise due to the fact that the transformation in Eq.~\eqref{rotJ} is anomalous under $SU(2)_L\times U(1)_Y$. Their coefficients are defined as
\be
c_B=\sum_\psi \eta_\psi\, Y_\psi^2\, d(R_\psi)\qquad\text{and}\qquad
c_W=\sum_\psi \eta_\psi\, T(R_\psi)\,,
\label{eq:anomalyNORM}
\ee
where the sums run over all the Weyl fields rotated in Eq.~\eqref{rotJ}, with $\eta_\psi=+1\,(-1)$ for LH (RH) chiralities, $Y_\psi$ the hypercharge, $d(R_\psi)$ the dimension of the $SU(2)_L$ representation, and $T(R_\psi)$ its Dynkin index ($T=1/2$ for the fundamental). Summing over the $n_g=3$ generations yields
\be
(c_W,c_B)=\left(+\frac32,-\frac32\right)\,.
\ee
Note that the quarks are untouched by Eq.~\eqref{rotJ}, so no $J\,G\widetilde G$ coupling is generated and the Majoron is not a QCD axion. Since $c_W+c_B=0$, the photon projection cancels exactly and the Majoron has no coupling to two photons in the type~I case: as we will see in Sec.~\ref{sec:typeIII}, this is no longer true for type~III, and it constitutes the sharpest phenomenological discriminator among these two seesaw realizations.

The type~I Majoron Lagrangian is characterized by having two types of heavy d.o.f.: the radial mode $\rho$ and the HNLs $N$. The masses of these particles are both proportional to $v_\phi$, but, while $m_\rho$ is controlled by $\sqrt{\lambda_1}$, $M_N$ is instead weighted by $Y_{NN}$. Depending on the relative magnitude between these two quantities, we have a different hierarchy in the masses. Despite the fact that the low-energy effective Lagrangian is the same after integrating out both the radial mode and the HNLs, two different intermediate descriptions can be identified. In what follows, we will distinguish between the two cases $m_\rho > M_N$ and $M_N> m_\rho$. 

\boldmath
\subsubsection{The case $m_\rho > M_N$ and the NJSMEFT Lagrangian}
\label{sec:NJSMEFT}
\unboldmath

The radial mode $\rho$ is integrated out first, through its (nonlinear) equation of motion (EOM), solved perturbatively in inverse powers of $m_\rho$: the relevant $\rho$ Lagrangian, its EOM, and the perturbative solution $\rho=\rho^{(0)}+\rho^{(1)}+\dots$ are collected in Appendix~\ref{app:typeI}. Substituting the solution back, one obtains the NJSMEFT Lagrangian, valid at energies in the window $[M_N,\,m_\rho]$, including the SM spectrum, the Majoron $J$, and the HNLs, where with $M_N$ we refer here to the largest HNL mass, in a slight abuse of language. It can be written as
\begin{align}
\sL^\text{NJSMEFT}=&\sL_\text{SM}+\dfrac12\derp_\mu J\derp^\mu J+\dfrac12\ov{N}\left(i\slashed{\partial}-M_N\right)N
-\dfrac12\left(\overline{\cJ}_N N+\ov{N}\cJ_N\right)+\nn\\
&+\dfrac{\partial_\mu J}{2v_\phi}\left(\ov{L_L}\gamma^\mu L_L+\ov{e_R}\gamma^\mu e_R+\frac12\ov{N}\gamma^\mu\gamma_5 N\right)+
\label{NJSMEFTLag}\\
&-\dfrac{3\,J}{32\pi^2v_\phi}
\left(g^{\prime 2}B_{\mu\nu}\widetilde{B}^{\mu\nu}-g^2 W_{\mu\nu}^i\widetilde{W}^{i\mu\nu}\right)+\left(\sum_i \cC_i^{d}\, \cO_i^{d}+\hc\right)\,,\nn
\end{align}
where we introduced the leptonic source of the HNLs
\be
\label{eq:JNDEF}
\cJ_N\equiv Y_N^\dag\widetilde{H}^\dag L_L+Y_N^T\widetilde{H}^TL_L^c\,,
\ee
a shorthand adopted here and in the following to keep the equations compact. Wherever a current appears contracted directly with a fermion field rather than with its Dirac conjugate, as in the term $\overline{\cJ}_N N$ of Eq.~\eqref{NJSMEFTLag}, it is understood as $\overline{\cJ}\equiv\cJ^\dag\gamma^0$, so that the whole expression is a genuine Lorentz scalar. Here, $\cO_i^{d}$ is the operator of dimension $d$ that receives contributions from the integration out of $\rho$, and $\cC_i^{d}$ is the corresponding Wilson coefficient. It is worth commenting on the fact that the derivative Majoron couplings $\propto\derp_\mu J/2v_\phi$ in Eq.~\eqref{NJSMEFTLag} are also nonrenormalizable but are not listed within the sum of the $\cC_i^{d}$ operators, as they do not arise due to the integration out of the radial mode but as a consequence of the field redefinition. Therefore, we treat them separately.
Before listing the operators, it is worth fixing a convention that would otherwise be a recurrent source of factors of $2$. In this and in all the effective Lagrangians that follow, the operator sums are written as in the last line of Eq.~\eqref{NJSMEFTLag}, with the $+\hc$ always displayed explicitly, even when it is trivial. This is precisely the case for the self-Hermitian operators, $\cO_i^{d\dag}=\cO_i^d$, for which the $+\hc$ simply doubles the term: for those, and only for those, a compensating factor $1/2$ is included in the Wilson coefficient, so that $\cC_i^d$ is one-half of the physical matching value, $\cC_i^d=\kappa_i/2$, and $\kappa_i$ is the quantity to be compared with the literature. The lepton-number-violating operators, which are not self-Hermitian, carry instead a genuine and nonredundant $+\hc$, and their Wilson coefficients are the physical ones, $\cC_i^d=\kappa_i$, with no $1/2$. This is the case of the Weinberg operator and its dressings---that is, operators obtained by multiplying the Weinberg one by isosinglet quantities. We keep the redundant $+\hc$ rather than dropping it in the trivial cases, so that a single expression covers both classes of operators and the normalization of every $\cC_i^d$ is unambiguous. Where the comparison with the literature is most likely to be a source of confusion, we also indicate the physical value $\kappa_i$ explicitly.

The leading contributions to the $\cC_i^{d}$ operators are listed below.
\begin{itemize}
\item{\bf d=2}
\be
\begin{aligned}
\cO_{H^2}^{d=2} &= \left(H^\dag H\right)\,,\qquad\qquad&
\cC_{H^2}^{d=2}&=-\frac{\beta_1 m_\rho^2}{8\lambda_1} \,.
\end{aligned}
\label{eq:NJSMEFT-d2}
\ee
This operator is the threshold correction to the Higgs mass parameter $\mu_2^2$ of Eq.~\eqref{CompleteScalarPotential}: it is the explicit form of the $\phi$-VEV shift $\mu_2^2\to\mu_2^2-\beta_1 v_\phi^2/2$, usually reabsorbed into $\mu_2^2$. As for all self-Hermitian operators, $\cC_{H^2}^{d=2}$ carries the $1/2$ of the convention above, so that the physical shift is $2\,\cC_{H^2}^{d=2}=-\beta_1 v_\phi^2/2$. 
\item{\bf d=4}
\be
\begin{aligned}
\cO_{\lambda}^{d=4} &= \left(H^\dag H\right)^2\,,\qquad\qquad&
\cC_{\lambda}^{d=4}&=\frac{\beta_1^2}{8\lambda_1} \,.
\end{aligned}
\label{eq:NJSMEFT-d4}
\ee
This is trivially the quartic Higgs operator of the SM scalar potential, leading to a redefinition of the $\lambda_2$ coefficient in Eq.~\eqref{CompleteScalarPotential}.
\item{\bf d=5}
\be
\begin{aligned}
\cO_{NH}^{d=5} &= \left(\ov{N} N\right)_{\alpha\beta}\left(H^\dag H\right)\,,\qquad\qquad&
\cC_{NH}^{d=5}&= \frac{\beta_1}{8\sqrt{\lambda_1}m_\rho}\,\left(Y_{NN}\right)_{\alpha\beta}\,.
\end{aligned}
\label{eq:NJSMEFT-d5}
\ee
The Greek indices refer to the flavor contractions of the fermion bilinear $\ov{N}N$. This is one of the two $d=5$ of the $\nu$SMEFT Lagrangian~\cite{
Anisimov:2006hv,Graesser:2007yj,Graesser:2007pc,delAguila:2008ir,Aparici:2009fh,Liao:2016qyd,Bhattacharya:2015vja,Li:2021tsq} that is usually written in terms of the flavor RH neutrinos, $\left(\ov{N_R} N^c_R\right)\left(H^\dag H\right)$. 
\item{\bf d=6}
\be
\begin{aligned}
\cO_{H\Box}^{d=6} &= \left(H^\dag H\right)\Box\left(H^\dag H\right)\,,\qquad\qquad& 
\cC_{H\Box}^{d=6}&= -\dfrac{\beta_1^2}{8\lambda_1 m_\rho^2}\,,\\
\cO_{\Box JH}^{d=6} &= \left(\partial_\mu J\partial^\mu J\right)\left(H^\dag H\right)\,,\qquad\qquad&
\cC_{\Box JH}^{d=6}&=-\dfrac{\beta_1}{2m_\rho^2} \,,\\
\cO_{4N}^{d=6} &= \left(\ov{N} N\right)_{\alpha\beta}\left(\ov{N}N\right)_{\gamma\delta}\,,\qquad\qquad&
\cC_{4N}^{d=6}&=\dfrac{\left(Y_{NN}\right)_{\alpha\beta}\left(Y_{NN}\right)_{\gamma\delta}}{32m_\rho^2}\,.
\end{aligned}
\label{eq:NJSMEFT-d=6}
\ee
While the first operator, $\cO_{H\Box}^{d=6}$, belongs to the $d=6$ Warsaw basis of the SMEFT~\cite{Grzadkowski:2010es} and the last one, $\cO_{4N}^{d=6}$, can be found listed in the $\nu$SMEFT description~\cite{Liao:2016qyd}, the second operator, $\cO_{\Box JH}^{d=6}$, received much less attention. It appears in the specific scenario where the Majoron is considered a possible dark matter candidate, and it is referred as derivative Higgs portal~\cite{Balkin:2018tma} and more recently exploited for concrete Majoron dark matter phenomenology in~\cite{Akita:2026gzk,deGiorgi:2026jqn,Batell:2026avi}.
\item{\bf d=7}
\be
\begin{aligned}
\cO_{JNN}^{d=7} &= \left(\partial_\mu J\partial^\mu J\right)\left(\ov{N} N\right)_{\alpha\beta}\,,\qquad\qquad&
\cC_{JNN}^{d=7}&=-\dfrac{\sqrt{\lambda_1}}{4m_\rho^3}\left(Y_{NN}\right)_{\alpha\beta}\,,\\
\cO_{H\Box N}^{d=7} &= \left(H^\dag H\right)\Box\left(\ov{N}N\right)_{\alpha\beta}\,,\qquad\qquad&
\cC_{H\Box N}^{d=7}&=-\dfrac{\beta_1}{8\sqrt{\lambda_1}\,m_\rho^3}\left(Y_{NN}\right)_{\alpha\beta}\,,\\
\cO_{4H2N}^{d=7} &= \left(H^\dag H\right)^2\left(\ov{N}N\right)_{\alpha\beta}\,,\qquad\qquad&
\cC_{4H2N}^{d=7}&=\dfrac{\beta^2_1}{16\sqrt{\lambda_1}\,m_\rho^3}\left(Y_{NN}\right)_{\alpha\beta}\,.
\end{aligned}
\label{eq:NJSMEFT-d7}
\ee
To our knowledge, the first operator $\cO_{JNN}^{d=7}$ has never been considered before in the literature. On the other hand, the second operator $\cO_{H\Box N}^{d=7}$, that is obtained directly from the matching procedure, should be present in the basis of the $\nu$SMEFT Lagrangian in Ref.~\cite{Liao:2016qyd}, but this is not the case. Using integration by parts and the free equation of motion of $N$, it can be reexpressed in terms of the two independent operators of the $N^2H^2D^2$ class of the $\nu$SMEFT dimension-seven basis, up to a correction to $\cO_{NH}^{d=5}$ suppressed by $(M_N/m_\rho)^2$ relative to its original coefficient. Moreover, notice the pattern $\cC_{H\Box N}^{d=7}=-\cC_{NH}^{d=5}/m_\rho^2$, mirroring $\cC_{H\Box}^{d=6}=-\cC_{\lambda}^{d=4}/m_\rho^2$. 
\end{itemize} 

\boldmath
\subsubsection{The case $M_N > m_\rho$ and the $\rho$JSMEFT-I Lagrangian}
\label{sec:rhoJSMEFT-I}
\unboldmath

In this case, the HNLs are the most massive d.o.f., and we proceed along the same lines as the previous subsection but integrating out the $N$ fields and obtaining an effective description that includes the radial mode $\rho$. 

The relevant Lagrangian collects all the terms containing $N$. Using the current $\cJ_N$ of Eq.~\eqref{eq:JNDEF} and the axial Majoron current
\be
\cJ_J\equiv\dfrac{\slashed\derp J}{4v_\phi}\gamma_5\,,
\label{eq:JJDEF}
\ee
the $N$ EOM and its solution as a geometric series in $M_N^{-1}$ are collected in Appendix~\ref{app:typeI}. Inserting this solution back, we obtain the $\rho$JSMEFT-I effective description, valid in the energy window $[m_\rho,\,M_N]$, including the SM spectrum, the Majoron $J$, and the radial mode $\rho$; also here, in an abuse of notation, we refer with $M_N$ to the lightest HNL mass. The result reads
\begin{align}
\sL^\text{$\rho$JSMEFT-I}=&\sL_\text{SM}+\dfrac12\derp_\mu J\derp^\mu J+\dfrac12\derp_\mu \rho\derp^\mu \rho-\frac{\lambda_1}{4}\rho^4-\lambda_1v_\phi\rho^3-\frac{m_\rho^2}{2}\rho^2+
\label{rhoJSMEFTILag}\\
&+\frac{\left(\rho^2+2v_\phi\rho\right)}{2v_\phi^2}\derp_\mu J\derp^\mu J-\frac{\beta_1}{2}v_\phi^2\left(H^\dag H\right)-\beta_1v_\phi\rho\left(H^\dag H\right)-\frac{\beta_1}{2}\rho^2\left(H^\dag H\right)+\nn\\
&+\dfrac{\partial_\mu J}{2v_\phi}\left(\ov{L_L}\gamma^\mu L_L+\ov{e_R}\gamma^\mu e_R\right)+\sum_i \left(\cC_i^{d}\, \cO_i^{d}+\hc\right)\,,\nn
\end{align}
where we used the same notation and conventions as in Eq.~\eqref{NJSMEFTLag} (explicit $+\hc$ and the $1/2$ for self-Hermitian operators) for the effective operators originated by integrating out the HNLs.~\footnote{The $\beta_1(H^\dag H)$ terms appearing explicitly in Eq.~\eqref{rhoJSMEFTILag} are not part of this sum: they are not generated by integrating out $N$ but are inherited unchanged from the parent Lagrangian, exactly like the $\rho^3,\rho^4$ self-interactions alongside them. Instead, the same $\beta_1(H^\dag H)\phi^\dag\phi$ coupling reappeared as a genuine cataloged operator, $\cO_{H^2}^{d=2}$, in the complementary matching order of Sec.~\ref{sec:NJSMEFT}, where it is $\rho$ itself that has been integrated out. The same convention is used throughout, in the $\rho$JSMEFT-II and $\rho$JSMEFT-III Lagrangians below.} The leading operators and their corresponding Wilson coefficients are the following:
\begin{itemize}
\item{\bf d=5}
\be
\begin{aligned}
\cO_{W}^{d=5} &= \left(\ov{L_L}\widetilde{H}\right)_\alpha\left({\widetilde{H}}^TL^c_L\right)_\beta\,,\qquad&
\cC_{W}^{d=5} &= \dfrac12 \left(Y_N\,M_N^{-1}\,Y_N^T\right)_{\alpha\beta}\,.
\label{eq:rhoJSMEFT-I-d5}
\end{aligned}
\ee
This is the well-known Weinberg operator~\cite{Weinberg:1979sa} responsible for the active neutrino masses, $m_\nu=-\cC_{W}^{d=5}v^2$. Throughout, we adopt the standard convention in which the light-neutrino Majorana mass term reads $\sL\supset-\frac12(m_\nu)_{\alpha\beta}\,\ov{\nu_{L\alpha}^c}\,\nu_{L\beta}+\hc$, with $\vev{H^0}=v/\sqrt2$ and $v\simeq246\GeV$.
\item{\bf d=6}
\be
\begin{aligned}
\cO_{U}^{d=6} &= \left(\ov{L_L}\widetilde{H}\right)_\alpha i\slashed{\partial}\left({\widetilde{H}}^\dag L_L\right)_\beta\,,\qquad&
\cC_{U}^{d=6}&= \dfrac12\left(Y_N M_N^{-1}\right)_{\alpha\gamma}\left(M_N^{-1}Y_N^\dagger\right)_{\gamma\beta}\,,\\
\cO_{W\rho}^{d=6} &= \left(\ov{L_L}\widetilde{H}\right)_\alpha\left({\widetilde{H}}^TL^c_L\right)_\beta\rho\,,\qquad&
\cC_{W\rho}^{d=6} &=-\dfrac{1}{2v_\phi}\left(Y_N M_N^{-1} Y_N^T\right)_{\alpha\beta}=-\dfrac{\cC_{W}^{d=5}}{v_\phi}\,.
\end{aligned}
\label{eq:rhoJSMEFT-I-ds6}
\ee
The first operator, $\cO_{U}^{d=6}$, describes the violation of the unitarity associated to the Pontecorvo-Maki-Nakagawa-Sakata (PMNS) matrix~\cite{Broncano:2002rw}, and it has been deeply studied for its possible effects in electroweak precision observables, lepton-flavor-universality ratios, charged-lepton-flavor-violating processes, neutrino oscillation probabilities, and neutrinoless double beta decay (see Ref.~\cite{Blennow:2023mqx} for a recent review). This operator can be easily written in terms of the operators of the $d=6$ Warsaw SMEFT basis using EOMs and Fierz identities as also discussed in Ref.~\cite{Broncano:2002rw}. Being self-Hermitian, $\cC_U^{d=6}$ carries the $1/2$ of our convention: the physical value quoted in the literature is $2\,\cC_U^{d=6}=(Y_NM_N^{-1})(M_N^{-1}Y_N^\dag)$.

The second operator is the Weinberg one but with the additional insertion of the radial mode. Notice that its Wilson coefficient turns out to be suppressed by the active neutrino masses through the presence of $\cC_{W}^{d=5}$ coefficient.
\item{\bf d=7}
\begin{align}
\cO_{JHL}^{d=7} &= \left(\ov{L_L}\widetilde{H}\right)_\alpha\slashed\partial J \gamma_5 \left({\widetilde{H}}^\dag L_L\right)_\beta\,,\qquad&
\cC_{JHL}^{d=7}&=\dfrac{1}{4v_\phi}\left(Y_N M_N^{-1}\right)_{\alpha\gamma}\left(M_N^{-1} Y_N^\dag\right)_{\gamma\beta}\nn\,,\\
\cO_{DHL}^{d=7} &= \left(\ov{L_L}\widetilde{H}\right)_\alpha D^2 \left({\widetilde{H}}^T L^c_L\right)_\beta\,,\qquad&
\cC_{DHL}^{d=7}&=-\dfrac{\cC_W^{d=5}}{M_N^2}=-\dfrac{1}{2M_N^2}\left(Y_N M_N^{-1}Y_N^T\right)_{\alpha\beta}\nn\,,\\
\cO_{U\rho}^{d=7} &= \left(\ov{L_L}\widetilde{H}\right)_\alpha i\slashed\partial\left({\widetilde{H}}^\dag L_L\right)_\beta\rho\,,\qquad&
\cC_{U\rho}^{d=7}&=-\dfrac{2\,\cC_U^{d=6}}{v_\phi}\,,
\label{eq:rhoJSMEFT-I-d7}\\
\cO_{W\rho\rho}^{d=7} &= \left(\ov{L_L}\widetilde{H}\right)_\alpha\left({\widetilde{H}}^T L^c_L\right)_\beta\rho^2\,,\qquad&
\cC_{W\rho\rho}^{d=7}&=+\dfrac{\cC_W^{d=5}}{v_\phi^2}\,.\nn
\end{align}
The $\rho$-dressed coefficients are not independent. The heavy mass entering Eq.~\eqref{Jops-typeI} is not $M_N$ but the $\rho$-dependent combination
\be
\label{eq:MNrho}
M_N(\rho)=M_N\left(1+\dfrac{\rho}{v_\phi}\right)\,,
\ee
since the radial mode enters the HNL mass exactly as the $\phi$ VEV does. Keeping that dependence unexpanded, the coefficients above resum into
\be
\label{eq:Iresum}
\cC_W^{d=5}(\rho)=\dfrac12\dfrac{Y_NY_N^T}{M_N(\rho)}=\dfrac{\cC_W^{d=5}}{1+\rho/v_\phi}\,,\qquad
\cC_U^{d=6}(\rho)=\dfrac12\dfrac{Y_NY_N^\dag}{M_N^2(\rho)}=\dfrac{\cC_U^{d=6}}{\left(1+\rho/v_\phi\right)^2}\,,
\ee
where the different powers in the denominator simply count how many heavy propagators each operator contains: one for $\cO_W$, two for $\cO_U$. Expanding in $\rho$ reproduces every dressing at once,
$\cC_{W\rho}^{d=6}=-\cC_W^{d=5}/v_\phi$, $\cC_{W\rho\rho}^{d=7}=+\cC_W^{d=5}/v_\phi^2$, and $\cC_{U\rho}^{d=7}=-2\,\cC_U^{d=6}/v_\phi$, the relative factor $2$ in the last one being nothing but the second power in Eq.~\eqref{eq:Iresum}. The first operator, $\cO_{JHL}^{d=7}$, is instead the leading Majoron operator, generated by the single insertion of the axial Majoron current $2\cJ_J$. Being self-Hermitian, its coefficient carries the convention $1/2$, and it is related to the (equally $1/2$-carrying) unitarity coefficient by $\cC_{JHL}^{d=7}=\cC_{U}^{d=6}/(2v_\phi)$. The second operator, $\cO_{DHL}^{d=7}$, is the derivative ($D^2$) dressing of the Weinberg operator, generated by two insertions of $i\slashed\partial$ in the fermionic propagator; it is lepton number violating (hence, with the genuine $+\hc$, no $1/2$) and its coefficient is the Weinberg one further suppressed by $M_N^2$, $\cC_{DHL}^{d=7}=-\cC_W^{d=5}/M_N^2$, mirroring the type~II and type~III cases. The last two operators are the radial-mode dressings of the unitarity and Weinberg operators, respectively, generated by the insertions of $2\cJ_\rho\propto\rho$.
\end{itemize}

\subsection{The type III Majoron model}
\label{sec:typeIII}
The type~III Majoron model is the generalization of the traditional type~III seesaw mechanism~\cite{Foot:1988aq}, in which the active neutrino masses are generated through the exchange of fermionic $SU(2)_L$ triplets with vanishing hypercharge, and has been considered previously in Ref.~\cite{Cheng:2020rla}. As in the type~I case, we supplement it with the same LN-breaking complex scalar singlet $\phi$, charged under the global $U(1)_L$, whose spontaneous breaking gives rise to the Majoron~\cite{Chikashige:1980qk,Chikashige:1980ui}. Beyond the SM fields, the spectrum contains the scalar $\phi$ and the RH fermionic triplets $\Sigma_R$, whose quantum numbers are collected in Table~\ref{tab:typeIII-spectrum}. As anticipated in Sec.~\ref{sec:typeI}, and in strict analogy with the $n_N=3$ RH neutrinos of the type~I case, we explicitly consider $n_\Sigma=3$ fermionic triplets. This choice is immaterial for the tree-level Wilson coefficients and fixes only the normalization of the anomalous couplings.

\begin{table}[ht!]
\centering
\begin{tabular}{@{}lcccc@{}}
\toprule
Field & Spin & $SU(3)_c\times SU(2)_L\times U(1)_Y$ & $U(1)_L$ \\
\midrule
$\phi$      & $0$   & $(\mathbf{1},\mathbf{1},0)$ & $-2$ \\
$\Sigma_R$  & $1/2$ & $(\mathbf{1},\mathbf{3},0)$ & $+1$ \\
\bottomrule
\end{tabular}
\caption{\em Field content of the type~III Majoron model beyond the SM spectrum.}
\label{tab:typeIII-spectrum}
\end{table}

The Lagrangian we consider is given by
\begin{align}
    \sL^\text{type~III}=&\sL_\textrm{SM}+ \overline{\vec{\Sigma}_R}i\slashed{D} \vec{\Sigma}_R+\derp_\mu\phi^\ast\derp^\mu\phi-V(H,\phi)-\left(\frac{1}{2}\phi\,\overline{\vec{\Sigma}_R^c}Y_\Sigma \vec{\Sigma}_R+\overline{\vec{\Sigma}_R}Y_\chi \widetilde{H}^\dagger \vec{\sigma}L_L+\hc\right)\,,
    \label{eq:Lag-typeIII}
\end{align}
where $\vec{\Sigma}_R=(\Sigma_1,\Sigma_2,\Sigma_3)_R$ is the vector collecting the components of the triplet, $\vec{\sigma}$ is the vector of Pauli matrices, and $L_L$ is the SM lepton doublet. As in the previous cases, we work in the basis where the charged-lepton Yukawa is real and diagonal, and we take the Majorana coupling $Y_\Sigma$ to be real and diagonal, while $Y_\chi$ remains a generic complex $3\times 3$ matrix. $\sL_\textrm{SM}$ contains all the SM terms except the Higgs potential, which is subsumed, together with the terms involving $\phi$, into the complete scalar potential $V(H,\phi)$ that coincides with the one of the type~I case, Eq.~\eqref{CompleteScalarPotential}. The same considerations done before also apply here, and, thus, the minimization of the potential can be carried out in subsequent steps, with $\phi$ acquiring VEV before $H$ and leading to Eqs.~\eqref{eq:vphiDEF}--\eqref{mrhoDEF}. After $U(1)_L$ is spontaneously broken, the fermion triplets acquire the Majorana mass
\begin{equation}
    M_\Sigma = \dfrac{Y_\Sigma v_\phi}{\sqrt{2}}\,.
\end{equation}
The mass of the heavy mediator originates entirely from the LN SSB: above $v_\phi$ the triplet is massless, so that $\Sigma$ admits no bare-mass regime and only the two hierarchies $m_\rho\gtrless M_\Sigma$ are possible, both scales being set by $v_\phi$ and controlled, respectively, by $\lambda_1$ and $Y_\Sigma$.

The terms of the Lagrangian involving $\phi$ and $\Sigma$ can be rewritten, after the LN SSB but before removing the Majoron phase, as
\begin{align}
\sL^\text{type~III}\supset&\ \frac12\derp_\mu\rho\derp^\mu\rho+\frac{(v_\phi+\rho)^2}{2v_\phi^2}\derp_\mu J\derp^\mu J+\mu_1^2 \frac{(v_\phi+\rho)^2}{2}-\lambda_1 \frac{(v_\phi+\rho)^4}{4}+\label{eq:SSBLagTypeIII}\\
&-\beta_1 (H^\dagger H) \frac{(v_\phi+\rho)^2}{2}+\overline{\vec{\Sigma}_R}\,i\slashed{D}\,\vec{\Sigma}_R-\left(\frac12\frac{(v_\phi+\rho)}{\sqrt2}e^{iJ/v_\phi}\overline{\vec{\Sigma}_R^c}Y_\Sigma\vec{\Sigma}_R+\hc\right)+\nn\\
&-\left(\overline{\vec{\Sigma}_R}Y_\chi\widetilde{H}^\dagger\vec{\sigma}L_L+\hc\right)\,,\nn
\end{align}
where the first five terms coincide with the $\rho$-$J$ sector already found in the type~I case, Eq.~\eqref{eq:SSBLagTypeI} [and, as we will see, in the type~II one, too, Eq.~\eqref{eq:SSBLagTypeII}].

The $J$ dependence in the Yukawa term proportional to $Y_\Sigma$ can be rotated away by performing on the lepton fields and on the triplet the vectorial transformation
\begin{equation}
    \left\{L_L,\Sigma_R,{e}_R\right\}\rightarrow\left\{L_L,\Sigma_R,{e}_R\right\}e^{-iJ/2v_\phi} \, .
\end{equation}
As in the type~I case, this reintroduces the $J$ dependence in the chirality-preserving kinetic terms, and, using the Majorana notation for the heavy fermions $\vec{\Sigma}\equiv\vec{\Sigma}_R+\vec{\Sigma}_R^c$, the Lagrangian can be rewritten in terms of the new fields as
\begin{align}
\sL^\text{type~III}\supset &\ \mu^2_2 H^\dagger H -\lambda_2(H^\dagger H)^2 +\mu_1^2 \frac{(v_\phi+\rho)^2}{2}-\lambda_1 \frac{(v_\phi+\rho)^4}{4}-\beta_1 (H^\dagger H) \frac{(v_\phi+\rho)^2}{2}+\nn\\
    &+\frac{1}{2}\partial_\mu \rho \partial^\mu \rho +\frac{1}{2}\partial_\mu J \partial^\mu J+\frac{(\partial_\mu J)^2}{v_\phi}\Big(\rho + \frac{\rho^2}{2v_\phi}\Big)+\nn\\
    &+\frac{1}{2}\overline{\vec{\Sigma}}(i\slashed{D}-M_\Sigma)\vec{\Sigma}-\frac{\rho}{2v_\phi}\overline{\vec{\Sigma}}M_\Sigma\vec{\Sigma}-\frac{1}{2}\left(\overline{\vec{\Sigma}}\cdot\vec{\cJ}_\Sigma+\overline{\vec{\cJ}}_\Sigma\cdot\vec{\Sigma}\right)+\label{Jops-typeIII}\\
    &+\frac{\partial_\mu J}{2v_\phi}\left(\frac{1}{2}\overline{\vec{\Sigma}}\gamma^\mu \gamma^5\vec{\Sigma}+\overline{L_L}\gamma^\mu L_L +\overline{e_R}\gamma^\mu e_R\right)+\nn\\
    &-\frac{3\,J}{32\pi^2 v_\phi}\left(g^{\prime2}B_{\mu\nu}\widetilde{B}^{\mu\nu}+3\,g^2 W^i_{\mu\nu}\widetilde{W}^{i\mu\nu}\right)\,,\nn
\end{align}
where we introduced the leptonic source of the triplets, the type~III counterpart of the current $\cJ_N$ of Eq.~\eqref{eq:JNDEF},
\begin{equation}
\label{eq:JSigmaDEF}
    \cJ_\Sigma^I\equiv Y_\chi \widetilde{H}^\dagger \sigma^IL_L+Y_\chi^\ast \widetilde{H}^T\sigma^{I\,T} L_L^c\,,
\end{equation}
with $\vec{\cJ}_\Sigma=(\cJ_\Sigma^1,\cJ_\Sigma^2,\cJ_\Sigma^3)$ and $\ov{\vec\Sigma}\cdot\vec{\cJ}_\Sigma\equiv\ov{\Sigma}^I\cJ_\Sigma^I$, the adjoint index $I$ being summed. As in the type~I case, this shorthand notation is adopted throughout to keep the equations compact.

The terms in the fourth line arise from the fermion rotation, exactly as in Eq.~\eqref{Jops-typeI} of the type~I case: the leptonic couplings reproduce the same structure found there, while the heavy triplet contributes to the axial current $\frac12\overline{\vec\Sigma}\gamma^\mu\gamma^5\vec\Sigma$, the extra factor $1/2$ following from the Majorana nature of $\Sigma$. The last line is again the anomalous term but now with a crucial difference with respect to type~I: since the rotated $\Sigma_R$ is an $SU(2)_L$ \emph{triplet} (with $Y_\Sigma=0$) whose mass originates entirely from $v_\phi$, it contributes to the $SU(2)_L^2$ anomaly, so the coefficient of $g^2 W\widetilde W$ is modified, while the $g'^2 B\widetilde B$ one is unchanged with respect to type I.

In terms of the coefficients defined in Eq.~\eqref{eq:anomalyNORM}, the counting goes as follows. Being a gauge singlet under $U(1)_Y$, $\Sigma_R$ leaves $c_B$ untouched, so that $c_B=-3/2$ exactly as in type~I. For $c_W$, instead, each RH triplet contributes $\eta\,T(\text{adj})=(-1)\cdot2=-2$ [with $\eta=-1$ for the right-handed $\Sigma_R$, as defined below Eq.~\eqref{eq:anomalyNORM}, and $T(\text{adj})=2$ the Dynkin index of the $SU(2)_L$ adjoint representation], so that
\be
c_W=n_g\cdot\dfrac12-2\,n_\Sigma=\dfrac32-6=-\dfrac92\,,\qquad\qquad c_B=-n_g\cdot\dfrac12=-\dfrac32\,,
\ee
having set $n_g=n_\Sigma=3$. This is what Eq.~\eqref{Jops-typeIII} displays, and it makes the contrast with type~I sharp: the ratio $c_W/c_B=+3$, to be compared with $c_W/c_B=-1$ of types~I and~II, is independent of the overall generation counting and reads $c_W/c_B=-1+4n_\Sigma/n_g$ for a generic number of triplets. The physically relevant consequence is that now $c_{\gamma\gamma}\equiv c_W+c_B=-2n_\Sigma=-6\neq0$: projecting onto the photon ($g^{\prime2}\cos^2\theta_W=g^2\sin^2\theta_W=e^2$), the Majoron acquires a genuine two-photon coupling
\be
\sL\supset\dfrac{e^2}{16\pi^2}\dfrac{J}{v_\phi}\,c_{\gamma\gamma}\,F_{\mu\nu}\widetilde{F}^{\mu\nu}=-\dfrac{3\,e^2}{8\pi^2}\dfrac{J}{v_\phi}F_{\mu\nu}\widetilde{F}^{\mu\nu}=-\dfrac{\alpha}{2\pi}\dfrac{J}{v_\phi}\,n_\Sigma\,F_{\mu\nu}\widetilde{F}^{\mu\nu}\,,
\ee
which is absent in types~I and~II, where $c_W+c_B=0$. It originates from the charged components $\Sigma^\pm$ of the triplet, and it is a genuine physical discriminator among the three seesaw realizations. Note that there is no ambiguity in the light-heavy split here: the $\Sigma$, whose mass comes entirely from $v_\phi$, decouples leaving its full Wess-Zumino coefficient behind, while the light leptons contribute $c_{\gamma\gamma}=0$ on their own, a difference already noted in Ref.~\cite{Sun:2021jpw}.

The type~III Majoron Lagrangian is therefore characterized, as in the type~I case, by two heavy d.o.f.: the radial mode $\rho$ and the fermion triplets $\Sigma$. Since both masses are proportional to $v_\phi$, their ratio is controlled only by $\lambda_1$ and $Y_\Sigma$, and, depending on the relative magnitude, two mass hierarchies, $m_\rho > M_\Sigma$ or $M_\Sigma > m_\rho$, are possible. While the final low-energy EFT is the same, the intermediate regime allows for two different descriptions, corresponding to the two type~III entries in Table~\ref{tab:EFTsummary}, which we characterize in the following.

\boldmath
\subsubsection{The case $m_\rho > M_\Sigma$ and the $\Sigma$JSMEFT Lagrangian}
\label{sec:SigmaJSMEFT}
\unboldmath

When the radial mode is the heaviest d.o.f., $m_\rho > M_\Sigma$, it is integrated out first, following exactly the procedure of the NJSMEFT case in Sec.~\ref{sec:NJSMEFT}, with the role of the HNLs now played by the fermion triplets $\Sigma$. The terms of Eq.~\eqref{Jops-typeIII} involving $\rho$, the corresponding EOM, and its perturbative solution are collected in Appendix~\ref{app:typeIII}. Substituting the solution back yields the $\Sigma$JSMEFT Lagrangian, valid in the energy window $[M_\Sigma,\,m_\rho]$ and containing the SM spectrum, the Majoron $J$, and the fermion triplets $\Sigma$:
\begin{align}
\sL^\text{$\Sigma$JSMEFT}=&\sL_\text{SM}+\dfrac12\derp_\mu J\derp^\mu J+\dfrac12\ov{\vec{\Sigma}}\left(i\slashed{D}-M_\Sigma\right)\vec{\Sigma}
-\dfrac12\left(\ov{\vec{\Sigma}}\cdot\vec{\cJ}_\Sigma+\overline{\vec{\cJ}}_\Sigma\cdot\vec{\Sigma}\right)+\label{SigmaJSMEFTLag}\\
&+\dfrac{\partial_\mu J}{2v_\phi}\left(\frac12\ov{\vec{\Sigma}}\gamma^\mu\gamma_5\vec{\Sigma}+\ov{L_L}\gamma^\mu L_L+\ov{e_R}\gamma^\mu e_R\right)+\sum_i \left(\cC_i^{d}\, \cO_i^{d}+\hc\right)\,,\nn
\end{align}
where, as in the NJSMEFT case of Eq.~\eqref{NJSMEFTLag}, the source term $\ov{\vec{\Sigma}}\cdot\vec{\cJ}_\Sigma$ in the first line and the nonrenormalizable derivative Majoron couplings in the second are not generated by the integration out of $\rho$ and are, therefore, kept separate from the sum $\sum_i\cC_i^d\cO_i^d$. In particular, the leptonic Dirac Yukawa $Y_\chi$, which is responsible for the active neutrino masses, does not enter the $\rho$ dynamics. The generation of the Weinberg operator is, thus, postponed to the subsequent integration out of the triplet $\Sigma$. The leading operators induced by the $\rho$ exchange are the direct type~III analogs of the NJSMEFT ones, Eqs.~\eqref{eq:NJSMEFT-d4}--\eqref{eq:NJSMEFT-d7}, obtained with $N\to\Sigma$ and $Y_{NN}\to Y_\Sigma$.

As in the NJSMEFT and $\Delta$JSMEFT, the sum is understood with the explicit $+\hc$ and the $1/2$ for self-Hermitian operators. All the operators below are self-Hermitian, so their coefficients coincide (with $N\to\vec\Sigma$, $Y_{NN}\to Y_\Sigma$) with the NJSMEFT ones, Eqs.~\eqref{eq:NJSMEFT-d2}--\eqref{eq:NJSMEFT-d7}, and carry the same $1/2$.
\begin{itemize}
\item{\bf d=2}
\be
\begin{aligned}
\cO_{H^2}^{d=2} &= \left(H^\dag H\right),\qquad\qquad&
\cC_{H^2}^{d=2}&=-\frac{\beta_1m_\rho^2}{8\lambda_1} \,.
\end{aligned}
\label{eq:SigmaJSMEFT-d2}
\ee
\item{\bf d=4}
\be
\begin{aligned}
\cO_{\lambda}^{d=4} &= \left(H^\dag H\right)^2,\qquad\qquad&
\cC_{\lambda}^{d=4}&=\frac{\beta_1^2}{8\lambda_1}\,.
\end{aligned}
\label{eq:SigmaJSMEFT-d4}
\ee
As in all the previous cases, this is the quartic Higgs operator, redefining the $\lambda_2$ coefficient in Eq.~\eqref{CompleteScalarPotential}.
\item{\bf d=5}
\be
\begin{aligned}
\cO_{\Sigma H}^{d=5} &= \left(\ov{\vec{\Sigma}}\vec{\Sigma}\right)_{\alpha\beta}\left(H^\dag H\right),\qquad\qquad&
\cC_{\Sigma H}^{d=5}&= \frac{\beta_1}{8\sqrt{\lambda_1}m_\rho}\,\left(Y_\Sigma\right)_{\alpha\beta}\,,
\end{aligned}
\label{eq:SigmaJSMEFT-d5}
\ee
where the Greek indices refer to the flavor contraction and the $SU(2)_L$ adjoint index of the triplet bilinear is understood to be contracted, $\ov{\vec{\Sigma}}\vec{\Sigma}\equiv\ov{\Sigma}^I\Sigma^I$. This is the type~III counterpart of the $\nu$SMEFT operator $\cO_{NH}^{d=5}$ in Eq.~\eqref{eq:NJSMEFT-d5}~\cite{Liao:2016qyd}, now built out of the fermion triplet.
\item{\bf d=6}
\be
\begin{aligned}
\cO_{H\Box}^{d=6} &= \left(H^\dag H\right)\Box\left(H^\dag H\right),\qquad\qquad&
\cC_{H\Box}^{d=6}&= -\dfrac{\beta_1^2}{8\lambda_1 m_\rho^2},\\
\cO_{\Box JH}^{d=6} &= \left(\partial_\mu J\partial^\mu J\right)\left(H^\dag H\right),\qquad\qquad&
\cC_{\Box JH}^{d=6}&=-\dfrac{\beta_1}{2m_\rho^2},\\
\cO_{\Sigma\Sigma\Sigma\Sigma}^{d=6} &= \left(\ov{\vec{\Sigma}}\vec{\Sigma}\right)_{\alpha\beta}\left(\ov{\vec{\Sigma}}\vec{\Sigma}\right)_{\gamma\delta},\qquad\qquad&
\cC_{\Sigma\Sigma\Sigma\Sigma}^{d=6}&=\dfrac{\left(Y_\Sigma\right)_{\alpha\beta}\left(Y_\Sigma\right)_{\gamma\delta}}{32m_\rho^2}\,.
\end{aligned}
\label{eq:SigmaJSMEFT-d6}
\ee
The first operator, $\cO_{H\Box}^{d=6}$, belongs to the $d=6$ Warsaw basis of the SMEFT~\cite{Grzadkowski:2010es}, and the second, $\cO_{\Box JH}^{d=6}$, is the derivative Higgs portal~\cite{Balkin:2018tma} already found in the NJSMEFT and $\Delta$JSMEFT cases with the very same coefficient, while the four-fermion operator $\cO_{\Sigma\Sigma\Sigma\Sigma}^{d=6}$ is the triplet analog of $\cO_{NNNN}^{d=6}$ in Eq.~\eqref{eq:NJSMEFT-d=6}.
\item{\bf d=7}
\be
\begin{aligned}
\cO_{J\Sigma\Sigma}^{d=7} &= \left(\partial_\mu J\partial^\mu J\right)\left(\ov{\vec{\Sigma}}\vec{\Sigma}\right)_{\alpha\beta},\qquad\qquad&
\cC_{J\Sigma\Sigma}^{d=7}&=-\dfrac{\sqrt{\lambda_1}}{4m_\rho^3}\left(Y_\Sigma\right)_{\alpha\beta},\\
\cO_{H\Box\Sigma}^{d=7} &= \left(H^\dag H\right)\Box\left(\ov{\vec{\Sigma}}\vec{\Sigma}\right)_{\alpha\beta},\qquad\qquad&
\cC_{H\Box\Sigma}^{d=7}&=-\dfrac{\beta_1}{8\sqrt{\lambda_1}\,m_\rho^3}\left(Y_\Sigma\right)_{\alpha\beta},\\
\cO_{HH\Sigma\Sigma}^{d=7}&=(H^\dagger H)^2 \left(\ov{\vec{\Sigma}}\vec{\Sigma}\right)_{\alpha\beta},\qquad\qquad& \cC_{HH\Sigma\Sigma}^{d=7}&=\frac{\beta_1^2}{16\sqrt{\lambda_1}m_\rho^3}\left(Y_\Sigma\right)_{\alpha\beta}\,.
\end{aligned}
\label{eq:SigmaJSMEFT-d7}
\ee
Exactly as in the NJSMEFT case, and to the best of our knowledge, the first operator $\cO_{J\Sigma\Sigma}^{d=7}$ has never been considered before in the literature, while the second, $\cO_{H\Box\Sigma}^{d=7}$, follows the same pattern $\cC_{H\Box\Sigma}^{d=7}=-\cC_{\Sigma H}^{d=5}/m_\rho^2$, mirroring $\cC_{H\Box}^{d=6}=-\cC_{\lambda}^{d=4}/m_\rho^2$; the third, $\cO_{HH\Sigma\Sigma}^{d=7}$, is the triplet analog of $\cO_{4H2N}^{d=7}$.
\end{itemize}

\boldmath
\subsubsection{The case $M_\Sigma > m_\rho$ and the $\rho$JSMEFT-III Lagrangian}
\unboldmath
\label{sec:rhoJSMEFT-III}

When the fermion triplets are the heaviest d.o.f., $M_\Sigma>m_\rho$, they are integrated out first, keeping the radial mode $\rho$ dynamical. The procedure is the exact type~III analog of the $\rho$JSMEFT-I case in Sec.~\ref{sec:rhoJSMEFT-I}, with the HNLs replaced by the triplets and the axial Majoron current $\cJ_J$ in Eq.~\eqref{eq:JJDEF}. The Lagrangian collecting the terms in Eq.~\eqref{Jops-typeIII} involving $\Sigma$ [built out of the triplet current $\vec{\cJ}_\Sigma$ in Eq.~\eqref{eq:JSigmaDEF} and the radial-mode insertion $\cJ_\rho$], the $\Sigma$ EOM and its geometric-series solution in $M_\Sigma^{-1}$ are collected in Appendix~\ref{app:typeIII}. Inserting the solution back, we obtain the $\rho$JSMEFT-III Lagrangian, valid in the window $[m_\rho,\,M_\Sigma]$ and containing the SM spectrum, the Majoron $J$, and the radial mode $\rho$:
\begin{align}
\sL^\text{$\rho$JSMEFT-III}=&\sL_\text{SM}+\dfrac12\derp_\mu J\derp^\mu J+\dfrac12\derp_\mu \rho\derp^\mu \rho-\frac{\lambda_1}{4}\rho^4-\lambda_1v_\phi\rho^3-\frac{m_\rho^2}{2}\rho^2+\label{rhoJSMEFTIIILag}\\
&+\frac{\left(\rho^2+2v_\phi\rho\right)}{2v_\phi^2}\derp_\mu J\derp^\mu J-\frac{\beta_1}{2}v_\phi^2\left(H^\dag H\right)-\beta_1v_\phi\rho\left(H^\dag H\right)-\frac{\beta_1}{2}\rho^2\left(H^\dag H\right)+\nn\\
&+\dfrac{\partial_\mu J}{2v_\phi}\left(\ov{L_L}\gamma^\mu L_L+\ov{e_R}\gamma^\mu e_R\right)+\sum_i \left(\cC_i^{d}\, \cO_i^{d}+\hc\right)\,,\nn
\end{align}
which has exactly the same $\rho$-$J$-SM structure of the $\rho$JSMEFT-I Lagrangian, Eq.~\eqref{rhoJSMEFTILag}. The operators generated by the integration out of $\Sigma$ are the type~III analogs of the $\rho$JSMEFT-I ones, Eqs.~\eqref{eq:rhoJSMEFT-I-d5}--\eqref{eq:rhoJSMEFT-I-d7}, now built out of the $SU(2)_L$-triplet current $\cJ_\Sigma^I$ (the adjoint index $I$ is understood to be contracted). The leading ones read as follows.
\begin{itemize}
\item{\bf d=5}
\be
\begin{aligned}
\cO_{W}^{d=5} &= \left(\ov{L^c_L}\sigma^I\widetilde{H}^\ast\right)_\alpha\left({\widetilde{H}}^\dag\sigma^I L_L\right)_\beta,\qquad&
\cC_{W}^{d=5} &= \dfrac12 \left(Y_\chi^T\,M_\Sigma^{-1}\,Y_\chi\right)_{\alpha\beta}.
\end{aligned}
\label{eq:rhoJSMEFT-III-d5}
\ee
This is the Weinberg operator~\cite{Weinberg:1979sa} in its type~III realization, responsible for the active neutrino masses, $m_\nu=-\cC_W^{d=5}v^2=-\tfrac12 Y_\chi^TM_\Sigma^{-1}Y_\chi\,v^2$.
\item{\bf d=6}
\be
\begin{aligned}
\cO_{U}^{d=6} &= \left(\ov{L_L}\sigma^I\widetilde{H}\right)_\alpha i\slashed{D}\left({\widetilde{H}}^\dag\sigma^I L_L\right)_\beta,\qquad&
\cC_{U}^{d=6}&= \dfrac12\left(Y_\chi^\dagger M_\Sigma^{-1}\right)_{\alpha\gamma}\left(M_\Sigma^{-1}Y_\chi\right)_{\gamma\beta},\\
\cO_{W\rho}^{d=6} &= \left(\ov{L^c_L}\sigma^I\widetilde{H}^\ast\right)_\alpha\left({\widetilde{H}}^\dag\sigma^I L_L\right)_\beta\rho,\qquad&
\cC_{W\rho}^{d=6} &=-\dfrac{1}{v_\phi}\,\cC_{W}^{d=5}.
\end{aligned}
\label{eq:rhoJSMEFT-III-d6}
\ee
The first operator, $\cO_{U}^{d=6}$, is the type~III counterpart of the unitarity operator of Eq.~\eqref{eq:rhoJSMEFT-I-ds6}: it encodes the nonunitarity of the PMNS matrix induced by the heavy triplets, and, being built out of a charged $SU(2)_L$ triplet, it modifies both the neutral- and charged-lepton gauge couplings~\cite{Broncano:2002rw,Abada:2007ux,Blennow:2023mqx}. Being self-Hermitian, $\cC_U^{d=6}$ carries the convention $1/2$ (physical value $2\cC_U^{d=6}$), exactly as in the type~I case. The second operator is the Weinberg one dressed with a radial mode, with $\cC_{W\rho}^{d=6}=-\cC_W^{d=5}/v_\phi$.
\item{\bf d=7}
\be
\begin{aligned}
\cO_{JHL}^{d=7} &= \left(\ov{L_L}\sigma^I\widetilde{H}\right)_\alpha\slashed\partial J \gamma_5 \left({\widetilde{H}}^\dag\sigma^I L_L\right)_\beta,\qquad&
\cC_{JHL}^{d=7}&=\dfrac{1}{2v_\phi}\,\cC_{U}^{d=6},\\
\cO_{DHL}^{d=7} &= \left(\ov{L^c_L}\sigma^I\widetilde{H}^\ast\right)_\alpha D^2\left({\widetilde{H}}^\dag\sigma^I L_L\right)_\beta,\qquad&
\cC_{DHL}^{d=7}&=-\dfrac{\cC_{W}^{d=5}}{M_\Sigma^2},\\
\cO_{U\rho}^{d=7} &= \left(\ov{L_L}\sigma^I\widetilde{H}\right)_\alpha i\slashed{D}\left({\widetilde{H}}^\dag\sigma^I L_L\right)_\beta\rho,\qquad&
\cC_{U\rho}^{d=7}&=-\dfrac{2\,\cC_U^{d=6}}{v_\phi},\\
\cO_{W\rho\rho}^{d=7} &= \left(\ov{L^c_L}\sigma^I\widetilde{H}^\ast\right)_\alpha\left({\widetilde{H}}^\dag\sigma^I L_L\right)_\beta\rho^2,\qquad&
\cC_{W\rho\rho}^{d=7}&=+\dfrac{\cC_W^{d=5}}{v_\phi^2}.
\end{aligned}
\label{eq:rhoJSMEFT-III-d7}
\ee
The first operator, $\cO_{JHL}^{d=7}$, is the type~III analog of the Majoron operator in Eq.~\eqref{eq:rhoJSMEFT-I-d7}, related to the unitarity one by $\cC_{JHL}^{d=7}=\cC_U^{d=6}/(2v_\phi)$, exactly as in the type~I case, Eq.~\eqref{eq:rhoJSMEFT-I-d7}. The second, $\cO_{DHL}^{d=7}$, is the derivative dressing of the Weinberg operator, generated by the $D^2$ term of the $\Sigma$ propagator expansion, with $\cC_{DHL}^{d=7}=-\cC_W^{d=5}/M_\Sigma^2$. It is the exact counterpart of the $\cO_{DHL}^{d=7}$ found in the $\rho$JSMEFT-IIa case discussed in Sec.~\ref{sec:rhoJSMEFT-IIa} below, Eq.~\eqref{eq:rhoJSMEFT-IIa-d7}. The last two operators, $\cO_{U\rho}^{d=7}$ and $\cO_{W\rho\rho}^{d=7}$, are the radial-mode dressings of the unitarity and Weinberg operators, respectively, now written explicitly as in the $\rho$JSMEFT-I case, Eq.~\eqref{eq:rhoJSMEFT-I-d7}. As there, $\cO_{JHL}^{d=7}$ and $\cO_{U\rho}^{d=7}$ are self-Hermitian and carry the $1/2$, while $\cO_{DHL}^{d=7}$ and $\cO_{W\rho\rho}^{d=7}$ are lepton number violating (genuine $+\hc$, no $1/2$).
\end{itemize}

The same resummation of Eq.~\eqref{eq:Iresum} holds here, with $M_\Sigma(\rho)=M_\Sigma(1+\rho/v_\phi)$: the $\rho$ dressings above follow from expanding $\cC_W^{d=5}(\rho)$ and $\cC_U^{d=6}(\rho)$, which carry one and two powers of the heavy propagator, respectively.

\subsection{The type~II Majoron model}
\label{sec:typeII}

We focus on the type~II Majoron model studied in Ref.~\cite{Biggio:2023gtm} and references therein, where a scalar $SU(2)_L$ triplet is added on top of the same LN-breaking scalar singlet $\phi$ already introduced for the type~I and III cases. While in the standard type~II seesaw, with only the heavy triplet, the LN is broken explicitly by the simultaneous presence of the trilinear coupling to the Higgs doublet and the Yukawa coupling to the leptons, here the LN-breaking trilinear coupling is instead generated dynamically by the VEV of $\phi$, so that the LN symmetry is broken spontaneously rather than explicitly.

Beyond the SM fields listed in Table~\ref{tab:typeI-spectrum}, the field content of the type~II Majoron model is completed by the same LN-breaking scalar singlet $\phi$ of the type~I case and by a complex scalar $SU(2)_L$ triplet $\Delta$, whose quantum numbers are collected in Table~\ref{tab:typeII-spectrum}.
\begin{table}[ht!]
\centering
\begin{tabular}{@{}lcccc@{}}
\toprule
Field & Spin & $SU(3)_c\times SU(2)_L\times U(1)_Y$ & $U(1)_L$ \\
\midrule
$\phi$ & $0$ & $(\mathbf{1},\mathbf{1},0)$ & $-2$ \\
$\Delta$ & $0$ & $(\mathbf{1},\mathbf{3},+1)$ & $-2$ \\
\bottomrule
\end{tabular}
\caption{\em Field content of the type~II Majoron model beyond the SM spectrum.}
\label{tab:typeII-spectrum}
\end{table}

If we express the triplet in the adjoint representation, i.e., $\Delta=\sigma^I\Delta^I$, the Lagrangian reads
\begin{equation}
\sL^\text{type~II}=\sL_\text{SM}+D_\mu\Delta^\dag D^\mu\Delta+\derp_\mu\phi^\ast\derp^\mu\phi-V(\Delta,\phi,H)-\left(Y_\Delta\,\ov{L_L^c}\epsilon\sigma^I L_L\Delta^I+\hc\right)\,,
\end{equation}
where, as in the type~I case, $\sL_\text{SM}$ does not include the Higgs potential, which is subsumed, together with all the terms involving $\phi$ and $\Delta$, into the complete scalar potential
\begin{align}
V(\Delta,\phi,H)=&-\mu_1^2\phi^\ast\phi+\lambda_1\left(\phi^\ast\phi\right)^2-\mu_2^2H^\dag H+\lambda_2\left(H^\dag H\right)^2+\beta_1\,\phi^\ast\phi\,H^\dag H+\label{CompleteScalarPotentialTypeII}\\
&+\mu_3^2\Delta^\dag\Delta+\lambda_3\left(\Delta_I^\dag\Delta_I\right)^2+\lambda_4\left|\epsilon^{IJK}\Delta_J^\dag\Delta_K\right|^2\footnotemark+\beta_2\,\phi^\ast\phi\,\Delta_I^\dag\Delta_I+\beta_3\,H^\dag H\,\Delta_I^\dag\Delta_I+\nn\\
&-i\,\beta_4\left(H^\dag\sigma^IH\right)\epsilon^{IJK}\Delta_J^\dag\Delta_K-\beta_5\,\phi\left(H^T\epsilon\sigma^I H\right)\Delta_I^\dag+\beta_5\,\phi^\ast\left(H^\dag\sigma^I\epsilon H^\ast\right)\Delta_I\,.\nn
\end{align}

\footnotetext{The complex scalar $SU(2)_L$ triplet has two independent quartic self-couplings, which we take to be $\left(\Delta_I^\dag\Delta_I\right)^2$ and $\left|\epsilon^{IJK}\Delta_J^\dag\Delta_K\right|^2$. The latter equals $\left(\Delta_I^\dag\Delta_I\right)^2-\left|\Delta_I\Delta_I\right|^2$, so that any two of the three structures $\left(\Delta_I^\dag\Delta_I\right)^2$, $\left|\Delta_I\Delta_I\right|^2$, and $\left|\epsilon^{IJK}\Delta_J^\dag\Delta_K\right|^2$ form an equivalent basis. In the matrix basis of Refs.~\cite{Arhrib:2011uy,Moultaka:2020dmb} one has $\mathrm{Tr}(\Delta^\dag\Delta)=2\,\Delta_I^\dag\Delta_I$ and $\mathrm{Tr}[(\Delta^\dag\Delta)^2]=[\mathrm{Tr}(\Delta^\dag\Delta)]^2-2\left|\Delta_I\Delta_I\right|^2$. Being pure $\Delta^4$ self-interactions, these couplings do not enter the tree-level matching.}
Notice that, with respect to the Yukawa coupling $Y_\Delta$, the trilinear term has been written with $\phi$ (not $\phi^\ast$) coupled to $\Delta^\dag$ (not $\Delta$): this is the assignment required by LN conservation, given that $\phi$ and $\Delta$ carry LN $-2$ each; see Table~\ref{tab:typeII-spectrum}. As for $\beta_1$, the coupling $\beta_5$ can be taken real without loss of generality, since its phase can be reabsorbed through a redefinition of $\phi$. As done for $Y_{NN}$ in the type~I case, we work in the basis where the charged-lepton Yukawa is real and diagonal, while $Y_\Delta$ remains a generic complex symmetric $3\times3$ matrix.

The mixing couplings $\beta_1,\dots,\beta_5$ entangle the $\phi$, $H$, and $\Delta$ sectors. As discussed in Sec.~\ref{sec:typeI}, keeping $\beta_1$ generic would reintroduce, through the $\phi$ VEV, a hierarchy problem for the Higgs mass. The same reasoning applies to $\beta_2$ and $\beta_5$, which would otherwise feed $v_\phi$-suppressed contributions into the mass and interactions of $\Delta$ and spoil the perturbative matching performed below. We therefore take $\beta_1,\beta_2,$ and $\beta_5$ to be small parameters and work at leading order in them, so that the minimization of $V(\Delta,\phi,H)$ can again be carried out step by step, first along $\phi$ and then along $H$. The couplings $\beta_3$ and $\beta_4$, instead, involve only $H$ and $\Delta$: they play no role in this step and enter only once $\Delta$ itself is considered. Note the explicit $i$ multiplying $\beta_4$: the bilinear $\epsilon^{IJK}\Delta_J^\dag\Delta_K$ is anti-Hermitian, since under conjugation it picks up $\epsilon^{IKJ}=-\epsilon^{IJK}$, so the $i$ is what makes the term real for real $\beta_4$. The same structure is what one obtains by decomposing the standard type~II invariant~\cite{Arhrib:2011uy}, $H^\dag\Delta\Delta^\dag H=(\Delta^\dag\Delta)(H^\dag H)+i\,\epsilon^{IJK}\Delta_I\Delta_J^\ast(H^\dag\sigma^KH)$, into the $\beta_3$ and $\beta_4$ structures.

The $\phi$ sector is, therefore, identical to the type~I case: $\phi$ acquires the same VEV $v_\phi^2=\mu_1^2/\lambda_1$ of Eq.~\eqref{eq:vphiDEF} and can be written in terms of the Majoron $J$ and the radial mode $\rho$ as in Eq.~\eqref{eq:phidecomp}, with $m_\rho$ again given by Eq.~\eqref{mrhoDEF}.

Also $\Delta$ is a massive field, but the explicit expression of its mass and its relation to $m_\rho$ depend on further assumptions on the parameters.
\begin{enumerate}
    \item If $v_\phi>\mu_3$, the mass of $\Delta$ is affected by the LN SSB and reads $M_\Delta^2=\mu_3^2+\dfrac12\beta_2v_\phi^2$. In this case, both $\rho$ and $\Delta$ can be the heaviest d.o.f., depending on the relative size of $\beta_2$ and $\lambda_1$: if $\beta_2\ll\lambda_1\sim\cO(1)$, then $m_\rho>M_\Delta$, while if $\beta_2\gg\lambda_1$, then $M_\Delta>m_\rho$;
    \item if $\mu_3>v_\phi$, the situation is simpler, since $M_\Delta=\mu_3$ and, for $\lambda_1$ in the perturbative regime, $M_\Delta>m_\rho$.
\end{enumerate}

After the LN SSB, the terms of the Lagrangian involving $\phi$ and $\Delta$ can be rewritten in terms of $\rho$, $J$, and $\Delta$ as
\begin{align}
\sL^\text{type~II}\supset&\ \dfrac12\derp_\mu\rho\derp^\mu\rho+\dfrac{(\rho+v_\phi)^2}{2v_\phi^2}\derp_\mu J\derp^\mu J
+D_\mu\Delta^\dag D^\mu\Delta
-\left(Y_\Delta\,\ov{L_L^c}\epsilon\sigma^I L_L\Delta^I+\hc\right)-\dfrac{\lambda_1}{4}\rho^4+\nn\\
&
-\lambda_1v_\phi\rho^3-\frac{m_\rho^2}{2}\rho^2-\dfrac{\beta_1}{2}v_\phi^2\left(H^\dag H\right)-\beta_1v_\phi\rho\left(H^\dag H\right)-\dfrac{\beta_1}{2}\rho^2\left(H^\dag H\right)-\mu_3^2\Delta^\dag\Delta+\nn\\
&-\lambda_3\left(\Delta^\dag\Delta\right)^2
-\dfrac{(v_\phi+\rho)^2}{2}\beta_2\,\Delta_I^\dag\Delta_I-\beta_3H^\dag H\,\Delta_I^\dag\Delta_I+i\,\beta_4\left(H^\dag\sigma^IH\right)\epsilon^{IJK}\Delta_J^\dag\Delta_K+\nn\\
&+\beta_5\dfrac{(v_\phi+\rho)}{\sqrt2}e^{iJ/v_\phi}\left(H^T\epsilon\sigma^I H\right)\Delta_I^\dag-\beta_5\dfrac{(v_\phi+\rho)}{\sqrt2}e^{-iJ/v_\phi}\left(H^\dag\sigma^I\epsilon H^\ast\right)\Delta_I\,,\label{eq:SSBLagTypeII}
\end{align}
where the terms not involving $\Delta$ are the same $\rho$, $J$ kinetic, self-interaction, and Higgs-portal terms already found in the type~I case, Eq.~\eqref{eq:SSBLagTypeI}.

The $J$ dependence in the trilinear term proportional to $\beta_5$ can be rotated away by performing the following vectorial transformation on the lepton fields and on the triplet:
\be
\label{rotJII}
\left\{L_L,\,e_R\right\}\rightarrow\left\{L_L,\,e_R\right\}e^{-iJ/2v_\phi}\,,\qquad\qquad \Delta\rightarrow\Delta\,e^{iJ/v_\phi}\,,
\ee
where the transformation of the leptons is the same as in Eq.~\eqref{rotJ}, while the coefficient for $\Delta$ is twice as large, consistently with $\Delta$ carrying twice the LN charge of a lepton doublet. As in the previous cases, this procedure reintroduces the $J$ dependence in the kinetic terms: the leptonic kinetic terms reproduce exactly the same contributions already found in Eq.~\eqref{Jops-typeI}, while the triplet kinetic term produces the additional piece
\be
D_\mu\Delta^\dag D^\mu\Delta\ \supset\ -\dfrac{i\,\derp_\mu J}{v_\phi}\left(\Delta^\dag D^\mu\Delta-\left(D^\mu\Delta\right)^\dag\Delta\right)+\dfrac{\derp_\mu J\derp^\mu J}{v_\phi^2}\,\Delta^\dag\Delta\,.
\ee
Collecting all the $J$-dependent terms together with the (now phase-free) trilinear coupling, the relevant part of the Lagrangian reads
\begin{align}
\sL^\text{type~II}\supset&\ \dfrac{\derp_\mu J}{2v_\phi}\left(\ov{L_L}\gamma^\mu L_L+\ov{e_R}\gamma^\mu e_R\right)-\dfrac{i\,\derp_\mu J}{v_\phi}\left(\Delta^\dag D^\mu\Delta-\left(D^\mu\Delta\right)^\dag\Delta\right)+\nn\\
&-\frac{3J}{32\pi^2v_\phi}(g'^2 B\widetilde B-g^2 W\widetilde W)+\dfrac{\derp_\mu J\derp^\mu J}{v_\phi^2}\,\Delta^\dag\Delta+
\label{Jops-typeII}\\
&+\beta_5\dfrac{(v_\phi+\rho)}{\sqrt2}\left(H^T\epsilon\sigma^I H\right)\Delta_I^\dag-\beta_5\dfrac{(v_\phi+\rho)}{\sqrt2}\left(H^\dag\sigma^I\epsilon H^\ast\right)\Delta_I\,,\nn
\end{align}
where the last line is the analog, for the type~II model, of the trilinear coupling that in the ordinary (explicitly broken) type~II seesaw is usually associated to a dimensionful $\mu$ parameter, here dynamically generated as $\beta_5v_\phi/\sqrt2$ by the $\phi$ VEV.

Concerning the anomalous couplings, the type~II case reproduces exactly the type~I result: only the leptons $L_L$ and $e_R$ are rotated by a chiral phase, and, since $N_R$ was a gauge singlet, its absence here changes nothing, so that $(c_W,c_B)=(+3/2,-3/2)$ after summing over the $n_g=3$ generations, giving the same result of Eq.~\eqref{Jops-typeI}. The rotation of the triplet $\Delta$ in Eq.~\eqref{rotJII} is instead a phase rotation of a \emph{scalar} field, which does not generate any chiral anomaly (the anomaly is intrinsically a fermion-loop effect, requiring the $\gamma_5$ in the triangle); hence, $\Delta$ produces no additional $J\,F\widetilde F$ term.

The type~II Majoron Lagrangian is, thus, characterized, as in the type~I case, by two types of heavy d.o.f.: the radial mode $\rho$ and the triplet $\Delta$. Depending on the relative size of $m_\rho$ and $M_\Delta$, the integration procedure has to be separated into three different cases, listed here for convenience:
\begin{itemize}
    \item case 1a: $m_\rho^2 > M_\Delta^2 = \mu_3^2+\dfrac{1}{2}\beta_2v_\phi^2$,
    \item case 1b: $M_\Delta^2 = \mu_3^2+\dfrac{1}{2}\beta_2v_\phi^2 > m_\rho^2$,
    \item case 2: $M_\Delta = \mu_3 > m_\rho$\,.
\end{itemize}
In the following, we will derive the intermediate EFT Lagrangians integrating out the heaviest d.o.f.\ in each of these three cases.

A remark on the matching strategy is in order. The UV Lagrangian above is written in the exact phase, in which the global $U(1)_{LN}$ and the gauge symmetries are manifest and the singlet $\phi$ has not yet acquired its VEV. Only in the bare-mass regime of the triplet (case~2, $\mu_3>v_\phi$) is $\Delta$ already massive in this exact phase, so that one could, in principle, integrate it out before $\phi$ takes its VEV, defining an intermediate effective Lagrangian in which the singlet $\phi$ is still dynamical---a complex-scalar-singlet extension of the SMEFT. We do not do so. Also in that case we prefer to lower the scale down to $v_\phi$, where $\phi$ acquires its VEV and the lepton number is spontaneously broken, and to write the effective Lagrangian there, in terms of the radial mode $\rho$ and the Majoron $J$, with $\rho$ kept dynamical and integrated out only in the final step down to the low-energy effective description dubbed as JSMEFT, whose spectrum includes only the SM fields and the Majoron. For the HNLs $N$ and the triplets $\Sigma$ of type~I and~III, by contrast, this alternative does not even arise: their masses, $M_N=Y_{NN}v_\phi/\sqrt2$ and $M_\Sigma=Y_\Sigma v_\phi/\sqrt2$, are generated entirely by the LN SSB, so they are exactly massless in the exact phase and cannot be integrated out before $\phi$ acquires its VEV.

\boldmath
\subsubsection[The case $m_\rho^2 > M_\Delta^2 = \mu_3^2+\beta_2v_\phi^2/2$ and the $\Delta$JSMEFT Lagrangian]{The case $m_\rho^2 > M_\Delta^2 = \mu_3^2+\dfrac{1}{2}\beta_2v_\phi^2$ and the $\Delta$JSMEFT Lagrangian}
\label{sec:DeltaJSMEFT}
\unboldmath

In this case, the radial mode $\rho$ is the heaviest d.o.f.\ and is integrated out first, following exactly the procedure of the NJSMEFT case in Sec.~\ref{sec:NJSMEFT}. Introducing the notation
\be
\cJ_{H\Delta}\equiv\left(H^T\epsilon\sigma^I H\right)\Delta_I^\dag+\hc\,,
\label{eq:JHDeltaDEF}
\ee
the terms of Eq.~\eqref{eq:SSBLagTypeII} involving $\rho$ [with the phase-free trilinear obtained after the rotation of Eq.~\eqref{rotJII}], the corresponding EOM, and its perturbative solution are collected in Appendix~\ref{app:typeII}. Substituting the solution back yields the $\Delta$JSMEFT Lagrangian, valid in the energy window $[M_\Delta,\,m_\rho]$ and containing only the SM spectrum, the Majoron $J$, and the triplet scalar $\Delta$:
\begin{align}
\sL^\text{$\Delta$JSMEFT}=&\sL_\text{SM}+\dfrac12\derp_\mu J\derp^\mu J+D_\mu\Delta^\dag D^\mu\Delta-M_\Delta^2\,\Delta^\dag\Delta-\lambda_3\left(\Delta^\dag\Delta\right)^2-\beta_3\left(H^\dag H\right)\left(\Delta^\dag\Delta\right)+\nn\\
&+i\,\beta_4\left(H^\dag\sigma^IH\right)\epsilon^{IJK}\Delta_J^\dag\Delta_K+\frac{\beta_5 m_\rho}{2\sqrt{\lambda_1}}\cJ_{H\Delta}-\left(Y_\Delta\,\ov{L_L^c}\epsilon\sigma^I L_L\Delta^I+\hc\right)+\nn\\
&-\dfrac{i\,\derp_\mu J}{v_\phi}\left(\Delta^\dag D^\mu\Delta-\left(D^\mu\Delta\right)^\dag\Delta\right)+\dfrac{\derp_\mu J\derp^\mu J}{v_\phi^2}\,\Delta^\dag\Delta+
\label{DeltaJSMEFTLag}\\
&+\dfrac{\derp_\mu J}{2v_\phi}\left(\ov{L_L}\gamma^\mu L_L+\ov{e_R}\gamma^\mu e_R\right)+\sum_i\left(\cC_i^d\,\cO_i^d+\hc\right)\,,
\nn
\end{align}
where $M_\Delta^2=\mu_3^2+\frac12\beta_2v_\phi^2$ is the triplet mass, already including the shift induced by the $\phi$ VEV, while the analogous $\phi$-VEV shift of the Higgs mass has been reabsorbed into $\mu_2^2$. As in the type~I and III cases, the nonrenormalizable terms suppressed by powers of $v_\phi$ are not generated by the integration out of $\rho$ and are, therefore, kept separate from the sum $\sum_i\cC_i^d\cO_i^d$. Among them, the trilinear $\cJ_{H\Delta}$ plays a special role: it is the dynamical counterpart of the dimensionful $\mu$ term of the ordinary type~II seesaw, and, together with $M_\Delta$, it induces the triplet VEV responsible for the active neutrino masses.

The leading contributions to the $\cC_i^d$ operators are listed below. As in the NJSMEFT, the sum is understood with the explicit $+\hc$ and the $1/2$ for self-Hermitian operators. All operators generated here are self-Hermitian ($\cJ_{H\Delta}$ is itself Hermitian by definition), so all coefficients below carry the convention $1/2$.
\begin{itemize}
\item{\bf d=2}
\be
\begin{aligned}
\cO_{H^2}^{d=2}&=\left(H^\dag H\right),\qquad&
\cC_{H^2}^{d=2}&=-\frac{\beta_1 m_\rho^2}{8\lambda_1}.
\end{aligned}
\label{eq:DeltaJSMEFT-d2}
\ee
As in the NJSMEFT, Eq.~\eqref{eq:NJSMEFT-d2}, this renormalizes $\mu_2^2$ (physical shift $2\cC_{H^2}^{d=2}=-\beta_1 v_\phi^2/2$); the analogous triplet-mass shift $-\beta_2 v_\phi^2/2$ is instead already included in $M_\Delta^2=\mu_3^2+\frac12\beta_2v_\phi^2$.
\item{\bf d=4}
\be
\begin{aligned}
\cO_{\lambda}^{d=4}&=\left(H^\dag H\right)^2,\qquad&
\cC_{\lambda}^{d=4}&=\frac{\beta_1^2}{8\lambda_1},\\
\cO_{\lambda\Delta}^{d=4}&=\left(\Delta^\dag\Delta\right)^2,\qquad&
\cC_{\lambda\Delta}^{d=4}&=\frac{\beta_2^2}{8\lambda_1},\\
\cO_{H\Delta}^{d=4}&=\left(H^\dag H\right)\left(\Delta^\dag\Delta\right),\qquad&
\cC_{H\Delta}^{d=4}&=\frac{\beta_1\beta_2}{4\lambda_1}.
\end{aligned}
\label{eq:DeltaJSMEFT-d4}
\ee
These three operators simply renormalize the quartic couplings $\lambda_2$, $\lambda_3$, and $\beta_3$ of the scalar potential, Eq.~\eqref{CompleteScalarPotentialTypeII}, respectively. The first coincides in form and coefficient with $\cO_\lambda^{d=4}$ of the NJSMEFT, Eq.~\eqref{eq:NJSMEFT-d4}.\item{\bf d=5}
\be
\begin{aligned}
\cO_{H\cJ}^{d=5}&=\left(H^\dag H\right)\left[\left(H^T\epsilon\sigma^I H\right)\Delta_I^\dag+\hc\right],\qquad&
\cC_{H\cJ}^{d=5}&=-\frac{\beta_1\beta_5}{4\sqrt{\lambda_1}\,m_\rho},\\
\cO_{\Delta \cJ}^{d=5}&=\left(\Delta^\dag\Delta\right)\left[\left(H^T\epsilon\sigma^I H\right)\Delta_I^\dag+\hc\right],\qquad&
\cC_{\Delta \cJ}^{d=5}&=-\frac{\beta_2\beta_5}{4\sqrt{\lambda_1}\,m_\rho}.
\end{aligned}
\label{eq:DeltaJSMEFT-d5}
\ee
\item{\bf d=6}
\be
\begin{aligned}
\cO_{\Box JH}^{d=6}&=\left(\derp_\mu J\derp^\mu J\right)\left(H^\dag H\right),\qquad&
\cC_{\Box JH}^{d=6}&=-\frac{\beta_1}{2m_\rho^2},\\
\cO_{\Box J\Delta}^{d=6}&=\left(\derp_\mu J\derp^\mu J\right)\left(\Delta^\dag\Delta\right),\qquad&
\cC_{\Box J\Delta}^{d=6}&=-\frac{\beta_2}{2m_\rho^2},\\
\cO_{\cJ\cJ}^{d=6}&=\left[\left(H^T\epsilon\sigma^I H\right)\Delta_I^\dag+\hc\right]^2,\qquad&
\cC_{\cJ\cJ}^{d=6}&=\frac{\beta_5^2}{8m_\rho^2},\\
\cO_{H\Box}^{d=6}&=\left(H^\dag H\right)\Box\left(H^\dag H\right),\qquad&
\cC_{H\Box}^{d=6}&=-\frac{\beta_1^2}{8\lambda_1 m_\rho^2},\\
\cO_{\Delta\Box}^{d=6}&=\left(\Delta^\dag\Delta\right)\Box\left(\Delta^\dag\Delta\right),\qquad&
\cC_{\Delta\Box}^{d=6}&=-\frac{\beta_2^2}{8\lambda_1 m_\rho^2},\\
\cO_{H\Box\Delta}^{d=6}&=\left(H^\dag H\right)\Box\left(\Delta^\dag\Delta\right),\qquad&
\cC_{H\Box\Delta}^{d=6}&=-\frac{\beta_1\beta_2}{4\lambda_1 m_\rho^2}.
\end{aligned}
\label{eq:DeltaJSMEFT-d6}
\ee
The first operator, $\cO_{\Box JH}^{d=6}$, is exactly the derivative Higgs portal~\cite{Balkin:2018tma} already found in the NJSMEFT case, Eq.~\eqref{eq:NJSMEFT-d=6}, with the very same Wilson coefficient, while $\cO_{\Box J\Delta}^{d=6}$ is its triplet counterpart. Among the $\Box$ operators, $\cO_{H\Box}^{d=6}$ belongs to the $d=6$ Warsaw basis of the SMEFT~\cite{Grzadkowski:2010es}, while $\cO_{\Delta\Box}^{d=6}$ and $\cO_{H\Box\Delta}^{d=6}$ are its triplet-sector analogs.
\item{\bf d=7}
\be
\begin{aligned}
\cO_{\Box J\cJ}^{d=7}&=\left(\derp_\mu J\derp^\mu J\right)\left[\left(H^T\epsilon\sigma^I H\right)\Delta_I^\dag+\hc\right],\qquad&
\cC_{\Box J\cJ}^{d=7}&=\frac{\sqrt{\lambda_1}\,\beta_5}{2m_\rho^3},\\
\cO_{H^2\cJ}^{d=7}&=\left(H^\dag H\right)^2\left[\left(H^T\epsilon\sigma^I H\right)\Delta_I^\dag+\hc\right],\qquad&
\cC_{H^2\cJ}^{d=7}&=-\frac{\beta_1^2\beta_5}{8\sqrt{\lambda_1}\,m_\rho^3},\\
\cO_{\Delta^2\cJ}^{d=7}&=\left(\Delta^\dag\Delta\right)^2\left[\left(H^T\epsilon\sigma^I H\right)\Delta_I^\dag+\hc\right],\qquad&
\cC_{\Delta^2\cJ}^{d=7}&=-\frac{\beta_2^2\beta_5}{8\sqrt{\lambda_1}\,m_\rho^3},\\
\cO_{H\Delta\cJ}^{d=7}&=\left(H^\dag H\right)\left(\Delta^\dag\Delta\right)\left[\left(H^T\epsilon\sigma^I H\right)\Delta_I^\dag+\hc\right],\qquad&
\cC_{H\Delta\cJ}^{d=7}&=-\frac{\beta_1\beta_2\beta_5}{4\sqrt{\lambda_1}\,m_\rho^3},\\
\cO_{H\Box\cJ}^{d=7}&=\left(H^\dag H\right)\Box\left[\left(H^T\epsilon\sigma^I H\right)\Delta_I^\dag+\hc\right],\qquad&
\cC_{H\Box\cJ}^{d=7}&=+\frac{\beta_1\beta_5}{4\sqrt{\lambda_1}\,m_\rho^3},\\
\cO_{\Delta\Box\cJ}^{d=7}&=\left(\Delta^\dag\Delta\right)\Box\left[\left(H^T\epsilon\sigma^I H\right)\Delta_I^\dag+\hc\right],\qquad&
\cC_{\Delta\Box\cJ}^{d=7}&=+\frac{\beta_2\beta_5}{4\sqrt{\lambda_1}\,m_\rho^3}.
\end{aligned}
\label{eq:DeltaJSMEFT-d7}
\ee
The first is the Majoron $d=7$ operator, the following three are the trilinear source $\cJ_{H\Delta}$ dressed by two scalar bilinears, and the last two are its derivative ($\Box$) dressings, obeying $\cC_{H\Box\cJ}^{d=7}=-\cC_{H\cJ}^{d=5}/m_\rho^2$ and $\cC_{\Delta\Box\cJ}^{d=7}=-\cC_{\Delta\cJ}^{d=5}/m_\rho^2$.
\end{itemize}
All the operators built out of the triplet $\Delta$ as a dynamical field---namely, the two $d=5$ operators, together with $\cO_{\Box J\Delta}^{d=6}$, $\cO_{\cJ\cJ}^{d=6}$, $\cO_{\Delta\Box}^{d=6}$, $\cO_{H\Box\Delta}^{d=6}$, and the $d=7$ operators---do not appear in the SMEFT description of the type~II seesaw, where the triplet is instead integrated out~\cite{Li:2022ipc}. They are specific to the intermediate $\Delta$JSMEFT and, together with those involving the Majoron, are considered here for the first time to the best of our knowledge. Restricted to the operators built solely out of $H$ and $\Delta$---that is, dropping the explicit Majoron dependence---this list falls within the general classification of \emph{triplet-dynamical} operators built out of the SM Higgs doublet and $\Delta$, worked out independently, for the ordinary and explicitly broken type~II seesaw, by Ponce D\'iaz~\cite{PonceDiaz:2021}. The remaining operators, built solely out of SM fields, are instead already known, as pointed out above.

\boldmath
\subsubsection[The case $M_\Delta^2 = \mu_3^2+\beta_2v_\phi^2/2 > m_\rho^2$ and the $\rho$JSMEFT-IIa Lagrangian]{The case $M_\Delta^2 = \mu_3^2+\dfrac{1}{2}\beta_2v_\phi^2 > m_\rho^2$ and the $\rho$JSMEFT-IIa Lagrangian}
\label{sec:rhoJSMEFT-IIa}
\unboldmath

In this case, the triplet $\Delta$ is the heaviest d.o.f.\ and is integrated out first, keeping the radial mode $\rho$ dynamical; the procedure is the exact type~II analog of the $\rho$JSMEFT-I case in Sec.~\ref{sec:rhoJSMEFT-I}. Contrary to the type~I and III cases, where the seesaw fields coupled to a single (Yukawa) source, the triplet couples to \emph{two} independent currents:
\be
\cJ_\Delta^I\equiv\beta_5\dfrac{(v_\phi+\rho)}{\sqrt2}\left(H^T\epsilon\sigma^I H\right)\,,\qquad\qquad
\cJ_L^I\equiv Y_\Delta\left(\ov{L_L^c}\epsilon\sigma^I L_L\right)\,,
\label{eq:JDeltaJLDEF}
\ee
the gauge trilinear $\cJ_\Delta^I$, built out of two Higgs doublets and proportional to $\beta_5(v_\phi+\rho)$, and the lepton bilinear $\cJ_L^I$, proportional to the triplet Yukawa $Y_\Delta$. It is precisely the interplay of these two sources that generates the active neutrino masses. The Lagrangian collecting the terms involving $\Delta$ [from Eqs.~\eqref{eq:SSBLagTypeII} and~\eqref{Jops-typeII}], together with the leading-order solution of its EOM, is given in Appendix~\ref{app:typeII}. Inserting the solution back yields the $\rho$JSMEFT-IIa Lagrangian, valid in the window $[m_\rho,\,M_\Delta]$ and containing the SM spectrum, the Majoron $J$, and the radial mode $\rho$:
\begin{align}
\sL^\text{$\rho$JSMEFT-IIa}=&\sL_\text{SM}+\dfrac12\derp_\mu J\derp^\mu J+\dfrac12\derp_\mu\rho\derp^\mu\rho-\frac{\lambda_1}{4}\rho^4-\lambda_1v_\phi\rho^3-\frac{m_\rho^2}{2}\rho^2+
\label{rhoJSMEFTIIaLag}\\
&+\frac{\left(\rho^2+2v_\phi\rho\right)}{2v_\phi^2}\derp_\mu J\derp^\mu J-\frac{\beta_1}{2}v_\phi^2\left(H^\dag H\right)-\beta_1v_\phi\rho\left(H^\dag H\right)-\frac{\beta_1}{2}\rho^2\left(H^\dag H\right)+\nn\\
&+\dfrac{\derp_\mu J}{2v_\phi}\left(\ov{L_L}\gamma^\mu L_L+\ov{e_R}\gamma^\mu e_R\right)+\sum_i\left(\cC_i^d\,\cO_i^d+\hc\right)\,,\nn
\end{align}
which has exactly the same $\rho$-$J$-SM structure of the $\rho$JSMEFT-I Lagrangian, Eq.~\eqref{rhoJSMEFTILag}. The only difference resides in the effective operators $\sum_i\cC_i^d\cO_i^d$, now generated by the integration out of $\Delta$. In listing the effective operators, it is convenient to keep together, on the one hand, the lepton-number-violating operators sourced by the interplay of the trilinear and the Yukawa currents (built from the cross term $\cJ_\Delta^\dag\cJ_L$ and from $\cJ_L^\dag\cJ_L$) and, on the other hand, the purely bosonic operators sourced by the trilinear alone ($\cJ_\Delta^\dag\cJ_\Delta$).~\footnote{We use the $SU(2)_L$ identity $\sum_I|H^T\epsilon\sigma^I H|^2=2(H^\dag H)^2$, so that $\cJ_\Delta^\dag\cJ_\Delta/M_\Delta^2=\beta_5^2(v_\phi+\rho)^2(H^\dag H)^2/M_\Delta^2$.} The leading operators, organized by dimension, are the following.
\begin{itemize}
\item{\bf d=4}
\be
\begin{aligned}
\cO_{\lambda}^{d=4}&=\left(H^\dag H\right)^2,\qquad&
\cC_{\lambda}^{d=4}&=\frac{\beta_5^2 v_\phi^2}{2M_\Delta^2}.
\end{aligned}
\label{eq:rhoJSMEFT-IIa-d4}
\ee
This operator renormalizes the Higgs quartic coupling $\lambda_2$, in full analogy with $\cO_\lambda^{d=4}$ of the NJSMEFT and $\Delta$JSMEFT cases. Being self-Hermitian, it carries the convention $1/2$.
\item{\bf d=5}
\be
\begin{aligned}
\cO_W^{d=5}&=\left(\ov{L_L}\widetilde H\right)_\alpha\left(\widetilde H^TL_L^c\right)_\beta,\qquad&
\cC_W^{d=5}&=\dfrac{\sqrt2\,\beta_5 v_\phi}{M_\Delta^2}\left(Y_\Delta\right)_{\alpha\beta},\\
\cO_{\rho H}^{d=5}&=\rho\left(H^\dag H\right)^2,\qquad&
\cC_{\rho H}^{d=5}&=\dfrac{\beta_5^2 v_\phi}{M_\Delta^2}\left(1-\dfrac{\beta_2v_\phi^2}{2M_\Delta^2}\right).
\end{aligned}
\label{eq:rhoJSMEFT-IIa-d5}
\ee

The first operator is the Weinberg operator~\cite{Weinberg:1979sa}, arising from the cross term between the two sources, $\cJ_\Delta^\dag\cJ_L$. It reproduces the well-known type~II seesaw prediction for the active neutrino masses, $m_\nu=-\cC_W^{d=5}v^2=-\sqrt2\,\beta_5\,Y_\Delta\,v_\phi\,v^2/M_\Delta^2$, with the dynamically generated trilinear $\beta_5 v_\phi/\sqrt2$ playing the role of the usual $\mu$ parameter. The second operator, $\cO_{\rho H}^{d=5}$, is the radial-mode dressing of the Higgs-quartic renormalization.
\item{\bf d=6}
\be
\begin{aligned}
\cO_{4L}^{d=6}&=\left(\ov{L_L^c}\epsilon\sigma^I L_L\right)_{\alpha\beta}\left(\ov{L_L}\sigma^I\epsilon L_L^c\right)_{\gamma\delta},\qquad&
\cC_{4L}^{d=6}&=\dfrac{\left(Y_\Delta\right)_{\alpha\beta}\left(Y_\Delta^\dag\right)_{\gamma\delta}}{2M_\Delta^2},\\
\cO_{W\rho}^{d=6}&=\left(\ov{L_L}\widetilde H\right)_\alpha\left(\widetilde H^TL_L^c\right)_\beta\rho,\qquad&
\cC_{W\rho}^{d=6}&=\dfrac{\cC_W^{d=5}}{v_\phi}\left(1-\dfrac{\beta_2v_\phi^2}{M_\Delta^2}\right),\\
\cO_{\rho\rho H}^{d=6}&=\rho^2\left(H^\dag H\right)^2,\qquad&
\cC_{\rho\rho H}^{d=6}&=\dfrac{\beta_5^2}{2M_\Delta^2}\left(1-\dfrac{5\beta_2v_\phi^2}{2M_\Delta^2}\right).
\end{aligned}
\label{eq:rhoJSMEFT-IIa-d6}
\ee

The first operator, $\cO_{4L}^{d=6}$, is the purely leptonic four-fermion operator arising from $\cJ_L^\dag\cJ_L$ and it contributes to lepton-flavor-violating processes such as $\mu\to3e$. Unlike the type~I unitarity operator $\cO_U^{d=6}$, which came with a derivative from the fermionic propagator, here the scalar propagator of $\Delta$ yields a genuine four-lepton contact interaction. The second operator is the Weinberg one dressed with an extra radial mode, exactly as $\cO_{W\rho}^{d=6}$ in the $\rho$JSMEFT-I case, Eq.~\eqref{eq:rhoJSMEFT-I-ds6}, with $\cC_{W\rho}^{d=6}=\cC_W^{d=5}/v_\phi$ up to the $\beta_2$ correction discussed below, while the third one is the second radial-mode dressing of the Higgs-quartic renormalization.

\item{\bf d=7}
\begin{align}
\cO_{DHL}^{d=7}&=\left(\ov{L_L}\widetilde H\right)_\alpha D^2\left(\widetilde H^TL_L^c\right)_\beta,\qquad&
\cC_{DHL}^{d=7}&=-\dfrac{\sqrt2\,\beta_5 v_\phi}{M_\Delta^4}\left(Y_\Delta\right)_{\alpha\beta}=-\dfrac{\cC_W^{d=5}}{M_\Delta^2},\nn\\
\cO_{WH}^{d=7}&=\left(\ov{L_L}\widetilde H\right)_\alpha\left(\widetilde H^TL_L^c\right)_\beta\left(H^\dag H\right),\qquad&
\cC_{WH}^{d=7}&=-\dfrac{\sqrt2\,(\beta_3+\beta_4)\,\beta_5 v_\phi}{M_\Delta^4}\left(Y_\Delta\right)_{\alpha\beta},\nn\\
\cO_{W\rho\rho}^{d=7}&=\left(\ov{L_L}\widetilde H\right)_\alpha\left(\widetilde H^TL_L^c\right)_\beta\rho^2,\qquad&
\cC_{W\rho\rho}^{d=7}&=-\dfrac{3\sqrt2\,\beta_2\beta_5 v_\phi}{2M_\Delta^4}\left(Y_\Delta\right)_{\alpha\beta}=-\dfrac{3\beta_2}{2M_\Delta^2}\cC_W^{d=5},\nn\\
\cO_{\rho\rho\rho H}^{d=7}&=\rho^3\left(H^\dag H\right)^2,\qquad&
\cC_{\rho\rho\rho H}^{d=7}&=-\dfrac{\beta_2\beta_5^2 v_\phi}{M_\Delta^4},
\label{eq:rhoJSMEFT-IIa-d7}\\
\cO_{\rho H^6}^{d=7}&=\rho\left(H^\dag H\right)^3,\qquad&
\cC_{\rho H^6}^{d=7}&=-\dfrac{(\beta_3+\beta_4)\,\beta_5^2 v_\phi}{M_\Delta^4}.\nn
\end{align}

The $\rho$-dressed coefficients above are not independent: they all follow from a single observation. The triplet mass in Eq.~\eqref{LagDelta2} is not $M_\Delta^2$ but the \emph{$\rho$-dependent} combination
\be
\label{eq:MDrho}
M_\Delta^2(\rho)\equiv\mu_3^2+\dfrac{\beta_2}{2}\left(v_\phi+\rho\right)^2=M_\Delta^2+\beta_2v_\phi\rho+\dfrac{\beta_2}{2}\rho^2\,,
\ee
since the radial mode enters the $\Delta$ mass exactly as the $\phi$ VEV does. Solving the $\Delta$ EOM at leading order in $1/M_\Delta^2(\rho)$ and keeping the $\rho$ dependence unexpanded, the whole trilinear-induced sector of the $\rho$JSMEFT-IIa Lagrangian resums into the two compact expressions
\be
\label{eq:IIaresum}
\sL_\text{eff}\supset\dfrac{\beta_5^2\left(v_\phi+\rho\right)^2\left(H^\dag H\right)^2}{M_\Delta^2(\rho)}
+\left[\dfrac{\sqrt2\,\beta_5\left(v_\phi+\rho\right)}{M_\Delta^2(\rho)}\left(Y_\Delta\right)_{\alpha\beta}\cO_W^{d=5\,\dag}+\hc\right]\,,
\ee
where we used $\sum_I|H^T\epsilon\sigma^I H|^2=2(H^\dag H)^2$ and $\left(\ov{L_L^c}\epsilon\sigma^I L_L\right)\left(H^T\epsilon\sigma^I H\right)=-2\,\cO_W^{d=5\,\dag}$. Expanding Eq.~\eqref{eq:IIaresum} in powers of $\rho$ reproduces, order by order, every coefficient listed above: $\cC_\lambda^{d=4}$, $\cC_{\rho H}^{d=5}$, and $\cC_{\rho\rho H}^{d=6}$ from the first term and $\cC_W^{d=5}$, $\cC_{W\rho}^{d=6}$, and $\cC_{W\rho\rho}^{d=7}$ from the second, together with the $d=7$ coefficients $\cC_{\rho\rho\rho H}^{d=7}$ and, at the order of $\beta_3$ and $\beta_4$, $\cC_{WH}^{d=7}$ and $\cC_{\rho H^6}^{d=7}$. We quote all the coefficients in this unreduced form, as we already did for $\cO_U^{d=6}$ and $\cO_{DHL}^{d=7}$, neither of which belongs to the Warsaw basis.~\footnote{This is worth keeping in mind when comparing with lists written directly in the Warsaw basis, since the same physics then appears distributed over several operators. The relevant example here is $\left(H^\dag H\right)\Box\left(H^\dag H\right)$, generated by the $D^2$ dressing of the trilinear source with coefficient $\propto\beta_5^2v_\phi^2/M_\Delta^4$. Using the Higgs EOM, $D^2H=\mu_2^2H-2\lambda_2(H^\dag H)H-\left(\ov{L_L}Y_\ell e_R+\cdots\right)$, one finds $\left(H^\dag H\right)\Box\left(H^\dag H\right)=2\mu_2^2\left(H^\dag H\right)^2-4\lambda_2\left(H^\dag H\right)^3-\left(H^\dag H\right)\left[\ov{L_L}H Y_\ell e_R+\hc\right]+2\left(H^\dag H\right)\left(D_\mu H^\dag D^\mu H\right)$, so that a single operator redistributes into $\left(H^\dag H\right)^3$---with the characteristic $4\lambda_2$---into the charged-lepton Yukawa structure $\left(H^\dag H\right)\left(\ov{L_L}He_R\right)$---together with the analogous quark structures $\left(H^\dag H\right)\left(\ov{Q_L}Hd_R\right)$ and $\left(H^\dag H\right)\left(\ov{Q_L}\widetilde H u_R\right)$---and, through a further $SU(2)$ rearrangement, into $\left(H^\dag D_\mu H\right)^\dag\left(H^\dag D^\mu H\right)$, all sharing the same $\beta_5^2v_\phi^2/M_\Delta^4$ prefactor.} In particular, the apparently accidental rational factors have a transparent origin: the $5/2$ in $\cC_{\rho\rho H}^{d=6}$ and the $3/2$ in $\cC_{W\rho\rho}^{d=7}$ are simply the $\rho^2$ Taylor coefficients of $(v_\phi+\rho)^2/M_\Delta^2(\rho)$ and $(v_\phi+\rho)/M_\Delta^2(\rho)$. It also fixes the expansion in which the list above is written: the $\rho$ dressings organize themselves in powers of $\beta_2v_\phi^2/M_\Delta^2$, small in the regime discussed below Eq.~\eqref{CompleteScalarPotentialTypeII}, and we quote every coefficient to first order in that ratio, with $M_\Delta$ the physical mass of Eq.~\eqref{eq:MDrho}. Writing the coefficients in terms of $M_\Delta$ does not, however, remove $\beta_2$ from them. What the resummed mass absorbs is the value $M_\Delta^2(0)$, while the slope $\partial_\rho M_\Delta^2=\beta_2(v_\phi+\rho)$ is an independent piece of information: any coefficient obtained by differentiating Eq.~\eqref{eq:IIaresum} therefore retains an explicit $\beta_2$, as $\cC_{\rho H}^{d=5}$, $\cC_{W\rho}^{d=6}$, and $\cC_{\rho\rho H}^{d=6}$ above already show. The same mechanism survives the removal of $\rho$ itself and reappears in Eqs.~\eqref{eq:JSMEFT-II-lep} and~\eqref{eq:JSMEFT-II-new}.

The first $d=7$ operator, $\cO_{DHL}^{d=7}$, is the derivative dressing of the Weinberg operator, generated by the $D^2$ term of the $\Delta$ propagator expansion, and its coefficient is simply the Weinberg one further suppressed by $M_\Delta^2$, $\cC_{DHL}^{d=7}=-\cC_W^{d=5}/M_\Delta^2$. The second, $\cO_{WH}^{d=7}$, is the Weinberg operator dressed with an additional Higgs bilinear, generated by the $\beta_3$ and $\beta_4$ corrections to the $\Delta$ mass in the propagator expansion.
The last three are the radial-mode dressings: $\cO_{W\rho\rho}^{d=7}$ is the Weinberg operator with two extra radial modes, while $\cO_{\rho\rho\rho H}^{d=7}$ and $\cO_{\rho H^6}^{d=7}$ dress the Higgs-quartic renormalization and the $(H^\dag H)^3$ interaction. All three are controlled by the $\rho$ dependence of the triplet mass and are, therefore, proportional to $\beta_2$ or to $\beta_3+\beta_4$, as Eq.~\eqref{eq:IIaresum} makes manifest.
\end{itemize}
We stress that the Majoron does not generate any new operator up to $d=7$ beyond the derivative lepton coupling already displayed in Eq.~\eqref{rhoJSMEFTIIaLag}.

\boldmath
\subsubsection{The case $M_\Delta = \mu_3 > m_\rho$ and the $\rho$JSMEFT-IIb Lagrangian}
\label{sec:rhoJSMEFT-IIb}
\unboldmath

When the triplet mass is dominated by the bare term, $\mu_3>v_\phi$, one has $M_\Delta=\mu_3$ and, for $\lambda_1$ in the perturbative regime, $M_\Delta>m_\rho$, so that the triplet is again the heaviest d.o.f. and is integrated out first. The procedure follows step by step the one of the $\rho$JSMEFT-IIa case in Sec.~\ref{sec:rhoJSMEFT-IIa}: the currents $\cJ_\Delta^I$ and $\cJ_L^I$, the $\Delta$ EOM, and the resulting $\rho$JSMEFT-IIb Lagrangian are exactly those of Eqs.~\eqref{LagDelta2} and~\eqref{rhoJSMEFTIIaLag}, and the effective operators coincide with the basis of Eqs.~\eqref{eq:rhoJSMEFT-IIa-d4}--\eqref{eq:rhoJSMEFT-IIa-d7}. The only difference is that the triplet mass is now the bare, LN-independent scale $\mu_3$. The Wilson coefficients are, therefore, obtained through the replacement $M_\Delta^2=\mu_3^2+\frac12\beta_2v_\phi^2\to\mu_3^2$, and read, for the operators $\cO_i^d$ defined in Eqs.~\eqref{eq:rhoJSMEFT-IIa-d4}--\eqref{eq:rhoJSMEFT-IIa-d7} of the $\rho$JSMEFT-IIa case,
\begin{align}
&d=4:\quad&& \cC_\lambda^{d=4}=\dfrac{\beta_5^2 v_\phi^2}{2\mu_3^2}\,;\\[3pt]
&d=5:\quad&& \cC_W^{d=5}=\dfrac{\sqrt2\,\beta_5 v_\phi}{\mu_3^2}\left(Y_\Delta\right)_{\alpha\beta}\,,
\qquad\qquad
\cC_{\rho H}^{d=5}=\dfrac{\beta_5^2 v_\phi}{\mu_3^2}\left(1-\dfrac{\beta_2v_\phi^2}{2\mu_3^2}\right)\,;\\[3pt]
&d=6:\quad&& \cC_{4L}^{d=6}=\dfrac{\left(Y_\Delta\right)_{\alpha\beta}\left(Y_\Delta^\dag\right)_{\gamma\delta}}{2\mu_3^2}\,,
\qquad\qquad
\cC_{W\rho}^{d=6}=\dfrac{\cC_W^{d=5}}{v_\phi}\left(1-\dfrac{\beta_2v_\phi^2}{\mu_3^2}\right)\,,\label{eq:rhoJSMEFT-IIb-WC}\\[3pt]
&&&\cC_{\rho\rho H}^{d=6}=\dfrac{\beta_5^2}{2\mu_3^2}\left(1-\dfrac{5\beta_2v_\phi^2}{2\mu_3^2}\right)\,;\nn
\end{align}
\begin{align}
&d=7:\quad&& \cC_{DHL}^{d=7}=-\dfrac{\sqrt2\,\beta_5 v_\phi}{\mu_3^4}\left(Y_\Delta\right)_{\alpha\beta}=-\dfrac{\cC_W^{d=5}}{\mu_3^2}\,,
\qquad
\cC_{WH}^{d=7}=-\dfrac{\sqrt2\,(\beta_3+\beta_4)\,\beta_5 v_\phi}{\mu_3^4}\left(Y_\Delta\right)_{\alpha\beta}\,,\nn\\[3pt]
&&&\cC_{W\rho\rho}^{d=7}=-\dfrac{3\beta_2}{2\mu_3^2}\cC_W^{d=5}\,,
\qquad\qquad
\cC_{\rho\rho\rho H}^{d=7}=-\dfrac{\beta_2\beta_5^2 v_\phi}{\mu_3^4}\,,
\qquad\qquad
\cC_{\rho H^6}^{d=7}=-\dfrac{(\beta_3+\beta_4)\,\beta_5^2 v_\phi}{\mu_3^4}\,.
\end{align}
The physical content mirrors that of Sec.~\ref{sec:rhoJSMEFT-IIa}, with one key difference: since $\mu_3$ is an independent scale, decoupled from the LN-breaking one, the active neutrino mass scales as $m_\nu=-\cC_W^{d=5}v^2\sim\beta_5\,Y_\Delta\,v_\phi\,v^2/\mu_3^2$, i.e., as $v^2/\mu_3^2$ instead of the $v^2/v_\phi$ behavior of the $\rho$JSMEFT-IIa case, with the dynamically generated trilinear $\beta_5 v_\phi/\sqrt2$ playing the role of the $\mu$ parameter of the traditional type~II description. As before, the Majoron generates no new operator up to $d=7$, and the status in the literature is the same as in Sec.~\ref{sec:rhoJSMEFT-IIa}. The Weinberg operator $\cO_W^{d=5}$~\cite{Weinberg:1979sa} and the four-lepton $\cO_{4L}^{d=6}$, together with the $d=7$ corrections, belong to the SMEFT of the type~II seesaw~\cite{Li:2022ipc}, while the radial-mode operators are specific to the intermediate $\rho$JSMEFT and appear here for the first time. The resummation of Eq.~\eqref{eq:IIaresum} holds here as well, with $M_\Delta^2(\rho)=\mu_3^2+\frac{\beta_2}{2}(v_\phi+\rho)^2$. Expanding it in $\rho$ reproduces the $\rho$-dressed coefficients above, which in this case are further suppressed since $\mu_3>v_\phi$.

Concerning the status in the literature, the pattern is the same as in the $\rho$JSMEFT-IIa case. The Weinberg operator $\cO_W^{d=5}$~\cite{Weinberg:1979sa} and the four-lepton operator $\cO_{4L}^{d=6}$ are well known---the latter being the standard $d=6$ operator of the SMEFT description of the type~II seesaw~\cite{Li:2022ipc}, relevant for charged-lepton-flavor-violating observables---together with their $d=7$ corrections $\cO_{DHL}^{d=7}$ and $\cO_{WH}^{d=7}$~\cite{Li:2022ipc}. The operators dressed by the radial mode $\rho$ ($\cO_{\rho H}^{d=5}$, $\cO_{W\rho}^{d=6}$, $\cO_{\rho\rho H}^{d=6}$, and at $d=7$ $\cO_{W\rho\rho}^{d=7}$, $\cO_{\rho\rho\rho H}^{d=7}$, and $\cO_{\rho H^6}^{d=7}$) are instead specific to the intermediate $\rho$JSMEFT and are considered here, to the best of our knowledge, for the first time.

\section{Low-Energy EFTs: the JSMEFT Lagrangian}
\label{sec:JSMEFT}

Below both heavy scales, the theory reduces to an effective Lagrangian containing only the SM fields and the Majoron $J$, which we dub the JSMEFT. As anticipated in Sec.~\ref{sec:intro}, it can be reached along the two complementary matching orders: integrating out the radial mode $\rho$ first and then the seesaw mediator, or vice versa. By EFT consistency, the two must yield the same result, and we have checked that they do, operator by operator. The purely bosonic operators are generated by the $\rho$ exchange and are common to the two orders, the lepton-number-violating and Majoron operators are generated by the seesaw-mediator exchange, and the mixed operator $\cQ_{HW}$ arises identically in the two orders---from the $\rho$ dressing of the Weinberg operator when $\rho$ is removed last and from the Higgs dressing of the mediator mass when the mediator is removed last. We collect here the resulting JSMEFT for each seesaw realization, the starting point for the phenomenological analysis in Sec.~\ref{sec:pheno}. All coefficients follow the conventions of Eq.~\eqref{NJSMEFTLag}.

\subsection{The type~I Majoron model}
\label{sec:JSMEFT-I}

Integrating out both $\rho$ and the HNLs $N$---either removing $\rho$ from the $\rho$JSMEFT-I Lagrangian, Eq.~\eqref{rhoJSMEFTILag}, or removing $N$ from the NJSMEFT Lagrangian, Eq.~\eqref{NJSMEFTLag}---yields the same JSMEFT Lagrangian, valid below $\min(m_\rho,M_N)$:
\be\begin{split}
\sL^\text{JSMEFT}_\text{I}=&\sL_\text{SM}+\dfrac12\derp_\mu J\derp^\mu J+\dfrac{\partial_\mu J}{2v_\phi}\left(\ov{L_L}\gamma^\mu L_L+\ov{e_R}\gamma^\mu e_R\right)+\\
&-\dfrac{3\,J}{32\pi^2v_\phi}\left(g^{\prime 2} B_{\mu\nu}\widetilde{B}^{\mu\nu}-g^2 W^i_{\mu\nu}\widetilde{W}^{i\mu\nu}\right)+\sum_i\left(\cC_i^d\,\cQ_i^d+\hc\right)\,.
\end{split}
\label{JSMEFTLag-I}
\ee
The purely bosonic operators, generated by the $\rho$ exchange and inherited unchanged from the NJSMEFT, are
\be
\begin{aligned}
\cQ_{H^2}^{d=2}&=\left(H^\dag H\right),\qquad&
\cC_{H^2}^{d=2}&=-\frac{\beta_1 m_\rho^2}{8\lambda_1},\\
\cQ_{\lambda}^{d=4}&=\left(H^\dag H\right)^2,\qquad&
\cC_{\lambda}^{d=4}&=\frac{\beta_1^2}{8\lambda_1},\\
\cQ_{H\Box}^{d=6}&=\left(H^\dag H\right)\Box\left(H^\dag H\right),\qquad&
\cC_{H\Box}^{d=6}&=-\frac{\beta_1^2}{8\lambda_1 m_\rho^2},\\
\cQ_{JH}^{d=6}&=\left(\derp_\mu J\derp^\mu J\right)\left(H^\dag H\right),\qquad&
\cC_{JH}^{d=6}&=-\frac{\beta_1}{2m_\rho^2}\,,
\end{aligned}
\label{eq:JSMEFT-I-bos}
\ee
while the lepton-number-violating and Majoron operators, generated by the $N$ exchange, are
\be
\begin{aligned}
\cQ_W^{d=5}&=\left(\ov{L_L}\widetilde H\right)_\alpha\left(\widetilde H^TL_L^c\right)_\beta,\qquad&
\cC_W^{d=5}&=\dfrac12\left(Y_N M_N^{-1}Y_N^T\right)_{\alpha\beta},\\
\cQ_U^{d=6}&=\left(\ov{L_L}\widetilde H\right)_\alpha i\slashed\partial\left(\widetilde H^\dag L_L\right)_\beta,\qquad&
\cC_U^{d=6}&=\dfrac12\left(Y_N M_N^{-1}\right)_{\alpha\gamma}\left(M_N^{-1}Y_N^\dag\right)_{\gamma\beta},\\
\cQ_{JHL}^{d=7}&=\left(\ov{L_L}\widetilde H\right)_\alpha\slashed\partial J\gamma_5\left(\widetilde H^\dag L_L\right)_\beta,\qquad&
\cC_{JHL}^{d=7}&=\dfrac{\cC_U^{d=6}}{2v_\phi},\\
\cQ_{HW}^{d=7}&=\left(\ov{L_L}\widetilde H\right)_\alpha\left(\widetilde H^TL_L^c\right)_\beta\left(H^\dag H\right),\qquad&
\cC_{HW}^{d=7}&=\dfrac{\beta_1}{m_\rho^2}\,\cC_W^{d=5},\\
\cQ_{DHL}^{d=7}&=\left(\ov{L_L}\widetilde H\right)_\alpha D^2\left(\widetilde H^TL_L^c\right)_\beta,\qquad&
\cC_{DHL}^{d=7}&=-\dfrac{\cC_W^{d=5}}{M_N^2}\,.
\end{aligned}
\label{eq:JSMEFT-I-lep}
\ee
The Weinberg operator $\cQ_W^{d=5}$ reproduces the active neutrino masses, $m_\nu=-\cC_W^{d=5}v^2$; $\cQ_U^{d=6}$ is the unitarity operator; $\cQ_{JHL}^{d=7}$ the leading Majoron-lepton coupling, tied to the unitarity one by $\cC_{JHL}^{d=7}=\cC_U^{d=6}/(2v_\phi)$; $\cQ_{HW}^{d=7}$ the Higgs dressing of the Weinberg operator, with $\cC_{HW}^{d=7}=\beta_1\cC_W^{d=5}/m_\rho^2$ reproduced identically in the two matching orders; and $\cQ_{DHL}^{d=7}$ its derivative dressing, $\cC_{DHL}^{d=7}=-\cC_W^{d=5}/M_N^2$. The Majoron couples to the light leptons through the derivative current in the first line in Eq.~\eqref{JSMEFTLag-I} and to the electroweak gauge bosons through the anomalous term, which yields no coupling to two photons since $c_W+c_B=0$.

\subsection{The type~III Majoron model}
\label{sec:JSMEFT-III}

Integrating out both $\rho$ and the fermion triplets $\Sigma$---either removing $\rho$ from the $\rho$JSMEFT-III Lagrangian, Eq.~\eqref{rhoJSMEFTIIILag}, or removing $\Sigma$ from the $\Sigma$JSMEFT Lagrangian, Eq.~\eqref{SigmaJSMEFTLag}---yields the same JSMEFT Lagrangian, valid below $\min(m_\rho,M_\Sigma)$:
\be
\begin{split}
\sL^\text{JSMEFT}_\text{III}=&\sL_\text{SM}+\dfrac12\derp_\mu J\derp^\mu J+\dfrac{\partial_\mu J}{2v_\phi}\left(\ov{L_L}\gamma^\mu L_L+\ov{e_R}\gamma^\mu e_R\right)+\\
&-\dfrac{3\,J}{32\pi^2v_\phi}\left(g^{\prime 2} B_{\mu\nu}\widetilde{B}^{\mu\nu}+3\,g^2 W^i_{\mu\nu}\widetilde{W}^{i\mu\nu}\right)+\sum_i\left(\cC_i^d\,\cQ_i^d+\hc\right)\,.
\end{split}
\label{JSMEFTLag-III}
\ee
The whole construction is the exact analog of the type~I case, with the HNLs replaced by the triplets, $N\to\Sigma$, $Y_N\to Y_\chi$, and $M_N\to M_\Sigma$. The purely bosonic operators are generated by the $\rho$ exchange alone---the type~III model contains no extra scalar---and coincide, operator by operator and coefficient by coefficient, with those of the type~I JSMEFT, Eq.~\eqref{eq:JSMEFT-I-bos}:
\be
\cC_{H^2}^{d=2}=-\frac{\beta_1 m_\rho^2}{8\lambda_1}\,,\qquad
\cC_\lambda^{d=4}=\frac{\beta_1^2}{8\lambda_1}\,,\qquad
\cC_{H\Box}^{d=6}=-\frac{\beta_1^2}{8\lambda_1 m_\rho^2}\,,\qquad
\cC_{JH}^{d=6}=-\frac{\beta_1}{2m_\rho^2}\,.
\label{eq:JSMEFT-III-bos}
\ee

The lepton-number-violating and Majoron operators are instead generated by the exchange of the fermion triplets. Being built out of the $SU(2)_L$-triplet current $\cJ_\Sigma^I$ of Eq.~\eqref{eq:JSigmaDEF}, they carry the adjoint structure $\sigma^I\!\otimes\sigma^I$ absent in the type~I case and read
\be
\begin{aligned}
\cQ_W^{d=5}&=\left(\ov{L_L^c}\sigma^I\widetilde H^\ast\right)_\alpha\left(\widetilde H^\dag\sigma^I L_L\right)_\beta,\qquad&
\cC_W^{d=5}&=\dfrac12\left(Y_\chi^T M_\Sigma^{-1}Y_\chi\right)_{\alpha\beta},\\
\cQ_U^{d=6}&=\left(\ov{L_L}\sigma^I\widetilde H\right)_\alpha i\slashed D\left(\widetilde H^\dag\sigma^I L_L\right)_\beta,\qquad&
\cC_U^{d=6}&=\dfrac12\left(Y_\chi^\dag M_\Sigma^{-1}\right)_{\alpha\gamma}\left(M_\Sigma^{-1}Y_\chi\right)_{\gamma\beta},\\
\cQ_{JHL}^{d=7}&=\left(\ov{L_L}\sigma^I\widetilde H\right)_\alpha\slashed\partial J\gamma_5\left(\widetilde H^\dag\sigma^I L_L\right)_\beta,\qquad&
\cC_{JHL}^{d=7}&=\dfrac{\cC_U^{d=6}}{2v_\phi},\\
\cQ_{HW}^{d=7}&=\left(\ov{L_L^c}\sigma^I\widetilde H^\ast\right)_\alpha\left(\widetilde H^\dag\sigma^I L_L\right)_\beta\left(H^\dag H\right),\qquad&
\cC_{HW}^{d=7}&=\dfrac{\beta_1}{m_\rho^2}\,\cC_W^{d=5},\\
\cQ_{DHL}^{d=7}&=\left(\ov{L_L^c}\sigma^I\widetilde H^\ast\right)_\alpha D^2\left(\widetilde H^\dag\sigma^I L_L\right)_\beta,\qquad&
\cC_{DHL}^{d=7}&=-\dfrac{\cC_W^{d=5}}{M_\Sigma^2}\,.
\end{aligned}
\label{eq:JSMEFT-III-lep}
\ee
The Weinberg operator reproduces the type~III seesaw prediction for the active neutrino masses, $m_\nu=-\cC_W^{d=5}v^2=-\tfrac12 Y_\chi^T M_\Sigma^{-1}Y_\chi\,v^2$. The unitarity operator $\cQ_U^{d=6}$ encodes the nonunitarity of the PMNS matrix and, being built out of a charged triplet, modifies both the neutral- and charged-lepton gauge couplings~\cite{Broncano:2002rw,Abada:2007ux,Blennow:2023mqx}. $\cQ_{JHL}^{d=7}$ is its Majoron dressing, tied to it by $\cC_{JHL}^{d=7}=\cC_U^{d=6}/(2v_\phi)$, exactly as in type~I. The Higgs dressing $\cQ_{HW}^{d=7}$ arises identically in the two matching orders---from the $\rho$ dressing of the Weinberg operator when $\rho$ is removed last and from the Higgs dependence of the triplet mass when $\Sigma$ is removed last---with $\cC_{HW}^{d=7}=\beta_1\cC_W^{d=5}/m_\rho^2$, the same sign as in type~I, Eq.~\eqref{eq:JSMEFT-I-lep}, since here too the radial mode couples to the Weinberg operator through the heavy-mass insertion; and $\cQ_{DHL}^{d=7}$ is its derivative dressing.

We have written these operators in Eq.~\eqref{eq:JSMEFT-III-lep} with the explicit $\sigma^I\otimes\sigma^I$ structure, as it makes their $SU(2)_L$-triplet origin transparent, but this form is redundant and calls for a comment. The Weinberg operator is the clearest instance: the dimension-five operator is \emph{unique}~\cite{Weinberg:1979sa}, so the triplet writing is not an independent operator. Using the identity $\sigma^I_{ab}\sigma^I_{cd}=2\delta_{ad}\delta_{cb}-\delta_{ab}\delta_{cd}$, it reduces to the standard Weinberg operator of Eq.~\eqref{eq:JSMEFT-I-lep}, precisely the situation used in Ref.~\cite{Grzadkowski:2010es} to illustrate the elimination of redundant structures from the operator basis. Its contribution is, therefore, reported directly on that operator, with the very same coefficient $\cC_W^{d=5}=\tfrac12 Y_\chi^T M_\Sigma^{-1}Y_\chi$---unchanged because the two forms yield the same neutrino mass, $m_\nu=-\cC_W^{d=5}v^2$. The same holds for its dressings $\cQ_{HW}^{d=7}$ and $\cQ_{DHL}^{d=7}$, which reduce to the corresponding type~I operators. In the minimal basis, these operators are, thus, \emph{identical} to the type~I ones. Indeed, in the summary of Table~\ref{tab:JSMEFTsummary} they appear as a single, unambiguously identified operator, and they cannot be used to distinguish the fermionic-singlet from the fermionic-triplet realization. The one genuine exception among the triplet operators is the unitarity operator $\cQ_U^{d=6}$: it is redundant, too, but its reduction [Sec.~\ref{sec:JSMEFT-summary}, Eq.~\eqref{eq:OU-warsaw}] distributes onto $\cQ_{Hl}^{(1)}$ and $\cQ_{Hl}^{(3)}$ with a coefficient \emph{ratio} $3:1$ that differs from the type~I one, $-1:1$, and additionally generates $\cQ_{eH}$. It is this different pattern of Warsaw coefficients, and not the presence of the operator itself, that carries genuine type~III information at the level of the effective interactions.

Beyond it, the sharpest distinctive feature of the type~III realization is in the anomalous couplings of the Majoron. The charged components $\Sigma^\pm$ of the triplets contribute to the $SU(2)_L^2$ anomaly, so that $c_W/c_B=+3$ instead of the $-1$ of types~I and~II, and $c_W+c_B=-6\neq0$. As shown in Eq.~\eqref{Jops-typeIII}, projecting the anomalous term onto the photon then yields a genuine two-photon coupling of the Majoron, $\sL\supset-(\alpha/2\pi)(J/v_\phi)\,n_\Sigma\,F_{\mu\nu}\widetilde F^{\mu\nu}$, absent in the other two realizations. This $J\to\gamma\gamma$ vertex is the sharpest phenomenological discriminator among the three seesaw mechanisms, as we discuss in Sec.~\ref{sec:pheno}.

\subsection{The type~II Majoron model}
\label{sec:JSMEFT-II}

Integrating out both $\rho$ and the triplet $\Delta$---either removing $\rho$ from the $\rho$JSMEFT-IIa Lagrangian, Eq.~\eqref{rhoJSMEFTIIaLag} (which, with $M_\Delta^2\to\mu_3^2$, also covers the $\rho$JSMEFT-IIb case), or removing $\Delta$ from the $\Delta$JSMEFT Lagrangian, Eq.~\eqref{DeltaJSMEFTLag}---yields the same JSMEFT Lagrangian, valid below $\min(m_\rho,M_\Delta)$:
\be
\begin{split}
\sL^\text{JSMEFT}_\text{II}=&\sL_\text{SM}+\dfrac12\derp_\mu J\derp^\mu J+\dfrac{\partial_\mu J}{2v_\phi}\left(\ov{L_L}\gamma^\mu L_L+\ov{e_R}\gamma^\mu e_R\right)+\\
&-\dfrac{3\,J}{32\pi^2v_\phi}\left(g^{\prime 2} B_{\mu\nu}\widetilde{B}^{\mu\nu}-g^2 W^i_{\mu\nu}\widetilde{W}^{i\mu\nu}\right)+\sum_i\left(\cC_i^d\,\cQ_i^d+\hc\right)\,.
\end{split}
\label{JSMEFTLag-II}
\ee
The Majoron couplings are the very same as in the type~I case, Eq.~\eqref{JSMEFTLag-I}: the derivative coupling to the light-lepton current and the electroweak anomaly, the latter yielding again no $J\gamma\gamma$ vertex since $c_W+c_B=0$. Indeed, the type~II model adds no charged fermion to the SM spectrum, so the anomaly coefficients coincide with those of type~I. Contrary to that case, however, the Majoron does not couple to the leptons through any $\cQ_{JHL}$-like operator. In fact, the unitarity operator $\cQ_U^{d=6}$ and its Majoron dressing are absent here, as they originated from the fermionic propagator of the heavy $N$, whereas the scalar $\Delta$ yields instead a genuine four-lepton contact interaction.

Most of the operators coincide with those already introduced for the type~I JSMEFT, generated by the same $\rho$ exchange and, in addition, by the $\Delta$ exchange. We do not repeat their definitions and quote only the Wilson coefficients. The bosonic operators $\cQ_{H^2}^{d=2}$, $\cQ_\lambda^{d=4}$, $\cQ_{H\Box}^{d=6}$, and $\cQ_{JH}^{d=6}$ of Eq.~\eqref{eq:JSMEFT-I-bos} reappear with
\be
\cC_{H^2}^{d=2}=-\frac{\beta_1 m_\rho^2}{8\lambda_1}\,,\qquad
\cC_{JH}^{d=6}=-\frac{\beta_1}{2m_\rho^2}\,,\qquad
\cC_{\lambda}^{d=4}=\frac{\beta_1^2}{8\lambda_1}+\frac{\beta_5^2 v_\phi^2}{2M_\Delta^2}\,,\qquad
\cC_{H\Box}^{d=6}=-\frac{\beta_1^2}{8\lambda_1 m_\rho^2}+\frac{\beta_5^2 v_\phi^2}{2M_\Delta^4}\,,
\label{eq:JSMEFT-II-bos}
\ee
where $\cQ_{H^2}^{d=2}$ and $\cQ_{JH}^{d=6}$ keep exactly the type~I coefficients, while $\cQ_\lambda^{d=4}$ and $\cQ_{H\Box}^{d=6}$ acquire, on top of the type~I $\rho$ contribution, the $\Delta$-induced $\beta_5$ pieces. Likewise, the lepton-number-violating $\cQ_W^{d=5}$, $\cQ_{DHL}^{d=7}$, and $\cQ_{HW}^{d=7}$ of Eq.~\eqref{eq:JSMEFT-I-lep} reappear, now generated by the $\Delta$ exchange, with
\be
\begin{gathered}
\cC_W^{d=5}=\dfrac{\sqrt2\,\beta_5 v_\phi}{M_\Delta^2}\left(Y_\Delta\right)_{\alpha\beta}\,,\qquad
\cC_{DHL}^{d=7}=-\dfrac{\cC_W^{d=5}}{M_\Delta^2}\,,\\
\cC_{HW}^{d=7}=-\dfrac{\sqrt2\,(\beta_3+\beta_4)\beta_5 v_\phi}{M_\Delta^4}\left(Y_\Delta\right)_{\alpha\beta}-\dfrac{\beta_1}{m_\rho^2}\cC_W^{d=5}\left(1-\dfrac{\beta_2v_\phi^2}{M_\Delta^2}\right)\,.
\end{gathered}
\label{eq:JSMEFT-II-lep}
\ee
The Weinberg operator reproduces the standard type~II seesaw prediction for the active neutrino masses, $m_\nu=-\cC_W^{d=5}v^2=-\sqrt2\,\beta_5\,Y_\Delta\,v_\phi\,v^2/M_\Delta^2$, with the dynamically generated trilinear $\beta_5 v_\phi/\sqrt2$ playing the role of the ordinary $\mu$ parameter, and $\cQ_{DHL}^{d=7}$ is again its derivative dressing, $\cC_{DHL}^{d=7}=-\cC_W^{d=5}/M_\Delta^2$. The Higgs dressing $\cQ_{HW}^{d=7}$ instead receives two contributions: the $(\beta_3+\beta_4)$ piece from the Higgs dependence of the triplet mass, already present in the $\rho$JSMEFT-IIa, and the $-\beta_1\cC_W^{d=5}/m_\rho^2$ piece from the $\rho$ dressing of the Weinberg operator when $\rho$ is removed last---with a sign opposite to the type~I one, Eq.~\eqref{eq:JSMEFT-I-lep}, since here the radial mode couples to the Weinberg operator through the trilinear source rather than through the heavy-mass insertion. The $\beta_2$ correction multiplying that piece is the one already carried by $\cC_{W\rho}^{d=6}$, Eq.~\eqref{eq:rhoJSMEFT-IIa-d6}, and originates in the $\rho$ dependence of the triplet mass.

Genuinely new to the type~II realization are the four-lepton contact operator, which replaces the type~I unitarity operator, and a set of purely bosonic operators descending from the $\Delta$ exchange:
\be
\begin{aligned}
\cQ_{ll}^{d=6}&=\left(\ov{L_L}\gamma^\mu L_L\right)_{\alpha\beta}\left(\ov{L_L}\gamma_\mu L_L\right)_{\gamma\delta},\qquad&
\cC_{ll}^{d=6}&=-\dfrac{\left(Y_\Delta^\dag\right)_{\alpha\gamma}\left(Y_\Delta\right)_{\beta\delta}}{2M_\Delta^2},\\
\cQ_{HD}^{d=6}&=\left(H^\dag D_\mu H\right)^\dag\left(H^\dag D^\mu H\right),\qquad&
\cC_{HD}^{d=6}&=\frac{\beta_5^2 v_\phi^2}{M_\Delta^4},\\
\cQ_{H}^{d=6}&=\left(H^\dag H\right)^3,\qquad&
\cC_{H}^{d=6}&=\frac{2\lambda_2\beta_5^2 v_\phi^2}{M_\Delta^4}-\frac{(\beta_3+\beta_4)\beta_5^2 v_\phi^2}{2M_\Delta^4}+\\
&&&-\frac{\beta_1\beta_5^2 v_\phi^2}{m_\rho^2M_\Delta^2}\left(1-\frac{\beta_2v_\phi^2}{2M_\Delta^2}\right),\\
\cQ_{eH}^{d=6}&=\left(H^\dag H\right)\left(\ov{L_L}H e_R\right),\qquad&
\cC_{eH}^{d=6}&=\frac{\beta_5^2 v_\phi^2}{M_\Delta^4}\,Y_\ell
\end{aligned}
\label{eq:JSMEFT-II-new}
\ee
together with the quark analogs of the last operator, $\cQ_{uH}^{d=6}=\left(H^\dag H\right)\left(\ov{Q_L}\widetilde H u_R\right)$ and $\cQ_{dH}^{d=6}=\left(H^\dag H\right)\left(\ov{Q_L}H d_R\right)$, with $\cC_{uH}^{d=6}=\beta_5^2 v_\phi^2\,Y_u/M_\Delta^4$ and $\cC_{dH}^{d=6}=\beta_5^2 v_\phi^2\,Y_d/M_\Delta^4$. The four-lepton operator, arising from the square of the Yukawa current $\cJ_L^\dag\cJ_L$, contributes to charged-lepton-flavor-violating processes such as $\mu\to3e$. Unlike the type~I unitarity operator, which came with a derivative from the fermionic propagator, here the scalar propagator of $\Delta$ yields a genuine contact interaction. Fierzing the $SU(2)_L$-triplet contraction, it takes the standard Warsaw form $\cQ_{ll}^{d=6}$~\cite{Grzadkowski:2010es}, the same $d=6$ operator of the SMEFT description of the type~II seesaw~\cite{Li:2022ipc}. The bosonic operators $\cQ_{HD}^{d=6}$, $\cQ_{H}^{d=6}$ and the Yukawa ones all descend from the $D^2$ dressing of the trilinear source $\cJ_\Delta$, which generates $\left(H^\dag H\right)\left(D_\mu H^\dag D^\mu H\right)+|H^\dag D_\mu H|^2$ with coefficient $\beta_5^2 v_\phi^2/M_\Delta^4$: the second structure is directly $\cQ_{HD}^{d=6}$, while the first is redistributed onto the Warsaw basis through the Higgs EOM, as detailed below Eq.~\eqref{eq:rhoJSMEFT-IIa-d7}. That reduction feeds the $\lambda_2$ piece of $\cC_H^{d=6}$ and the Yukawa operators and further shifts $\cC_\lambda^{d=4}$ by $-\mu_2^2\beta_5^2 v_\phi^2/M_\Delta^4$, which merely renormalizes $\lambda_2$ and we do not display; the $(\beta_3+\beta_4)$ piece of $\cC_H^{d=6}$ comes instead from the Higgs dependence of the triplet mass---the very combination entering $\cC_{HW}^{d=7}$---and the last piece from the $\rho$-$\Delta$ cross term, whose $\beta_2$ correction is inherited from $\cC_{\rho H}^{d=5}$, Eq.~\eqref{eq:rhoJSMEFT-IIa-d5}. As in the intermediate EFTs, the Weinberg and four-lepton operators, together with their $d=7$ corrections, belong to the SMEFT of the type~II seesaw~\cite{Li:2022ipc}, while the Majoron and radial-mode operators are specific to the present construction.

\subsection{Summary and comparison}
\label{sec:JSMEFT-summary}

We collect in Table~\ref{tab:JSMEFTsummary} the operators of the three low-energy JSMEFTs derived above, Eqs.~\eqref{JSMEFTLag-I}, \eqref{JSMEFTLag-II}, and~\eqref{JSMEFTLag-III}, indicating which seesaw realization generates each of them, with the Wilson coefficients being collected in the respective subsections. In the table, we highlight in red the operators generated purely by the radial-mode $\rho$ exchange: the $\beta_1$ Higgs-portal operators $\cQ_{H^2}$, $\cQ_\lambda$, $\cQ_{H\Box}$, and the Majoron kinetic dressing $\cQ_{JH}$. These are the ingredients genuinely specific to the Majoron extension, since the mutual comparison of the ordinary type~I, II, and III seesaws is by itself standard, and it is only the addition of the lepton-number-breaking scalar $\phi$---and, hence, of the radial mode $\rho$ and the Majoron $J$---that brings them in. Notice that in type~II $\cQ_\lambda$ and $\cQ_{H\Box}$ receive in addition a contribution from the triplet $\Delta$, already present in the ordinary type~II seesaw, and are, therefore, not flagged in that column. The comparison makes the disentangling power of the effective description explicit. All three realizations share the bosonic Higgs-portal operators inherited from the radial mode $\rho$ and the derivative Majoron couplings to the light leptons, so these are blind to the underlying mechanism. The discriminating information sits instead in a few well-defined places. The Higgs-current operators $\cQ_{Hl}^{(1)}$ and $\cQ_{Hl}^{(3)}$---the Warsaw form of the unitarity operator; cf.~Eq.~\eqref{eq:OU-warsaw}---and the Majoron dressing $\cQ_{JHL}^{d=7}$, which stem from the derivative expansion of a heavy \emph{fermionic} propagator, are generated in types~I and~III but not in type~II, where the scalar triplet yields instead the four-lepton contact operator $\cQ_{ll}^{d=6}$ and the additional operators $\cQ_{HD}^{d=6}$ and $\cQ_{H}^{d=6}$ and the Higgs-dressed Yukawas $\cQ_{eH},\cQ_{uH}$, and $\cQ_{dH}$. The two fermionic realizations are distinguished in three complementary ways: the two-photon coupling $J\,F_{\mu\nu}\widetilde F^{\mu\nu}$, generated by the charged components of the $SU(2)_L$ triplet and present only in type~III; the different coefficient ratio $\cC_{Hl}^{(1)}\!:\!\cC_{Hl}^{(3)}$, equal to $-1:1$ in type~I and to $3:1$ in type~III; and the Yukawa operator $\cQ_{eH}$, produced in type~III by the same reduction of the unitarity operator but forbidden in type~I by $\widetilde H^\dag H=0$. No single operator identifies a mechanism, but the pattern of presences, absences, and coefficient ratios does, and it is the starting point of the phenomenological discussion in Sec.~\ref{sec:pheno}.

\begin{table}[ht!]
\centering
\renewcommand{\arraystretch}{1.35}
\begin{tabular}{@{}lccc@{}}
\toprule
Operator & Type~I & Type~II & Type~III \\
\midrule
$\cQ_{H^2}=\left(H^\dag H\right)$                                  & {\color{red}\checkmark} & {\color{red}\checkmark} & {\color{red}\checkmark} \\
$\cQ_{\lambda}=\left(H^\dag H\right)^2$                            & {\color{red}\checkmark} & \checkmark & {\color{red}\checkmark} \\
$\cQ_{H\Box}=\left(H^\dag H\right)\Box\left(H^\dag H\right)$       & {\color{red}\checkmark} & \checkmark & {\color{red}\checkmark} \\
$\cQ_{HD}=\left(H^\dag D_\mu H\right)^\dag\left(H^\dag D^\mu H\right)$ &         & \checkmark &            \\
$\cQ_{H}=\left(H^\dag H\right)^3$                                  &            & \checkmark &            \\
$\cQ_{eH}=\left(H^\dag H\right)\left(\ov{L_L}H e_R\right)$         &            & \checkmark & \checkmark \\
$\cQ_{uH},\,\cQ_{dH}=\left(H^\dag H\right)\left(\ov{Q_L}\widetilde H u_R\right),\,\left(H^\dag H\right)\left(\ov{Q_L}H d_R\right)$ & & \checkmark & \\
\midrule
$\cQ_{JL}=\partial_\mu J\left(\ov{L_L}\gamma^\mu L_L\right)$       & \checkmark & \checkmark & \checkmark \\
$\cQ_{Je}=\partial_\mu J\left(\ov{e_R}\gamma^\mu e_R\right)$       & \checkmark & \checkmark & \checkmark \\
$\cQ_{JH}=\left(\partial_\mu J\partial^\mu J\right)\left(H^\dag H\right)$ & {\color{red}\checkmark} & {\color{red}\checkmark} & {\color{red}\checkmark} \\
$\cQ_{JHL}=\left(\ov{L_L}\widetilde H\right)\slashed\partial J\gamma_5\left(\widetilde H^\dag L_L\right)$ & \checkmark & & \checkmark \\
$J\,F_{\mu\nu}\widetilde F^{\mu\nu}$\quad($J\to\gamma\gamma$)      &            &            & \checkmark \\
\midrule
$\cQ_{W}=\left(\ov{L_L}\widetilde H\right)\left(\widetilde H^T L_L^c\right)$ & \checkmark & \checkmark & \checkmark \\
$\cQ_{HW}=\cQ_W\left(H^\dag H\right)$                              & \checkmark & \checkmark & \checkmark \\
$\cQ_{DHL}=\left(\ov{L_L}\widetilde H\right)D^2\left(\widetilde H^T L_L^c\right)$ & \checkmark & \checkmark & \checkmark \\
\midrule
$\cQ_{ll}=\left(\ov{L_L}\gamma^\mu L_L\right)\left(\ov{L_L}\gamma_\mu L_L\right)$ & & \checkmark & \\
$\cQ_{Hl}^{(1)}=\left(H^\dag i\!\overleftrightarrow{D}_\mu H\right)\left(\ov{L_L}\gamma^\mu L_L\right)$ & \checkmark & & \checkmark \\
$\cQ_{Hl}^{(3)}=\left(H^\dag i\!\overleftrightarrow{D}^I_\mu H\right)\left(\ov{L_L}\sigma^I\gamma^\mu L_L\right)$ & \checkmark & & \checkmark \\
\bottomrule
\end{tabular}
\caption{\em Operators of the low-energy JSMEFT in the minimal (Warsaw) basis and the seesaw realization that generates each of them, grouped into the bosonic and Higgs-Yukawa, Majoron, lepton-number-violating, and lepton-number-conserving sectors. Check marks in {\color{red}red} flag the operators generated purely by the radial-mode ($\rho$) exchange. These owe their existence to the Majoron sector, whereas all the others already appear in the effective description of the ordinary seesaws. The redundant type~III triplet operators of Eq.~\eqref{eq:JSMEFT-III-lep} have been reduced to this basis, as discussed in the text. The Wilson coefficients are listed in Secs.~\ref{sec:JSMEFT-I}, \ref{sec:JSMEFT-III}, and~\ref{sec:JSMEFT-II}.}
\label{tab:JSMEFTsummary}
\end{table}

A comment on the unitarity operator $\cQ_U^{d=6}$ is in order, given its central phenomenological role. As anticipated, it does not belong to the Warsaw basis of the SMEFT~\cite{Grzadkowski:2010es}. However, through the fermion equations of motion and Fierz identities~\cite{Broncano:2002rw}, it can be traded for the Higgs-current operators
\be
\cQ_{Hl}^{(1)}=\left(H^\dag i\!\overleftrightarrow{D}_\mu H\right)\left(\ov{L_L}\gamma^\mu L_L\right)\,,\qquad
\cQ_{Hl}^{(3)}=\left(H^\dag i\!\overleftrightarrow{D}^I_\mu H\right)\left(\ov{L_L}\sigma^I\gamma^\mu L_L\right)\,,
\ee
where $H^\dag i\!\overleftrightarrow{D}_\mu H\equiv iH^\dag(D_\mu H)-i(D_\mu H^\dag)H$ and $H^\dag i\!\overleftrightarrow{D}^I_\mu H\equiv iH^\dag\sigma^I(D_\mu H)-i(D_\mu H^\dag)\sigma^I H$. Carrying out the reduction, one finds
\be
\text{type~I:}\quad \cC_{Hl}^{(1)}=-\cC_{Hl}^{(3)}=\dfrac14\,\cC_U^{d=6}\,,\qquad\qquad
\text{type~III:}\quad \cC_{Hl}^{(1)}=3\,\cC_{Hl}^{(3)}=\dfrac34\,\cC_U^{d=6}\,,
\label{eq:OU-warsaw}
\ee
with, in the type~III case only, an additional Higgs-dressed charged-lepton Yukawa operator \be
\cQ_{eH}=(H^\dag H)(\ov{L_L}H e_R)\qquad \text{with} \qquad \cC_{eH}=\cC_U^{d=6}\,Y_\ell\,.
\ee
The latter is absent in type~I, where it is forbidden by $\widetilde H^\dag H=0$, while in type~III it is generated because the charged components of the triplet make $\widetilde H^\dag\sigma^I H\neq0$. The different ratio $\cC_{Hl}^{(1)}:\cC_{Hl}^{(3)}$---equal to $-1:1$ for the fermionic singlet and to $3:1$ for the fermionic triplet---together with the presence or absence of $\cQ_{eH}$---encodes the distinct way the two fermionic realizations modify the neutral- and charged-current couplings of the light leptons and provides a further handle, beyond the $J\to\gamma\gamma$ coupling, to disentangle type~I from type~III in electroweak-precision and nonunitarity observables, as we discuss next.

\section{Phenomenological Analysis}
\label{sec:pheno}

The three low-energy Lagrangians in Sec.~\ref{sec:JSMEFT} differ from one another in ways that, taken by themselves, reproduce the familiar comparison among the ordinary type~I, II, and III seesaws: the nonunitarity of the leptonic mixing matrix, the four-lepton contact interactions, and the custodial-breaking effects of a scalar triplet. That comparison is standard, and we refer to the citations already mentioned for this part of the analysis. The question specific to the present construction is a different one: the models in Sec.~\ref{sec:HEEFTs} do not merely realize a seesaw, they realize it through the \emph{spontaneous} breaking of the lepton number, and the price---or the reward---is a pair of new states, the radial mode $\rho$ and the Majoron $J$. Can one establish that this is what happened? And, if so, can one identify which Majoron model is at work?

The answer rests on a single structural fact, which the effective description makes manifest. The VEV $v_\phi$ is the \emph{only} scale that simultaneously sets the mass of the seesaw mediator ($M_N=Y_{NN}v_\phi/\sqrt2$, $M_\Sigma=Y_\Sigma v_\phi/\sqrt2$, or the $\beta_2$ part of $M_\Delta^2$), the mass and couplings of the radial mode ($m_\rho^2=2\lambda_1v_\phi^2$), and the couplings of the Majoron, which is the Goldstone boson of the very symmetry whose breaking generates the neutrino mass. Nothing in the Majoron models is free to move independently: every signature is a \emph{correlation} that collapses onto $v_\phi$. Generic new physics, in contrast, populates the same operators with unrelated Wilson coefficients. It is, therefore, the pattern of relations among observables, and not the size of any single one, that carries the identification power, and the phenomenological program is to find the places where that pattern is sharp enough to be tested.

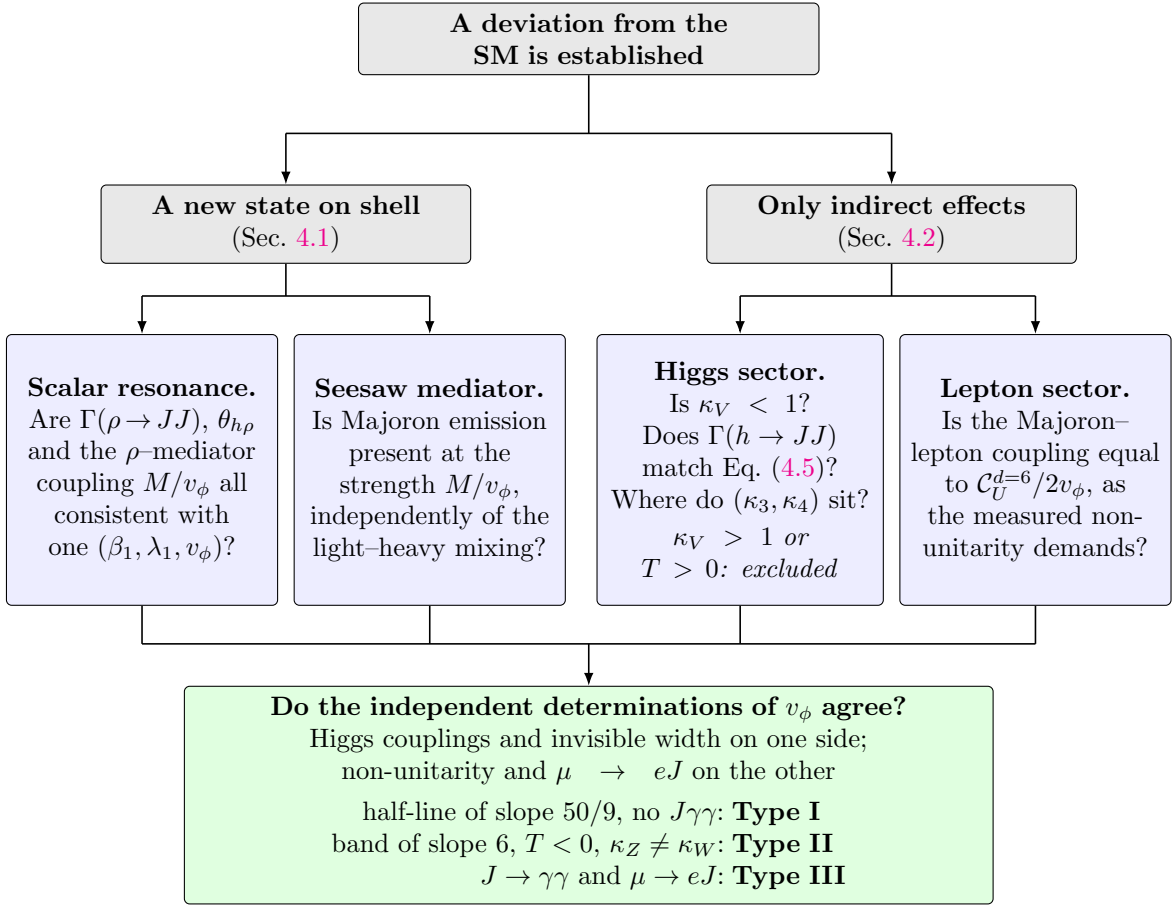
\begin{figure}[ht!]
\centering
\begin{tikzpicture}[
  font=\footnotesize,
  bx/.style ={draw, rounded corners=2pt, align=center, inner sep=4pt, text width=46mm},
  hd/.style ={bx, fill=gray!18},
  ts/.style ={bx, fill=blue!7, minimum height=35.8mm},
  gd/.style ={bx, fill=green!12, text width=104mm},
  ar/.style ={-{Latex[length=1.7mm]}, semithick}
]
\node[hd, text width=58mm] (top) at (0,0) {\textbf{A deviation from the SM is established}};

\draw[semithick] (top.south) -- (0,-1.25);
\draw[semithick] (-4,-1.25) -- (4,-1.25);
\node[hd] (res) at (-4,-2.45) {\textbf{A new state on shell}\\ (Sec.~\ref{sec:pheno-high})};
\node[hd] (eft) at ( 4,-2.45) {\textbf{Only indirect effects}\\ (Sec.~\ref{sec:pheno-low})};
\draw[ar] (-4,-1.25) -- (res.north);
\draw[ar] ( 4,-1.25) -- (eft.north);

\node[ts, text width=33mm] (rho) at (-5.9,-5.7) {\textbf{Scalar resonance.}\\ Are $\Gamma(\rho\!\to\!JJ)$, $\theta_{h\rho}$ and the $\rho$--mediator coupling $M/v_\phi$ all consistent with one $(\beta_1,\lambda_1,v_\phi)$?};
\node[ts, text width=33mm] (med) at (-2.1,-5.7) {\textbf{Seesaw mediator.}\\ Is Majoron emission present at the strength $M/v_\phi$, independently of the light--heavy mixing?};
\draw[semithick] (res.south) -- (-4,-3.4);
\draw[semithick] (-5.9,-3.4) -- (-2.1,-3.4);
\draw[ar] (-5.9,-3.4) -- (rho.north);
\draw[ar] (-2.1,-3.4) -- (med.north);

\node[ts, text width=35mm] (hig) at (2.0,-5.7) {\textbf{Higgs sector.}\\ Is $\kappa_V<1$?\\ Does $\Gamma(h\!\to\!JJ)$ match Eq.~\eqref{eq:hJJmaster}? Where do $(\kappa_3,\kappa_4)$ sit?\\[2pt] \emph{$\kappa_V>1$ or $T>0$: excluded}};
\node[ts, text width=33mm] (lep) at (5.9,-5.7) {\textbf{Lepton sector.}\\
Is the Majoron--lepton coupling equal to $\cC_U^{d=6}/2v_\phi$, as the measured non-unitarity demands?};
\draw[semithick] (eft.south) -- (4,-3.4);
\draw[semithick] (2.0,-3.4) -- (5.9,-3.4);
\draw[ar] (2.0,-3.4) -- (hig.north);
\draw[ar] (5.9,-3.4) -- (lep.north);

\draw[semithick] (rho.south) -- (-5.9,-8.0);
\draw[semithick] (med.south) -- (-2.1,-8.0);
\draw[semithick] (hig.south) -- (2.0,-8.0);
\draw[semithick] (lep.south) -- (5.9,-8.0);
\draw[semithick] (-5.9,-8.0) -- (5.9,-8.0);
\node[gd] (fin) at (0,-10.) {\textbf{Do the independent determinations of $v_\phi$ agree?}\\ Higgs couplings and invisible width on one side; non-unitarity and $\mu\to eJ$ on the other\\[4pt]
\begin{tabular}{@{}r@{\;}l@{}}
half-line of slope $50/9$, no $J\gamma\gamma$: & \textbf{Type~I}\\
band of slope $6$, $T<0$, $\kappa_Z\neq\kappa_W$: & \textbf{Type~II}\\
$J\to\gamma\gamma$ and $\mu\to eJ$: & \textbf{Type~III}
\end{tabular}};
\draw[ar] (0,-8.0) -- (fin.north);
\end{tikzpicture}
\caption{\em Logical structure of the phenomenological analysis. The two branches correspond to the two regimes in Secs.~\ref{sec:pheno-high} and~\ref{sec:pheno-low}. In each case, the identification does not rest on the size of a single observable but on whether independent measurements return a consistent value of the lepton-number-breaking scale $v_\phi$, which is what the Majoron models---and not generic new physics---predict.}
\label{fig:flowchart}
\end{figure}

Two regimes must be distinguished, mirroring the two-step matching in Sec.~\ref{sec:HEEFTs}. Either one of the heavy states lies within kinematical reach and can be produced as a resonance, in which case the correlations can be tested directly on the spectrum (Sec.~\ref{sec:pheno-high}); or both lie far above, only the JSMEFT is accessible, and the model can be probed solely through the correlated pattern of its Wilson coefficients (Sec.~\ref{sec:pheno-low}). The logic of the two regimes, and the specific tests associated to each of them, are summarized in Fig.~\ref{fig:flowchart}, which may serve as a map of the rest of this section.

Both regimes, however, are probes of $\rho$ and of the seesaw mediators. A direct search targets one of them on shell; an indirect one constrains the pattern of Wilson coefficients they leave behind once removed. The Majoron itself enters only through these same channels, never as the target of a dedicated search of its own. This is a choice of focus, not an approximation. Throughout this work, $J$ is the Goldstone boson of an exactly unbroken continuous $U(1)_L$ and is accordingly treated as exactly massless. A nonzero $m_J$ would open genuinely Majoron-specific channels, among them a two-loop Majoron-diphoton coupling and an enhanced charged-lepton-flavor-violating vertex, both proportional to $m_J^2$~\cite{Garcia-Cely:2017oco,Heeck:2019guh}. Such a mass is, in fact, generically expected, since nonperturbative gravitational effects are believed to explicitly break every global continuous symmetry~\cite{Akhmedov:1992hi,Rothstein:1992rh}, but its size is set by the free coefficient of the corresponding Planck-suppressed operator and calls for a dedicated study across the possible mass regimes~\cite{deGiorgi:2023tvn}, which we leave for future work.

Before concluding, it may be convenient to phrase part of the discussion in the standard $\kappa$ framework~\cite{LHCHiggsCrossSectionWorkingGroup:2012nn,LHCHiggsCrossSectionWorkingGroup:2016ypw}, in which every Higgs coupling is normalized to its SM value, $\kappa_i\equiv g_i/g_i^\text{SM}$, so that $\kappa_i=1$ reproduces the SM. This parametrization assumes that new physics presents the same tensor structure of the couplings as in the SM. In the following, we use $\kappa_V$ for the universal coupling modifier to gauge bosons $W$ and $Z$, while $\kappa_f$ refers to the modification of the couplings involving fermions. On the other hand, $\kappa_3$ and $\kappa_4$ stand for the trilinear and quartic Higgs self-couplings modifiers, respectively. In the specific case of type~II, we additionally need $\kappa_W$ and $\kappa_Z$, the coupling to $W$ or $Z$ bosons alone, since custodial-symmetry breaking decouples the coupling to $W$ from that to $Z$.

\subsection{Higher energies}
\label{sec:pheno-high}

This regime refers to the energy scales in which only one of the heavier degrees of freedom has been integrated out.
The discussion that follows is deliberately kept qualitative. No resonance attributable to either the radial mode or a seesaw mediator has been observed, and, in the absence of any experimental indication of its mass, of its mixing with the Higgs, or of the size of the invisible Higgs decay, a numerical analysis would require choosing those quantities by hand. Any such choice would be arbitrary, and quoting rates or exclusion regions derived from it would suggest a predictivity that the framework does not have in this regime. What the effective description does provide, and what we therefore emphasize here, is the \emph{structure} of the relations that a discovery would have to satisfy: these are fixed by the model and independent of where the resonance happens to sit.

Suppose then that a new state is discovered. Two cases arise, according to whether it belongs to the symmetry-breaking sector or to the seesaw sector, and in both the diagnostic question is the same: are the couplings of the new state related to $v_\phi$ in the way the Majoron models demand?

\paragraph{\boldmath A scalar resonance compatible with $\rho$\unboldmath} A new neutral scalar mixing with the Higgs is by itself a generic signal: any singlet coupled through a Higgs portal produces one. What distinguishes the radial mode is that its couplings are not independent parameters. It couples to the Higgs through $\beta_1$, which, after electroweak symmetry breaking, induces an off-diagonal mass term $\beta_1v_\phi v\,h\rho$ between the CP-even fluctuations $h$ and $\rho$. Diagonalizing the resulting $2\times2$ mass matrix defines the $h$-$\rho$ mixing angle:
\be
\tan2\theta_{h\rho}=\dfrac{2\beta_1v_\phi v}{m_\rho^2-m_h^2}\,,
\label{eq:thetahrhoDEF}
\ee
and the light mass eigenstate, identified with the observed Higgs boson, inherits only the $h$ component of every SM coupling, uniformly rescaled by $\cos\theta_{h\rho}$. This is precisely $\kappa_V$, the universal modification of the Higgs couplings to gauge bosons (and, by the same mechanism, to fermions), which reduces to $\kappa_V=1+\cC_{H\Box}^{d=6}v^2$ of Eq.~\eqref{eq:kappaV} once $\rho$ is integrated out, in the EFT limit $m_\rho\gg m_h$. The same $\beta_1$ also fixes the whole set of bosonic operators in Sec.~\ref{sec:JSMEFT}. $\rho$ couples to a pair of Majorons through the scalar potential, with a strength controlled by $m_\rho^2/v_\phi$, so that a $\rho\to JJ$ channel---invisible, or displaced, depending on the Majoron lifetime---is unavoidable and not optional; and, crucially, it couples to the seesaw mediator through the $\rho$ dependence of the heavy mass, $M(\rho)=M(1+\rho/v_\phi)$ for the fermionic realizations, so that the $\rho$-mediator vertex is fixed to $M/v_\phi$ once the mediator mass is known.

The consequence is an overconstrained system. The observables $\{m_\rho,\ \theta_{h\rho},\ \Gamma(\rho\to JJ)\}$, and, if a mediator is also within reach, $\{M_\text{med},\ \Gamma(\rho\to\text{med}\,\text{med})\}$, are all determined by the three underlying parameters $\beta_1$, $\lambda_1$, and $v_\phi$. Measuring more of them than the total number of parameters constitutes a genuine test: a generic singlet passes it only by accident. This is the sense in which a $\rho$ discovery could be attributed to a Majoron model rather than merely be compatible with one. Which realization is at work is then read from the accompanying Majoron couplings and, in particular, from the two-photon vertex of Eq.~\eqref{Jops-typeIII}, present in type~III and absent in types~I and~II. 

\paragraph{A seesaw mediator} Suppose instead that a seesaw mediator is found. The seesaw interpretation is then immediate, but it does not by itself reveal how lepton number is broken: the ordinary seesaws break it explicitly and reproduce the same neutrino masses. The discriminating observable is the coupling of the mediator to the Majoron. We restrict what follows to the two fermionic mediators, the HNL of type~I and the fermion triplet of type~III, whose mass comes entirely from $v_\phi$. The same Yukawa that generates their mass also couples them to the Goldstone direction, with a strength fixed by $M/v_\phi$, and a mediator whose lepton number is broken explicitly has no such vertex. Observing the seesaw production and decay modes \emph{and} a Majoron emission of the predicted relative strength would establish the spontaneous nature of the breaking; observing the former without the latter, at a sensitivity where the latter is predicted, would exclude it. The scalar triplet of type~II admits the same test but not the same formula: part of its mass is the lepton-number-conserving $\mu_3^2$, and its Majoron coupling descends from the kinetic term, Eq.~\eqref{Jops-typeII}, so it is derivative and set by the LN charge of $\Delta$ rather than proportional to its mass.

The collider phenomenology of precisely this vertex has been studied in Refs.~\cite{deGiorgi:2022oks,Marcos:2024yfm}, in the closely related language of a heavy neutral lepton coupled to an axionlike particle, of which the Majoron is an instance. Two features of those analyses are worth importing here. The first is that, because a Goldstone couples to fermions proportionally to their mass, the vertex is \emph{enhanced} for heavy mediators rather than suppressed, so that TeV-scale heavy neutral leptons give a phenomenologically relevant signal. This is the same mass proportionality that will reappear, with the opposite consequence, when we discuss the light-lepton couplings in Sec.~\ref{sec:pheno-low}. The second, and more important, is that the sensitivity of the relevant channels---a clean final state with four jets and two charged leptons at the LHC in Ref.~\cite{deGiorgi:2022oks}, extended to muon colliders in Ref.~\cite{Marcos:2024yfm}---does \emph{not} depend on the light-heavy mixing, whose overall scale drops out. This matters because the mixing is the parameter that the ordinary seesaw makes small: a probe that is independent of it can reach down to the genuine seesaw regime, and it tests the Majoron coupling directly rather than through the suppressed mixing. Concretely, the mediators are produced in pairs through the Majoron vertex itself, Eq.~\eqref{NJSMEFTLag}, and each of them decays, via its mixing with the active neutrinos, into a charged lepton and two jets, which is where the four jets and two leptons come from. Because the mediators are on shell, the overall scale of the mixing cancels between the partial and the total width, and what the rate measures is the Majoron-mediator coupling alone. In the Majoron models that coupling is not free but equal to $M/v_\phi$, so a signal in this channel is a determination of $v_\phi$ once the mediator mass has been reconstructed. It should be stressed that in these searches the mediator is produced on shell but is never observed as such: what is reconstructed are the decay products, as is always the case for heavy states, and the Majoron itself contributes as missing energy.

The same logic answers the complementary question of whether an observed deviation is due to the radial mode or to the mediator itself: the two contributions carry different $v_\phi$ dependences, and the presence of the scalar partner with the mass and mixing predicted from $m_\rho^2=2\lambda_1v_\phi^2$ is what settles it. In all cases, the identification comes from the pattern of relations among the mediator, the radial mode, and the Majoron, never from a single rate.

\subsection{Lower energies}
\label{sec:pheno-low}

If both heavy scales are out of reach, only the JSMEFT survives, and the Majoron models must be identified indirectly. This is the situation in which the correlations are most valuable, because the Wilson coefficients in Sec.~\ref{sec:JSMEFT} are not independent: those generated by the $\rho$ exchange all descend from $\beta_1$ and $m_\rho$, and their ratios are free of the underlying parameters.

\subsubsection{The Higgs-Majoron correlation}

This correlation is specific to the fermionic realizations, type~I and~III, where the radial mode $\rho$ is the only source of the relevant bosonic operators. In type~II, the scalar triplet $\Delta$ contributes to them as well, and the lock relation derived below does not hold there, as discussed separately below.

Consider first the operators generated purely by the radial mode. Of the four listed in Eq.~\eqref{eq:JSMEFT-I-bos}, two are not observable on their own: $\cQ_{H^2}^{d=2}$ and $\cQ_\lambda^{d=4}$ merely renormalize $\mu_2^2$ and $\lambda_2$ and are reabsorbed into the measured values of $v$ and $m_h$. The physical content is carried by the remaining two, $\cQ_{H\Box}^{d=6}$ and $\cQ_{JH}^{d=6}$, and these are tied together. Using $m_\rho^2=2\lambda_1v_\phi^2$ in the coefficients of Eq.~\eqref{eq:JSMEFT-I-bos}, one finds the parameter-free relation
\be
\cC_{H\Box}^{d=6}=-v_\phi^2\left(\cC_{JH}^{d=6}\right)^2\,,
\label{eq:HiggsMajoronLock}
\ee
which holds identically in $\beta_1$ and $\lambda_1$. The two operators control very different observables. The Majoron dressing $\cQ_{JH}^{d=6}=(\derp_\mu J\derp^\mu J)(H^\dag H)$ contains, after electroweak symmetry breaking, the vertex $\cC_{JH}^{d=6}\,v\,h\,\derp_\mu J\derp^\mu J$ and, therefore, an invisible Higgs decay into a Majoron pair,
\be
\Gamma(h\to JJ)=\dfrac{\left(\cC_{JH}^{d=6}\right)^2v^2m_h^3}{32\pi}\,,
\ee
while $\cQ_{H\Box}^{d=6}$ modifies the Higgs kinetic term and, hence, after canonical normalization, rescales all the Higgs couplings universally:
\be
\kappa_V=1+\cC_{H\Box}^{d=6}\,v^2\,.
\label{eq:kappaV}
\ee
Eliminating the Wilson coefficients between these expressions through Eq.~\eqref{eq:HiggsMajoronLock}, the electroweak VEV cancels, and one is left with
\be
\Gamma(h\to JJ)=\dfrac{\left(1-\kappa_V\right)m_h^3}{32\pi\,v_\phi^2}\,.
\label{eq:hJJmaster}
\ee
This is the sharpest statement the low-energy theory allows. The invisible width of the Higgs is not a free parameter, since, once the suppression of the Higgs couplings is measured, it is fixed by $v_\phi$ alone.
The relation is moreover consistent only in one direction, because $\cC_{H\Box}^{d=6}=-\beta_1^2/8\lambda_1m_\rho^2$ is negative definite, being proportional to $\beta_1^2$: the radial mode necessarily \emph{suppresses} the Higgs couplings, $\kappa_V<1$, and it is precisely this sign that makes the right-hand side of Eq.~\eqref{eq:hJJmaster} positive. A measured enhancement, $\kappa_V>1$, would exclude the whole class of models irrespective of any other observable. 

Numerically, with $\Gamma_\text{SM}\simeq4.1\MeV$, the current bound on the invisible branching ratio, $\text{BR}_\text{inv}<0.107$~\cite{ATLAS:2023tkt}, translates into
\be
v_\phi>\sqrt{1-\kappa_V}\;\times\;6.3\TeV\,,
\ee
that is, $v_\phi\gtrsim0.6$--$1.4\TeV$ for a coupling suppression in the range $1-\kappa_V=1\%$--$5\%$, while the projected high-luminosity sensitivity $\text{BR}_\text{inv}<2.5\%$ would push the same combination to $13.6\TeV$. The relevant window is, therefore, $v_\phi\sim\mathcal{O}(1$--$10)\TeV$, exactly where the precision Higgs program is most effective. The importance of Eq.~\eqref{eq:hJJmaster} is that $v_\phi$ is not a parameter local to the Higgs sector: it is the same scale that normalizes the derivative Majoron coupling to the lepton current, $\derp_\mu J/2v_\phi$, and the anomalous couplings to the gauge bosons, $J/v_\phi$, in Eqs.~\eqref{JSMEFTLag-I}--\eqref{JSMEFTLag-III}. Higgs coupling precision, invisible Higgs decays, and direct Majoron searches---astrophysical, cosmological, and in flavor experiments---therefore probe one and the same number, and the requirement that they agree is a closed and nontrivial test of the framework.

\subsubsection{Higgs self-couplings}

The Higgs self-interactions are distorted as well, and the way in which they are distorted differs sharply between the fermionic and the scalar realizations. We treat the two cases separately, since they are governed by different sets of operators.

\paragraph{Type~I and type~III.} Here, the radial mode is the only source of bosonic operators, and, among those, the single physical one is again $\cQ_{H\Box}^{d=6}$: no $(H^\dag H)^3$ and no $\cQ_{HD}^{d=6}$ are generated, as Table~\ref{tab:JSMEFTsummary} shows. Because one parameter controls everything, the self-couplings are not free. Writing $\kappa_3$ and $\kappa_4$ for the trilinear and quartic couplings normalized to their SM values, one finds, to linear order,
\be
\kappa_3-1=3\left(\kappa_V-1\right)\,,\qquad
\kappa_4-1=\dfrac{50}{3}\left(\kappa_V-1\right)\,,
\label{eq:kappa34lock}
\ee
so that the entire Higgs sector---couplings to gauge bosons and fermions, trilinear and quartic self-couplings---is governed by the single combination $\cC_{H\Box}^{d=6}v^2$, with the sign fixed as above.\footnote{The coefficient of $\kappa_3$ agrees with the standard SMEFT expression $\kappa_\lambda=1-2v^4\cC_H/m_h^2+3(\cC_{H\Box}-\cC_{HD}/4)v^2$ once $\cC_{HD}=0$, as is the case here. The coefficient of $\kappa_4$ does not seem to have been quoted in the literature.} Measuring $\kappa_V$ predicts $\kappa_3$ and $\kappa_4$ with no freedom left, and a violation of Eq.~\eqref{eq:kappa34lock} would signal that the radial mode is not the only contribution. Note that Eq.~\eqref{eq:kappa34lock} is a statement about the fermionic realizations only and does not apply to type~II.

\paragraph{Type~II.} The scalar triplet generates two further bosonic operators, Eq.~\eqref{eq:JSMEFT-II-new}, and both affect the self-couplings. The first is the genuine dimension-six Higgs potential operator $\cQ_H^{d=6}=(H^\dag H)^3$, which shifts the self-couplings without touching the single-Higgs ones; taken alone, it would give the well-known one-parameter line
\be
6\kappa_3=\kappa_4+5\,,
\label{eq:k3k4line}
\ee
equivalently, $\Delta\kappa_4=6\,\Delta\kappa_3$, the standard correlation of a $|H|^6$-dominated deformation. The second is $\cQ_{HD}^{d=6}$, which acts instead on the kinetic side: in unitary gauge it decomposes into a Higgs kinetic piece and a term involving the $Z$ boson only, never the $W$, and it therefore breaks custodial symmetry. Its net effect on the Higgs wave function is to replace $\cC_{H\Box}^{d=6}$ by the combination $\cC_{H\Box}^{d=6}-\cC_{HD}^{d=6}/4$, so that in type~II the analog of Eq.~\eqref{eq:kappa34lock} reads
\be
6\left(\kappa_3-1\right)-\left(\kappa_4-1\right)=\dfrac{4}{3}\left(\kappa_W-1\right)\,,
\label{eq:k3k4typeII}
\ee
where $\kappa_W$, and not $\kappa_V$, is the correct reference precisely because $\cQ_{HD}^{d=6}$ does not affect the $W$. Equation~\eqref{eq:k3k4typeII} has a transparent reading: the right-hand side is the universal wave function distortion, and once it is subtracted---using a measurement of $\kappa_W$---the pure $(H^\dag H)^3$ line in Eq.~\eqref{eq:k3k4line} is recovered. In the fermionic realizations, the same combination vanishes identically, by Eq.~\eqref{eq:kappa34lock}.

\begin{figure}[ht!]
\centering
\includegraphics[width=0.85\textwidth]{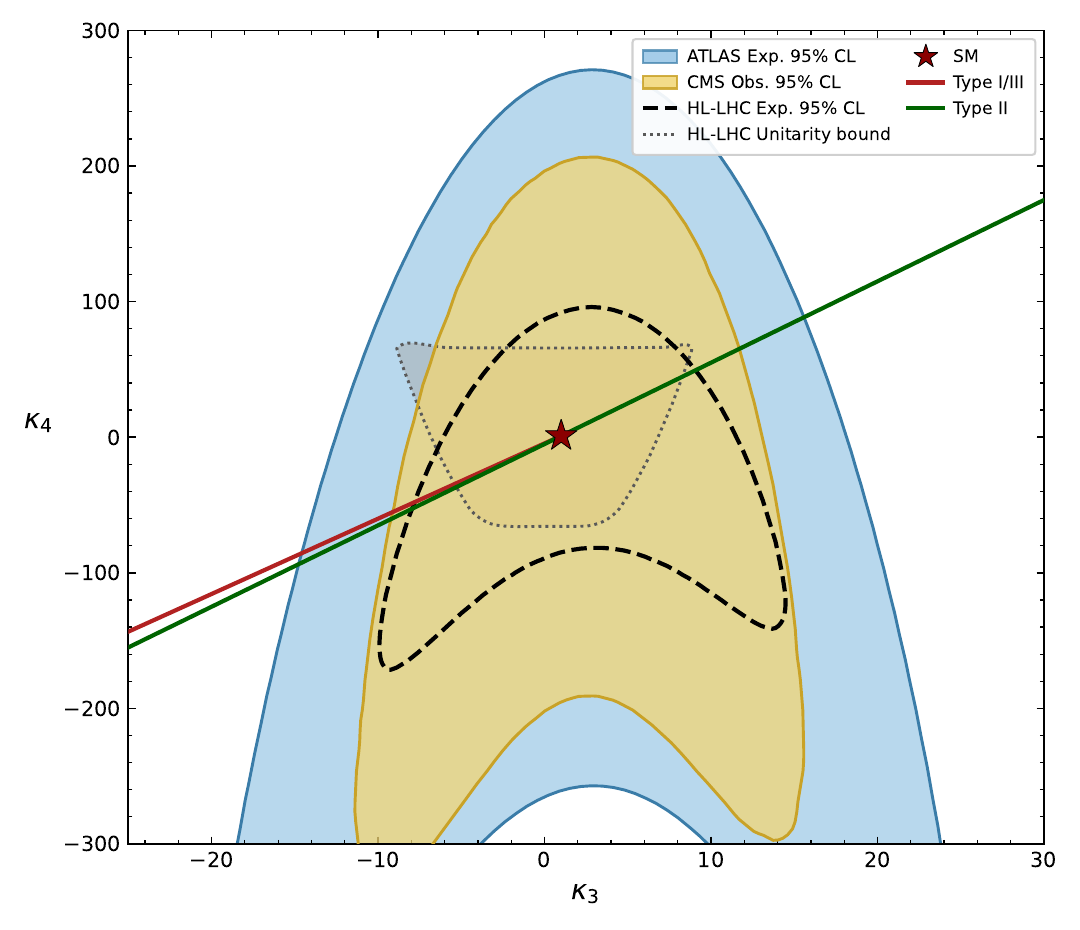}
\caption{\em Current and projected experimental constraints in the $(\kappa_3,\kappa_4)$ plane of the Higgs trilinear and quartic self-couplings, compared with the SM point. Shown are the ATLAS expected $95\%$ CL region from run-2 data~\cite{ATLAS:2025cae}, the CMS observed $95\%$ CL region from the same channel~\cite{CMS:2025gos}, the expected $95\%$ CL reach of the combined ATLAS and CMS HL-LHC program~\cite{ATLAS:2025eii}, and the perturbative-unitarity bound on $HH\to HH$ scattering~\cite{Stylianou:2023xit} as quoted in the same reference. Superimposed are the regions selected by the Majoron models: types~I and~III lie on the half line of Eq.~\eqref{eq:line-I-III}, of slope $50/9$ through the SM point and restricted to $\kappa_3<1$ by the sign of $\cC_{H\Box}^{d=6}$, shown as the red line; type~II fills the band of Eq.~\eqref{eq:band-II}, of slope $6$ and of width set by the current determination of $\kappa_W$, shown as the green line, the band itself being too narrow to be visible at the scale of the figure.}
\label{fig:k3k4}
\end{figure}

\paragraph{\boldmath The $(\kappa_3,\kappa_4)$ plane.\unboldmath} The two cases just discussed have a compact geometrical description, common to all three realizations, which is summarized in Fig.~\ref{fig:k3k4}. Eliminating the underlying parameters between the relations above, the Majoron models do not populate the plane of the two self-couplings freely: they occupy a set of measure zero in it and a different one in each case.

In types~I and~III, there is a single parameter, so the model traces a \emph{straight line} through the SM point. Dividing the two relations of Eq.~\eqref{eq:kappa34lock},
\be
\kappa_4-1=\dfrac{50}{9}\left(\kappa_3-1\right)\qquad\text{(types~I and~III)}\,,
\label{eq:line-I-III}
\ee
and, since $\cC_{H\Box}^{d=6}<0$ is sign definite, only the half line with $\kappa_3<1$ and $\kappa_4<1$ is physical. In type~II, there are two independent parameters, $\cC_H^{d=6}$ and the wave function combination $\cC_{H\Box}^{d=6}-\cC_{HD}^{d=6}/4$, and the model therefore fills a \emph{region}. Rewriting Eq.~\eqref{eq:k3k4typeII},
\be
\kappa_4-1=6\left(\kappa_3-1\right)-\dfrac43\left(\kappa_W-1\right)\qquad\text{(type~II)}\,,
\label{eq:band-II}
\ee
which is a family of parallel lines of slope $6$, whose offset from the origin is set by $\kappa_W$ alone. The width of the band is, thus, fixed not by theory but by how well $\kappa_W$ is measured: with the current determination $\kappa_W=1.05\pm0.06$~\cite{ATLAS:2022vkf}, that is, $0.93\lesssim\kappa_W\lesssim1.17$ at $95\%$ confidence level (CL), the allowed offsets span $-0.23\lesssim-\tfrac43(\kappa_W-1)\lesssim0.09$. Note also that, unlike in the fermionic realizations, the sign of $\kappa_W-1$ is \emph{not} fixed in type~II: the radial mode contributes negatively but the triplet positively, through $\cC_{H\Box}^{d=6}-\cC_{HD}^{d=6}/4=-\beta_1^2/8\lambda_1m_\rho^2+\beta_5^2v_\phi^2/4M_\Delta^4$, so the band straddles the SM point.

The two predictions are geometrically distinct---a half line of slope $50/9$ through the SM point on one side and a band of slope $6$ of experimentally determined width on the other---and this is what makes the $(\kappa_3,\kappa_4)$ plane a genuine discriminator between the fermionic and the scalar realizations, independent of, and complementary to, the $J\to\gamma\gamma$ and nonunitarity handles discussed elsewhere. Figure~\ref{fig:k3k4} collects the present and projected sensitivity in the plane. The CMS observed $95\%$ CL region~\cite{CMS:2025gos} is the tightest constraint available today, comparable in shape to the expected ATLAS reach from the same channel~\cite{ATLAS:2025cae}, and both leave the type~I and III half line and the (invisibly narrow) type~II band unconstrained near the SM point. The projected sensitivity of the combined ATLAS and CMS HL-LHC program~\cite{ATLAS:2025eii}, shown together with the perturbative-unitarity bound quoted in the same reference~\cite{Stylianou:2023xit}, narrows the allowed region enough to start probing the edge of that bound directly, without yet reaching either theoretical prediction. The point of the figure is not that the discrimination can be performed today but that it is well posed: it requires measuring $\kappa_3$, $\kappa_4$, and $\kappa_W$, and nothing else.

\subsubsection{Custodial breaking and the type~II pattern}

The custodial violation just mentioned deserves to be isolated, because in the type~II model it is again sign definite and correlated. From $\cC_{HD}^{d=6}=\beta_5^2v_\phi^2/M_\Delta^4$, manifestly positive, one obtains
\be
\alpha T=-\dfrac{\cC_{HD}^{d=6}v^2}{2}<0\,,\qquad
\kappa_Z-\kappa_W=-\alpha T\,,\qquad
\kappa_f-\kappa_W=2\,\alpha T\,,
\label{eq:typeIIcustodial}
\ee
so that the oblique parameter is necessarily negative and the departures of the $Z$, $W$, and fermionic couplings from one another are all fixed by it. The shift of the fermionic couplings deserves a comment: it originates from the Higgs-dressed Yukawa operators, whose coefficients obey $\cC_{eH}^{d=6}=\cC_{HD}^{d=6}Y_\ell$ and likewise for the quarks, so that the Yukawa matrix cancels in $\kappa_f$ and the shift is \emph{flavor universal}, with no accompanying flavor-changing neutral currents.

The sign is what makes this predictive. Global electroweak fits return a slightly \emph{positive} central value, $T=0.11\pm0.12$~\cite{ParticleDataGroup:2024cfk}, whereas Eq.~\eqref{eq:typeIIcustodial} forbids $T>0$ altogether. The type~II Majoron model can move the fit only in the disfavored direction, and the resulting constraint is one sided. Taking $|T|<0.13$ at the two-sigma level gives~\footnote{Since the model generates $T$ but not $S$, a fit conditioned on $S=0$ would be somewhat more constraining than the marginal bound used here; the refinement would not change the picture and we do not pursue it.}
\be
\cC_{HD}^{d=6}=\dfrac{\beta_5^2v_\phi^2}{M_\Delta^4}<3.4\times10^{-8}\GeV^{-2}\,,
\qquad\text{i.e.},\qquad
\dfrac{\beta_5v_\phi}{M_\Delta^2}<1.8\times10^{-4}\GeV^{-1}\,,
\label{eq:Tbound}
\ee
a bound on the dynamically generated trilinear $\beta_5v_\phi/\sqrt2$ relative to the triplet mass. It is instructive to rewrite it using the relations in Sec.~\ref{sec:JSMEFT-II}. Since $\cC_{HD}^{d=6}=\left(\cC_W^{d=5}\right)^2/2Y_\Delta^2$ and $m_\nu=-\cC_W^{d=5}v^2$, the custodial parameter is controlled by the neutrino mass and the triplet Yukawa alone:
\be
\alpha T=-\dfrac{m_\nu^2}{4\,Y_\Delta^2\,v^2}\,,
\ee
and Eq.~\eqref{eq:Tbound} becomes a bound on the combination $m_\nu/Y_\Delta$, which is precisely what plays the role of the induced triplet VEV in the ordinary type~II seesaw:
\be
\dfrac{m_\nu}{Y_\Delta}<2v\sqrt{\alpha|T|}\simeq16\GeV\,.
\ee
This reproduces the familiar electroweak-precision bound on the triplet VEV of the type~II seesaw, here derived entirely within the effective description and in the conventions in Sec.~\ref{sec:typeII}. For neutrino masses of the atmospheric scale, it implies only $Y_\Delta\gtrsim3\times10^{-12}$, so it does not constrain the Yukawa in any useful way; its real content is the upper bound on the trilinear-to-mass ratio in Eq.~\eqref{eq:Tbound}, which caps how much of the neutrino mass can come from a large induced VEV rather than from a sizable Yukawa. Together with the fact that the same $\cC_{HD}^{d=6}$ fixes $\kappa_Z-\kappa_W$ and $\kappa_f-\kappa_W$ through Eq.~\eqref{eq:typeIIcustodial}, this makes the type~II realization the most predictive of the three in the bosonic sector and, correspondingly, the most constrained.
\subsubsection{The leptonic sector and the absence of correlations}

In the fermionic realizations, the Majoron couples to the leptons through $\cQ_{JHL}^{d=7}$, whose coefficient is locked to the unitarity operator by $\cC_{JHL}^{d=7}=\cC_U^{d=6}/2v_\phi$, Eqs.~\eqref{eq:JSMEFT-I-lep} and~\eqref{eq:JSMEFT-III-lep}. This is the leptonic counterpart of Eq.~\eqref{eq:hJJmaster}: the strength with which the Majoron couples to leptons is not an independent quantity but is predicted by the nonunitarity of the leptonic mixing matrix, which is measured in electroweak precision and oscillation data, together with $v_\phi$.

Which observables this coupling actually feeds requires some care, because the Majoron is a genuine Goldstone boson and, therefore, couples derivatively. Upon electroweak symmetry breaking, $\cQ_{JHL}^{d=7}$ reduces, in the type~I case, to $\tfrac12\cC_{JHL}^{d=7}v^2\,\ov{\nu_L}\slashed\derp J\gamma_5\nu_L$, carrying the flavor structure of $\cC_U^{d=6}\propto Y_NM_N^{-2}Y_N^\dag$; in type~III, the $SU(2)_L$-triplet contraction makes the charged leptons participate as well, as the appearance of $\cQ_{eH}$ in the reduction of Eq.~\eqref{eq:OU-warsaw} already signals. Integrating by parts and using the equations of motion, a derivative coupling of this form becomes an effective pseudoscalar one, $g_{\alpha\beta}\,\ov{\ell_\alpha}\gamma_5\ell_\beta\,J$, whose strength is proportional to the \emph{mass of the fermion involved}. This single fact organizes the whole discussion, and it separates the observables into two very unequal groups.

For the neutrinos, one recovers the standard singlet-Majoron result $g_{\nu\nu}\sim m_\nu/v_\phi$. Numerically, this is minute: for $m_\nu$ at the atmospheric scale and $v_\phi\sim1\TeV$, one finds $g_{\nu\nu}\sim5\times10^{-14}$. Neutrinoless double beta decay with Majoron emission, $(A,Z)\to(A,Z+2)+2e^-+J$, whose distinctive electron sum-energy spectrum would separate it both from the standard $0\nu\beta\beta$ peak and from the two-neutrino continuum, is currently constrained at the level $\langle g_{ee}\rangle\lesssim10^{-5}$~\cite{KamLAND-Zen:2012uen}, 9 orders of magnitude above the prediction. It is important to be explicit about this, since $0\nu\beta\beta$ with Majoron emission is often presented as \emph{the} Majoron observable: in the models considered here, with a lepton-number-breaking scale at or above the TeV, it carries no sensitivity whatsoever.

The bounds from meson decays~\cite{Lessa:2007up} and the cosmological constraints on neutrino free streaming and on the effective number of relativistic species~\cite{Sandner:2023ptm} are likewise far from the relevant range for a massless Goldstone. The underlying reason is that every portal between $J$ and the SM plasma is suppressed by at least one power of $v_\phi$: the derivative lepton coupling by $m_\nu/v_\phi$; the anomalous gauge couplings by $1/v_\phi$; and the Higgs portal by $1/v_\phi^2$. As a result, for the TeV-scale $v_\phi$ required throughout this work, the Majoron never thermalizes with the SM plasma after the electroweak phase transition, and its contribution to $\Delta N_\text{eff}$ is automatically negligible. The bound would become relevant for $v_\phi$ only many orders of magnitude below the range considered here.

The situation is entirely different for the charged leptons and, hence, for the type~III realization. There, the same operator produces $\mu\to eJ$ with
\be
g_{\mu e}\simeq\dfrac{\eta_{\mu e}\,m_\mu}{2v_\phi}\,,
\ee
where $\eta$ is the usual nonunitarity matrix, $\eta=\cC_U^{d=6}v^2/2$. The enhancement factor relative to the neutrino case is $m_\mu/m_\nu\sim10^9$, which overwhelms the suppression by the off-diagonal $\eta_{\mu e}\lesssim10^{-5}$~\cite{Blennow:2023mqx}. The relevant experimental constraint comes from searches for the two-body decay $\mu^+\to e^+X$ into a light neutral boson, the strongest assumption-free one being that of the TWIST experiment at TRIUMF, a precision measurement of the muon decay spectrum, which gives $\text{BR}(\mu\to eX)<5.8\times10^{-5}$ at $90\%$ CL~\cite{TWIST:2014ymv}. This translates into $g_{\mu e}<9\times10^{-11}$, and, hence,
\be
v_\phi>\dfrac{\eta_{\mu e}\,m_\mu}{2\,g_{\mu e}^\text{max}}\simeq6\TeV
\qquad\text{for }\eta_{\mu e}\text{ at its current upper bound,}
\ee
and proportionally weaker for smaller $\eta_{\mu e}$.~\footnote{A nominally stronger limit on the same decay, $\text{BR}<2.6\times10^{-6}$, was obtained much earlier in Ref.~\cite{Jodidio:1986mz} but assuming isotropic emission of the light boson. That assumption fails here: for a Majoron coupled to a $V-A$ current, the emission from a polarized $\mu^+$ is anisotropic, so the bound does not apply as quoted and we use the weaker but assumption-free TWIST one.} This is the significant point of the section: the lower bound reached here, $v_\phi\gtrsim\mathcal{O}(1$--$10)\TeV$, sits in the \emph{same} window singled out by the invisible Higgs width in Eq.~\eqref{eq:hJJmaster}, reached through a completely unrelated measurement. Two observables as distant as a Higgs decay at the LHC and a muon decay at rest give comparable lower limits on the same lepton-number-breaking scale, and in the type~III model they are not independent: both are fixed by $v_\phi$, the first together with $\kappa_V$ and the second together with the nonunitarity.

The lock relation is what gives this its force. The matrix $\cC_U^{d=6}$ that sets the rate of Majoron emission is the very matrix whose entries are bounded by nonunitarity searches, so the \emph{flavor pattern} of Majoron emission is predicted rather than fitted: the ratio of $\mu\to eJ$ to $\tau\to\mu J$ follows from the same $Y_NM_N^{-2}Y_N^\dag$ that controls deviations from unitarity in oscillations and in electroweak precision data. A Majoron-lepton coupling observed at a strength or with a flavor structure incompatible with the measured nonunitarity would point to a different origin of the Goldstone boson. Conversely, nonunitarity established without the corresponding Majoron coupling, at a sensitivity where the latter is predicted, would indicate an explicitly broken lepton number.

It is worth stressing, finally, that the absence of a correlation is as informative as its presence. The operators in Sec.~\ref{sec:JSMEFT} span a much larger space than the handful of underlying parameters $\beta_1,\lambda_1,\beta_5,v_\phi$, and the Yukawas, and generic new physics has no reason to respect the resulting relations. Deviations found in the Higgs sector that do not satisfy Eqs.~\eqref{eq:hJJmaster} and~\eqref{eq:kappa34lock}, an enhancement rather than a suppression of the Higgs couplings, a positive $T$ parameter accompanied by the type~II operator pattern, or a Majoron-like state whose lepton couplings are unrelated to the nonunitarity would each exclude the corresponding realization without requiring the heavy states to be produced. In this sense, the effective description turns the Majoron models into a falsifiable framework even in the regime where the states responsible for the neutrino mass are entirely out of reach.

\section{Conclusions}
\label{sec:Conclusions}

We have constructed the effective field theory of the Majoron seesaw models, in which the lepton number of the type~I, II, and III seesaws is broken spontaneously by the VEV of a complex scalar rather than explicitly. Each realization admits two hierarchies between the radial mode and the seesaw mediator, and we have integrated the heavy states out in both orders, obtaining seven intermediate effective theories and, below all thresholds, the three low-energy Lagrangians that we have called the JSMEFT. The two matching orders agree operator by operator, which is a nontrivial check of the whole construction, and the resulting operator bases are collected in Table~\ref{tab:JSMEFTsummary}. The technical content is summarized there and in Sec.~\ref{sec:JSMEFT}. 

The analysis we presented is relevant for several reasons. The main point is that the Majoron sector is not a decoration that can be added to a seesaw and then tuned away. If the lepton number is broken spontaneously, the Goldstone boson and the radial mode are not optional, and their couplings are not free: the same VEV $v_\phi$ fixes the mediator mass, the mass and interactions of the radial mode, and the normalization of every Majoron coupling. The effective description makes this visible in the only place where it can be tested, namely, in relations among Wilson coefficients that survive the elimination of the underlying parameters. What the framework predicts is, therefore, not the size of any single observable---which always depends on unknown couplings---but a web of correlations, and it is falsified if the correlations fail. This changes what one has to do experimentally: establishing that lepton number is broken spontaneously does not require producing the heavy states; it requires measuring several ordinary quantities well enough to check whether they are consistent with a single $v_\phi$.

Concretely, the two sharpest consequences are of this kind. The Higgs sector gives $\Gamma(h\to JJ)=(1-\kappa_V)m_h^3/32\pi v_\phi^2$, in which the electroweak VEV cancels: the invisible width is fixed once the suppression of the Higgs couplings is measured, and the suppression is itself sign definite, so that an enhancement $\kappa_V>1$ would exclude the entire class of models. The leptonic sector gives, independently, the Majoron-lepton coupling in terms of the measured nonunitarity of the leptonic mixing matrix. That these two independent lower bounds on $v_\phi$ currently sit in the same window, $v_\phi\gtrsim\mathcal{O}(1\text{--}10)\TeV$---the first through the invisible Higgs branching ratio and the second through $\mu\to eJ$---is the most encouraging feature of the analysis, because it means the consistency test is not academic but is being performed, at low precision, already now. We have also found that the traditional flagship signature of Majoron models, neutrinoless double beta decay with Majoron emission, carries no sensitivity whatsoever in this class: the derivative nature of the Goldstone coupling suppresses it by the neutrino mass, leaving it 9 orders of magnitude below the experimental reach. The observables that matter are elsewhere, in Higgs precision and in charged-lepton flavor violation, and we regard this reallocation of experimental attention as one of the more useful outcomes of the effective analysis.

Several things would sharpen the picture. On the theory side, the matching has been performed at tree level. Going to one loop would generate the operators that are absent here and would allow a consistent treatment of the running between the matching scales, which for widely separated thresholds can matter. The same logic extends beyond the three canonical tree-level seesaws considered here. Any mechanism in which the lepton number is broken spontaneously by the same $\phi$---including inverse or linear seesaw variants or radiative constructions such as Zee-Babu---would generate an analogous JSMEFT. Each would differ in flavor structure and, in the loop-induced cases, in the anomalous-coupling pattern set by the particles running in the triangle. Working out these variants case by case is a natural extension of the present framework, left for future work. On the experimental side, the program is well defined and needs no new ideas: better determinations of $\kappa_V$ and $\kappa_{W,Z}$, a stronger bound on the invisible Higgs width, an improved search for $\mu\to eJ$, and eventually the Higgs self-couplings, whose position in the $(\kappa_3,\kappa_4)$ plane separates the fermionic realizations---confined to a half line through the Standard Model point---from the scalar one, which fills a band. None of these is a dedicated Majoron search, which is precisely the point: the question of whether the lepton number is a spontaneously broken symmetry is answerable with measurements that the field is already committed to making.

\section*{Acknowledgments}
We are grateful to Enrique Fern\'andez Mart\'inez, Daniel Naredo Tuero, and Xavier Ponce D\'iaz for valuable discussions. C.~B. and X.~L. acknowledge the Instituto de F\'isica Te\'orica (IFT) UAM/CSIC for hospitality while this project was carried out.

M.~F.~Z. and L.~M. acknowledge partial financial support by the Spanish Research Agency (Agencia Estatal de
Investigaci\'on) through the Instituto de F\'isica Te\'orica Centro de Excelencia Severo Ochoa Grant No.~CEX2025-001574-S---within the research line Particle Physics in the Standard Model and Beyond---and by Grant No.~PID2022-137127NB-I00 funded by the Ministerio de Ciencia, Innovaci\'on y Universidades (MICIU)/Agencia Estatal de Investigaci\'on (AEI)/10.13039/501100011033. The work of M.~F.~Z. is supported by the Spanish Ministerio de Universidades through the National Program Formaci\'on de Profesorado Universitario (FPU) (Grant No.~FPU22/03625).
This article is based upon work from COST Action COSMIC WISPers CA21106, supported by the European Cooperation in Science and Technology (COST).

\section*{Data availability}
There are no publicly available research data or software supporting this manuscript. Requests for further information or data should be sent to the authors.


\appendix

\section{Equations of motion and perturbative solutions}
\label{app:EOM}

For each hierarchy in Table~\ref{tab:EFTsummary}, the heaviest state is removed through a standard, if case-dependent, manipulation: one collects the terms of the corresponding post-SSB Lagrangian involving the field to be integrated out, derives its EOM and solves it perturbatively in the inverse heavy mass; substituting the solution back generates the effective operators order by order. We collect here these intermediate steps for all seven cases. In each case, the starting point is the post-SSB Lagrangian quoted in the main text, and the end point is the intermediate effective Lagrangian listed there.

\subsection{The type~I Majoron model}
\label{app:typeI}

\paragraph{Integrating out $\rho$ (NJSMEFT).}
The radial mode $\rho$ is integrated out first. Collecting from Eq.~\eqref{eq:SSBLagTypeI} the terms involving $\rho$, the relevant Lagrangian reads
\begin{align}
\sL_\rho=&\frac{1}{2}\derp_\mu\rho\derp^\mu\rho-\frac{\lambda_1}{4}\rho^4-\lambda_1v_\phi\rho^3-\frac{m_\rho^2}{2}\rho^2+\frac{\left(\rho^2+2v_\phi\rho\right)}{2v_\phi^2}\derp_\mu J\derp^\mu J-\beta_1v_\phi\rho\left(H^\dag H\right)+\nn\\
&-\frac{\beta_1}{2}\rho^2\left(H^\dag H\right)-\frac{\rho}{v_\phi}\frac{\ov{N}{M}_NN}{2}\,.
\label{Lagrho1}
\end{align}
Adopting the definition of $m_\rho$ in Eq.~\eqref{mrhoDEF}, we obtain the EOM
\begin{align}
-\Bigg(\square+m_\rho^2+\beta_1&\left(H^\dag H\right)-\frac{1}{v_\phi^2}\derp_\mu J\derp^\mu J\Bigg)\rho =\nn\\
&=\lambda_1\rho^3+3\lambda_1v_\phi\rho^2+\beta_1 v_\phi\left(H^\dag H\right)-\frac{1}{v_\phi}\derp_\mu J\derp^\mu J+\frac{1}{v_\phi}\dfrac{\ov{N}M_NN}{2}\, .
\label{eq:EOMrho}
\end{align}
Eq.~\eqref{eq:EOMrho} is nonlinear in $\rho$, due to the cubic and quartic self-interactions, and cannot be solved in closed form. We solve it perturbatively in inverse powers of $m_\rho$, writing $\rho=\rho^{(0)}+\rho^{(1)}+\dots$: at leading order, $\rho^{(0)}=\cO(1/m_\rho^2)$ follows from dropping both $\Box\rho$ and the nonlinear terms $\rho^2,\rho^3$, which are subleading once $\rho^{(0)}$ itself scales as $1/m_\rho^2$; each subsequent order resums these neglected terms evaluated on the previous-order solution. The leading and next-to-leading orders read
\begin{align}
\rho^{(0)}=&-\frac{\beta_1 v_\phi}{m_\rho^2}\left(H^\dag H\right)+\frac{1}{v_\phi m_\rho^2}\derp_\mu J\derp^\mu J-\frac{1}{2v_\phi m_\rho^2}\ov{N}M_N N\,,
\label{eq:rho0-NJ}\\
\rho^{(1)}=&\frac{1}{m_\rho^2}\left[-\left(\beta_1(H^\dag H)-\frac{1}{v_\phi^2}\derp_\mu J\derp^\mu J\right)\rho^{(0)}-3\lambda_1 v_\phi\big(\rho^{(0)}\big)^2-\square\rho^{(0)}\right]\,,
\label{eq:rho1-NJ}
\end{align}
and higher orders are obtained by iteration. Inserting the solution back into Eq.~\eqref{Lagrho1}---equivalently, using the compact form $\sL_{\rm eff}=\tfrac12 A\,\rho_{\rm cl}+\tfrac12\lambda_1 v_\phi\,\rho_{\rm cl}^3+\tfrac{\lambda_1}{4}\rho_{\rm cl}^4$ with $A$ the source linear in $\rho$---generates the effective operators of the NJSMEFT Lagrangian, Eq.~\eqref{NJSMEFTLag}.

\paragraph{Integrating out $N$ ($\rho$JSMEFT-I).}
The HNLs are integrated out first. The relevant Lagrangian collects all the terms of Eq.~\eqref{Jops-typeI} containing $N$, which, using the current $\cJ_N$ of Eq.~\eqref{eq:JNDEF}, reads compactly
\be
\sL_N=\dfrac12\ov{N}\left(i\slashed{\partial}-M_N\right) N+\ov{N}\cJ_JN+\ov{N}\cJ_\rho N
-\dfrac12\left(\overline{\cJ}_N N+\ov{N}\cJ_N\right)\,,
\label{eq:LagN-typeI}
\ee
where $\cJ_J$ is the axial Majoron current of Eq.~\eqref{eq:JJDEF} and we further introduced the radial-mode insertion
\be
\cJ_\rho\equiv-\dfrac{Y_{NN}\rho}{2\sqrt2}=-\dfrac{M_N\rho}{2v_\phi}\,,
\label{eq:JJJrhoDEF}
\ee
following from $M_N=Y_{NN}v_\phi/\sqrt2$. Varying with respect to $\ov{N}$, and recalling that for a Majorana field $N$ and $\ov{N}$ are not independent, so that each bilinear contributes twice, the EOM reads
\be
\left(i\slashed\partial-M_N+2\cJ_J+2\cJ_\rho\right)N=\cJ_N\,,
\label{eq:EOMN-typeI}
\ee
whose solution is the geometric series
\be
N=-\big[M_N-(i\slashed\partial+2\cJ_J+2\cJ_\rho)\big]^{-1}\cJ_N
=-\sum_{k\ge0}M_N^{-1}\big[(i\slashed\partial+2\cJ_J+2\cJ_\rho)M_N^{-1}\big]^k\cJ_N\,,
\label{eq:solN-typeI}
\ee
that is, order by order,
\be
\begin{aligned}
N^{(0)}&=-M_N^{-1}\cJ_N\,,\\
N^{(1)}&=-M_N^{-1}\left(i\slashed\partial+2\cJ_J+2\cJ_\rho\right)M_N^{-1}\cJ_N\,,\\
N^{(2)}&=-M_N^{-1}\left(i\slashed\partial+2\cJ_J+2\cJ_\rho\right)M_N^{-1}\left(i\slashed\partial+2\cJ_J+2\cJ_\rho\right)M_N^{-1}\cJ_N\,,
\end{aligned}
\ee
where the $d=5,6$ operators come from $N^{(0,1)}$ and the $d=7$ ones from $N^{(2)}$ or from a single $\cJ_J$ insertion. Inserting this solution back into Eq.~\eqref{eq:LagN-typeI} generates the effective operators of the $\rho$JSMEFT-I Lagrangian, Eq.~\eqref{rhoJSMEFTILag}.

\subsection{The type~III Majoron model}
\label{app:typeIII}

\paragraph{Integrating out $\rho$ ($\Sigma$JSMEFT).}
The radial mode $\rho$ is the heaviest d.o.f. and is integrated out first, exactly as in the NJSMEFT case, with the HNLs replaced by the fermion triplets $\Sigma$. Collecting from Eq.~\eqref{Jops-typeIII} the terms involving $\rho$, the relevant Lagrangian reads
\begin{align}
\sL_\rho=&\frac{1}{2}\derp_\mu\rho\derp^\mu\rho-\frac{\lambda_1}{4}\rho^4-\lambda_1v_\phi\rho^3-\frac{m_\rho^2}{2}\rho^2+\frac{\left(\rho^2+2v_\phi\rho\right)}{2v_\phi^2}\derp_\mu J\derp^\mu J-\beta_1v_\phi\rho\left(H^\dag H\right)+\nn\\
&-\frac{\beta_1}{2}\rho^2\left(H^\dag H\right)-\frac{\rho}{v_\phi}\frac{\ov{\vec{\Sigma}}M_\Sigma\vec{\Sigma}}{2}\,,
\label{Lagrho3}
\end{align}
the exact type~III analog of the type~I Lagrangian of Eq.~\eqref{Lagrho1}, with $\ov{N}M_NN\to\ov{\vec{\Sigma}}M_\Sigma\vec{\Sigma}$. Adopting the definition $m_\rho^2=2\lambda_1v_\phi^2$, the corresponding EOM reads
\begin{align}
-\Bigg(\square+m_\rho^2+\beta_1&\left(H^\dag H\right)-\frac{1}{v_\phi^2}\derp_\mu J\derp^\mu J\Bigg)\rho =\nn\\
&=\lambda_1\rho^3+3\lambda_1v_\phi\rho^2+\beta_1 v_\phi\left(H^\dag H\right)-\frac{1}{v_\phi}\derp_\mu J\derp^\mu J+\frac{1}{v_\phi}\dfrac{\ov{\vec{\Sigma}}M_\Sigma\vec{\Sigma}}{2}\,,
\label{eq:EOMrho3}
\end{align}
solved perturbatively in inverse powers of $m_\rho$ exactly as in Appendix~\ref{app:typeI}, with leading and next-to-leading orders
\begin{align}
\rho^{(0)}=&-\frac{\beta_1 v_\phi}{m_\rho^2}\left(H^\dag H\right)+\frac{1}{v_\phi m_\rho^2}\derp_\mu J\derp^\mu J-\frac{1}{2v_\phi m_\rho^2}\ov{\vec{\Sigma}}M_\Sigma\vec{\Sigma}\,,
\label{eq:rho0-SJ}\\
\rho^{(1)}=&\frac{1}{m_\rho^2}\left[-\left(\beta_1(H^\dag H)-\frac{1}{v_\phi^2}\derp_\mu J\derp^\mu J\right)\rho^{(0)}-3\lambda_1 v_\phi\big(\rho^{(0)}\big)^2-\square\rho^{(0)}\right]\,.
\label{eq:rho1-SJ}
\end{align}
Substituting the solution back into Eq.~\eqref{Lagrho3} generates the effective operators of the $\Sigma$JSMEFT Lagrangian, Eq.~\eqref{SigmaJSMEFTLag}.

\paragraph{Integrating out $\Sigma$ ($\rho$JSMEFT-III).}
The fermion triplets are integrated out first, keeping $\rho$ dynamical. Collecting from Eq.~\eqref{Jops-typeIII} the terms involving $\Sigma$, with the triplet current $\vec{\cJ}_\Sigma$ of Eq.~\eqref{eq:JSigmaDEF} and the axial Majoron current $\cJ_J$ of Eq.~\eqref{eq:JJDEF}, the relevant Lagrangian reads
\be
\sL_\Sigma=\dfrac12\ov{\vec{\Sigma}}\left(i\slashed{D}-M_\Sigma\right)\vec{\Sigma}+\ov{\vec{\Sigma}}\,\cJ_J\,\vec{\Sigma}+\ov{\vec{\Sigma}}\,\cJ_\rho\,\vec{\Sigma}-\dfrac12\left(\overline{\vec{\cJ}}_\Sigma\cdot\vec{\Sigma}+\ov{\vec{\Sigma}}\cdot\vec{\cJ}_\Sigma\right)\,,
\label{eq:LagSigma-typeIII}
\ee
where, in close analogy with the type~I case in Eq.~\eqref{eq:JJJrhoDEF}, we further introduced the radial-mode insertion
\be
\cJ_\rho\equiv-\dfrac{Y_\Sigma\rho}{2\sqrt2}=-\dfrac{M_\Sigma\rho}{2v_\phi}\,.
\label{eq:JrhoSigma}
\ee
Equation~\eqref{eq:LagSigma-typeIII} is then term by term the type~III transcription of the type~I Lagrangian, Eq.~\eqref{eq:LagN-typeI}, with $N\to\vec\Sigma$ and $\cJ_N\to\vec{\cJ}_\Sigma$; as for $\cJ_J$ of the type~I case, the Majoron current carries a factor $1/2$ relative to the vectorial lepton currents, following from the Majorana nature of $\Sigma$. Varying with respect to $\ov{\vec\Sigma}$, the EOM reads
\be
\left(i\slashed{D}-M_\Sigma+2\cJ_J+2\cJ_\rho\right)\vec\Sigma=\vec{\cJ}_\Sigma\,,
\label{eq:EOMSigma-typeIII}
\ee
whose solution is the geometric series
\be
\vec\Sigma=-\big[M_\Sigma-(i\slashed{D}+2\cJ_J+2\cJ_\rho)\big]^{-1}\vec{\cJ}_\Sigma
=-\sum_{k\ge0}M_\Sigma^{-1}\big[(i\slashed{D}+2\cJ_J+2\cJ_\rho)M_\Sigma^{-1}\big]^k\vec{\cJ}_\Sigma\,,
\label{eq:solSigma-typeIII}
\ee
that is, order by order,
\be
\begin{aligned}
\vec\Sigma^{(0)}&=-M_\Sigma^{-1}\vec{\cJ}_\Sigma\,,\\
\vec\Sigma^{(1)}&=-M_\Sigma^{-1}\left(i\slashed{D}+2\cJ_J+2\cJ_\rho\right)M_\Sigma^{-1}\vec{\cJ}_\Sigma\,,\\
\vec\Sigma^{(2)}&=-M_\Sigma^{-1}\left(i\slashed{D}+2\cJ_J+2\cJ_\rho\right)M_\Sigma^{-1}\left(i\slashed{D}+2\cJ_J+2\cJ_\rho\right)M_\Sigma^{-1}\vec{\cJ}_\Sigma\,.
\end{aligned}
\ee
Substituting the solution back into Eq.~\eqref{eq:LagSigma-typeIII} generates the effective operators of the $\rho$JSMEFT-III Lagrangian, Eq.~\eqref{rhoJSMEFTIIILag}.

\subsection{The type~II Majoron model}
\label{app:typeII}

\paragraph{Integrating out $\rho$ ($\Delta$JSMEFT).}
The radial mode $\rho$ is the heaviest d.o.f. and is integrated out first, exactly as in the NJSMEFT case above. Collecting from Eq.~\eqref{eq:SSBLagTypeII} the terms involving $\rho$, using the phase-free form of the trilinear coupling obtained after the rotation of Eq.~\eqref{rotJII} and the shorthand $\cJ_{H\Delta}$ of Eq.~\eqref{eq:JHDeltaDEF}, the relevant Lagrangian reads
\be
\begin{split}
\sL_\rho=&\frac{1}{2}\derp_\mu\rho\derp^\mu\rho-\frac{\lambda_1}{4}\rho^4-\lambda_1v_\phi\rho^3-\frac{m_\rho^2}{2}\rho^2+\frac{\left(\rho^2+2v_\phi\rho\right)}{2v_\phi^2}\derp_\mu J\derp^\mu J-\beta_1v_\phi\rho\left(H^\dag H\right)+\\
&-\frac{\beta_1}{2}\rho^2\left(H^\dag H\right)-\beta_2v_\phi\rho\left(\Delta^\dag\Delta\right)-\frac{\beta_2}{2}\rho^2\left(\Delta^\dag\Delta\right)+\frac{\beta_5}{\sqrt2}\rho\,\cJ_{H\Delta}\,.
\end{split}
\label{Lagrho2}
\ee
With respect to the type~I case, Eq.~\eqref{Lagrho1}, the source coupling linearly to $\rho$ now includes, besides the Higgs bilinear, the triplet bilinear $\Delta^\dag\Delta$ (weighted by $\beta_2$) and the gauge-invariant trilinear $\cJ_{H\Delta}$ (weighted by $\beta_5$), while the role played there by the HNL bilinear $\ov{N}M_NN$ is here absent. Adopting the definition of $m_\rho$ in Eq.~\eqref{mrhoDEF}, the corresponding EOM reads
\begin{align}
-\Bigg(\square+m_\rho^2+&\beta_1\left(H^\dag H\right)+\beta_2\left(\Delta^\dag\Delta\right)-\frac{1}{v_\phi^2}\derp_\mu J\derp^\mu J\Bigg)\rho=
\label{eq:EOMrho2}\\
&=\lambda_1\rho^3+3\lambda_1v_\phi\rho^2+\beta_1v_\phi\left(H^\dag H\right)+\beta_2v_\phi\left(\Delta^\dag\Delta\right)-\frac{1}{v_\phi}\derp_\mu J\derp^\mu J-\frac{\beta_5}{\sqrt2}\cJ_{H\Delta}\,.
\nn
\end{align}
Its perturbative solution $\rho=\rho^{(0)}+\rho^{(1)}+\cdots$ follows exactly as in the type~I case, with leading and next-to-leading orders
\begin{align}
\rho^{(0)}=&-\frac{\beta_1 v_\phi}{m_\rho^2}\left(H^\dag H\right)-\frac{\beta_2 v_\phi}{m_\rho^2}\left(\Delta^\dag\Delta\right)+\frac{1}{v_\phi m_\rho^2}\derp_\mu J\derp^\mu J+\frac{\beta_5}{\sqrt2\,m_\rho^2}\cJ_{H\Delta}\,,
\label{eq:rho0-DJ}\\
\rho^{(1)}=&\frac{1}{m_\rho^2}\left[-\left(\beta_1(H^\dag H)+\beta_2(\Delta^\dag\Delta)-\frac{1}{v_\phi^2}\derp_\mu J\derp^\mu J\right)\rho^{(0)}-3\lambda_1 v_\phi\big(\rho^{(0)}\big)^2-\square\rho^{(0)}\right]\,.
\label{eq:rho1-DJ}
\end{align}
Substituting the solution back into Eq.~\eqref{Lagrho2} generates the effective operators of the $\Delta$JSMEFT Lagrangian, Eq.~\eqref{DeltaJSMEFTLag}.

\paragraph{Integrating out $\Delta$ ($\rho$JSMEFT-IIa and IIb).}
The triplet $\Delta$ is integrated out first, keeping the radial mode $\rho$ dynamical. Collecting from Eqs.~\eqref{eq:SSBLagTypeII} and~\eqref{Jops-typeII} the terms involving $\Delta$, with the two currents $\cJ_\Delta^I$ and $\cJ_L^I$ in Eq.~\eqref{eq:JDeltaJLDEF}, the relevant Lagrangian reads
\begin{align}
\sL_\Delta=&\,D_\mu\Delta^\dag D^\mu\Delta-\left(M_\Delta^2+\beta_2v_\phi\rho+\frac{\beta_2}{2}\rho^2\right)\Delta^\dag\Delta-\lambda_3\left(\Delta^\dag\Delta\right)^2-\beta_3\left(H^\dag H\right)\Delta^\dag\Delta+\nn\\
&+i\,\beta_4\left(H^\dag\sigma^IH\right)\epsilon^{IJK}\Delta_J^\dag\Delta_K-\dfrac{i\,\derp_\mu J}{v_\phi}\left(\Delta^\dag D^\mu\Delta-\left(D^\mu\Delta\right)^\dag\Delta\right)+\dfrac{\derp_\mu J\derp^\mu J}{v_\phi^2}\Delta^\dag\Delta+\nn\\
&+\left(\cJ_\Delta^I\,\Delta_I^\dag+\hc\right)-\left(\cJ_L^I\,\Delta_I+\hc\right)\,,
\label{LagDelta2}
\end{align}
where $M_\Delta^2=\mu_3^2+\frac12\beta_2v_\phi^2$. The leading-order solution of the $\Delta$ EOM reads
\be
\Delta_I=\dfrac{1}{M_\Delta^2}\left(\cJ_\Delta^I-\cJ_L^{I\dag}\right)+\cO\!\left(\dfrac{1}{M_\Delta^4}\right)\,
\label{eq:solDelta}
\ee
and, substituting back into Eq.~\eqref{LagDelta2}, generates the effective operators of the $\rho$JSMEFT-IIa Lagrangian, Eq.~\eqref{rhoJSMEFTIIaLag}. The leading term reproduces the operators up to $d=6$; the $d=7$ ones---the derivative dressing $\cO_{DHL}^{d=7}$, the Higgs dressing $\cO_{WH}^{d=7}$, and the $\rho$ dressings---arise from the subleading $\cO(1/M_\Delta^4)$ term of the propagator expansion together with the $\rho$ dependence of $M_\Delta$ and are most compactly obtained from the resummation of Eq.~\eqref{eq:IIaresum}. The $\rho$JSMEFT-IIb case ($M_\Delta=\mu_3>m_\rho$) follows identically, with $M_\Delta^2\to\mu_3^2$.

\bibliographystyle{utphys}
\bibliography{references}

@article{Minkowski:1977sc,
    author = "Minkowski, Peter",
    title = "{$\mu \to e\gamma$ at a Rate of One Out of $10^{9}$ Muon Decays?}",
    reportNumber = "Print-77-0182 (BERN)",
    doi = "10.1016/0370-2693(77)90435-X",
    journal = "Phys. Lett. B",
    volume = "67",
    pages = "421--428",
    year = "1977"
}

@article{ATLAS:2025eii,
    author = "Aad, Georges and others",
    collaboration = "ATLAS, CMS",
    title = "{Highlights of the HL-LHC physics projections by ATLAS and CMS}",
    eprint = "2504.00672",
    archivePrefix = "arXiv",
    primaryClass = "hep-ex",
    reportNumber = "ATL-PHYS-PUB-2025-018, CMS-HIG-25-002",
    month = "4",
    year = "2025"
}

@misc{CMS:2025gos,
    author = "{CMS Collaboration}",
    title = "{Search for nonresonant triple Higgs boson production in the six b-quark final state in proton-proton collisions at $\sqrt{s}=13$ TeV}",
    howpublished = "{CMS Physics Analysis Summary CMS-PAS-HIG-24-012, CERN, Geneva}",
    url = "https://cds.cern.ch/record/2945361",
    year = "2025"
}

@misc{ATLAS:2025cae,
    author = "{ATLAS Collaboration}",
    title = "{HL-LHC prospects for the measurement of triple-Higgs production in the 6b final state at the ATLAS experiment}",
    howpublished = "{ATLAS PUB Note ATL-PHYS-PUB-2025-003, CERN, Geneva}",
    url = "https://cds.cern.ch/record/2924772",
    year = "2025"
}

@article{Stylianou:2023xit,
    author = "Stylianou, P. and Weiglein, G.",
    title = "{Constraints on the trilinear and quartic Higgs couplings from triple Higgs production at the LHC and beyond}",
    eprint = "2312.04646",
    archivePrefix = "arXiv",
    primaryClass = "hep-ph",
    reportNumber = "DESY-23-203",
    doi = "10.1140/epjc/s10052-024-12722-9",
    journal = "Eur. Phys. J. C",
    volume = "84",
    number = "4",
    pages = "366",
    year = "2024"
}

@article{Gell-Mann:1979vob,
    author = "Gell-Mann, Murray and Ramond, Pierre and Slansky, Richard",
    title = "{Complex Spinors and Unified Theories}",
    eprint = "1306.4669",
    archivePrefix = "arXiv",
    primaryClass = "hep-th",
    reportNumber = "PRINT-80-0576",
    journal = "Conf. Proc. C",
    volume = "790927",
    pages = "315--321",
    year = "1979"
}

@article{Yanagida:1979as,
    author = "Yanagida, Tsutomu",
    editor = "Sawada, Osamu and Sugamoto, Akio",
    title = "{Horizontal gauge symmetry and masses of neutrinos}",
    reportNumber = "KEK-79-18-95",
    journal = "Conf. Proc. C",
    volume = "7902131",
    pages = "95--99",
    year = "1979"
}

@article{Mohapatra:1979ia,
    author = "Mohapatra, Rabindra N. and Senjanovic, Goran",
    title = "{Neutrino Mass and Spontaneous Parity Nonconservation}",
    reportNumber = "MDDP-TR-80-060, MDDP-PP-80-105, CCNY-HEP-79-10",
    doi = "10.1103/PhysRevLett.44.912",
    journal = "Phys. Rev. Lett.",
    volume = "44",
    pages = "912",
    year = "1980"
}

@article{Malinsky:2005bi,
    author = "Malinsky, Michal and Romao, J. C. and Valle, J. W. F.",
    title = "{Novel supersymmetric SO(10) seesaw mechanism}",
    eprint = "hep-ph/0506296",
    archivePrefix = "arXiv",
    reportNumber = "IFIC-05-28",
    doi = "10.1103/PhysRevLett.95.161801",
    journal = "Phys. Rev. Lett.",
    volume = "95",
    pages = "161801",
    year = "2005"
}

@article{Kersten:2007vk,
    author = {Kersten, J{\"o}rn and Smirnov, Alexei Yu.},
    title = "{Right-Handed Neutrinos at CERN LHC and the Mechanism of Neutrino Mass Generation}",
    eprint = "0705.3221",
    archivePrefix = "arXiv",
    primaryClass = "hep-ph",
    doi = "10.1103/PhysRevD.76.073005",
    journal = "Phys. Rev. D",
    volume = "76",
    pages = "073005",
    year = "2007"
}

@article{Abada:2007ux,
    author = "Abada, A. and Biggio, C. and Bonnet, F. and Gavela, M. B. and Hambye, T.",
    title = "{Low energy effects of neutrino masses}",
    eprint = "0707.4058",
    archivePrefix = "arXiv",
    primaryClass = "hep-ph",
    reportNumber = "FTUAM-07-12, IFT-UAM-CSIC-07-41, LPT-ORSAY-07-34, ULB-TH-07-27",
    doi = "10.1088/1126-6708/2007/12/061",
    journal = "JHEP",
    volume = "12",
    pages = "061",
    year = "2007"
}

@article{Wyler:1982dd,
    author = "Wyler, D. and Wolfenstein, L.",
    title = "{Massless Neutrinos in Left-Right Symmetric Models}",
    reportNumber = "CERN-TH-3435",
    doi = "10.1016/0550-3213(83)90482-0",
    journal = "Nucl. Phys. B",
    volume = "218",
    pages = "205--214",
    year = "1983"
}

@article{Mohapatra:1986bd,
    author = "Mohapatra, R. N. and Valle, J. W. F.",
    title = "{Neutrino Mass and Baryon Number Nonconservation in Superstring Models}",
    reportNumber = "MdDP-PP-86-127",
    doi = "10.1103/PhysRevD.34.1642",
    journal = "Phys. Rev. D",
    volume = "34",
    pages = "1642",
    year = "1986"
}

@article{Bernabeu:1987gr,
    author = "Bernabeu, J. and Santamaria, A. and Vidal, J. and Mendez, A. and Valle, J. W. F.",
    title = "{Lepton Flavor Nonconservation at High-Energies in a Superstring Inspired Standard Model}",
    reportNumber = "FTUV-18-86",
    doi = "10.1016/0370-2693(87)91100-2",
    journal = "Phys. Lett. B",
    volume = "187",
    pages = "303--308",
    year = "1987"
}

@article{Magg:1980ut,
    author = "Magg, M. and Wetterich, C.",
    title = "{Neutrino Mass Problem and Gauge Hierarchy}",
    reportNumber = "CERN-TH-2829",
    doi = "10.1016/0370-2693(80)90825-4",
    journal = "Phys. Lett. B",
    volume = "94",
    pages = "61--64",
    year = "1980"
}

@article{Schechter:1980gr,
    author = "Schechter, J. and Valle, J. W. F.",
    title = "{Neutrino Masses in SU(2) x U(1) Theories}",
    reportNumber = "SU-4217-167, COO-3533-167",
    doi = "10.1103/PhysRevD.22.2227",
    journal = "Phys. Rev. D",
    volume = "22",
    pages = "2227",
    year = "1980"
}

@article{Cheng:1980qt,
    author = "Cheng, T. P. and Li, Ling-Fong",
    title = "{Neutrino Masses, Mixings and Oscillations in SU(2) x U(1) Models of Electroweak Interactions}",
    reportNumber = "PRINT-80-0511 (CARNEGIE-MELLON), COO-3066-152",
    doi = "10.1103/PhysRevD.22.2860",
    journal = "Phys. Rev. D",
    volume = "22",
    pages = "2860",
    year = "1980"
}

@article{Lazarides:1980nt,
    author = "Lazarides, George and Shafi, Q. and Wetterich, C.",
    title = "{Proton Lifetime and Fermion Masses in an SO(10) Model}",
    reportNumber = "FREIBURG-THEP-80-2",
    doi = "10.1016/0550-3213(81)90354-0",
    journal = "Nucl. Phys. B",
    volume = "181",
    pages = "287--300",
    year = "1981"
}

@article{Mohapatra:1980yp,
    author = "Mohapatra, Rabindra N. and Senjanovic, Goran",
    title = "{Neutrino Masses and Mixings in Gauge Models with Spontaneous Parity Violation}",
    reportNumber = "FERMILAB-PUB-80-061-THY, FERMILAB-PUB-80-061-T",
    doi = "10.1103/PhysRevD.23.165",
    journal = "Phys. Rev. D",
    volume = "23",
    pages = "165",
    year = "1981"
}

@article{Wetterich:1981bx,
    author = "Wetterich, C.",
    title = "{Neutrino Masses and the Scale of B-L Violation}",
    reportNumber = "FREIBURG-THEP-81-2",
    doi = "10.1016/0550-3213(81)90279-0",
    journal = "Nucl. Phys. B",
    volume = "187",
    pages = "343--375",
    year = "1981"
}

@article{Foot:1988aq,
    author = "Foot, Robert and Lew, H. and He, X. G. and Joshi, Girish C.",
    title = "{Seesaw Neutrino Masses Induced by a Triplet of Leptons}",
    reportNumber = "UM-P-88/89, OZ-P-88/7",
    doi = "10.1007/BF01415558",
    journal = "Z. Phys. C",
    volume = "44",
    pages = "441",
    year = "1989"
}

@article{Peccei:1977hh,
    author = "Peccei, R. D. and Quinn, Helen R.",
    title = "{CP Conservation in the Presence of Instantons}",
    reportNumber = "ITP-568-STANFORD",
    doi = "10.1103/PhysRevLett.38.1440",
    journal = "Phys. Rev. Lett.",
    volume = "38",
    pages = "1440--1443",
    year = "1977"
}

@article{Weinberg:1977ma,
    author = "Weinberg, Steven",
    title = "{A New Light Boson?}",
    reportNumber = "HUTP-77/A074",
    doi = "10.1103/PhysRevLett.40.223",
    journal = "Phys. Rev. Lett.",
    volume = "40",
    pages = "223--226",
    year = "1978"
}

@article{Biggio:2023gtm,
    author = "Biggio, Carla and Calibbi, Lorenzo and Ota, Toshihiko and Zanchini, Samuele",
    title = "{Majoron dark matter from a type II seesaw model}",
    eprint = "2304.12527",
    archivePrefix = "arXiv",
    primaryClass = "hep-ph",
    reportNumber = "IFT-UAM/CSIC-23-45",
    doi = "10.1103/PhysRevD.108.115003",
    journal = "Phys. Rev. D",
    volume = "108",
    number = "11",
    pages = "115003",
    year = "2023"
}

@article{Ardu:2026vsr,
    author = "Ardu, Marco and Calibbi, Lorenzo and Fedele, Marco and Mescia, Federico",
    title = "{ALP production in Lepton Flavour Violating meson, tau and gauge boson decays}",
    eprint = "2604.11889",
    doi = "10.1007/JHEP08(2026)217",
    journal = "JHEP",
    volume = "08",
    pages = "217",
    archivePrefix = "arXiv",
    primaryClass = "hep-ph",
    reportNumber = "MITP-26-018",
    month = "4",
    year = "2026"
}

@article{Marciano:2016yhf,
    author = "Marciano, W. J. and Masiero, A. and Paradisi, P. and Passera, M.",
    title = "{Contributions of axionlike particles to lepton dipole moments}",
    eprint = "1607.01022",
    archivePrefix = "arXiv",
    primaryClass = "hep-ph",
    doi = "10.1103/PhysRevD.94.115033",
    journal = "Phys. Rev. D",
    volume = "94",
    number = "11",
    pages = "115033",
    year = "2016"
}

@article{Cornella:2019uxs,
    author = "Cornella, Claudia and Paradisi, Paride and Sumensari, Olcyr",
    title = "{Hunting for ALPs with Lepton Flavor Violation}",
    eprint = "1911.06279",
    archivePrefix = "arXiv",
    primaryClass = "hep-ph",
    reportNumber = "ZU-TH 46/19",
    doi = "10.1007/JHEP01(2020)158",
    journal = "JHEP",
    volume = "01",
    pages = "158",
    year = "2020"
}

@article{DiLuzio:2020oah,
    author = {Di Luzio, Luca and Gr{\"o}ber, Ramona and Paradisi, Paride},
    title = "{Hunting for $CP$-violating axionlike particle interactions}",
    eprint = "2010.13760",
    archivePrefix = "arXiv",
    primaryClass = "hep-ph",
    reportNumber = "DESY 20-183, DESY-20-183",
    doi = "10.1103/PhysRevD.104.095027",
    journal = "Phys. Rev. D",
    volume = "104",
    number = "9",
    pages = "095027",
    year = "2021"
}

@article{DiLuzio:2023lmd,
    author = "Di Luzio, Luca and Gisbert, Hector and Levati, Gabriele and Paradisi, Paride and S{\o}rensen, Philip",
    title = "{CP-Violating Axions: A Theory Review}",
    eprint = "2312.17310",
    archivePrefix = "arXiv",
    primaryClass = "hep-ph",
    month = "12",
    year = "2023"
}

@article{DiLuzio:2023cuk,
    author = "Di Luzio, Luca and Levati, Gabriele and Paradisi, Paride",
    title = "{The chiral Lagrangian of CP-violating axion-like particles}",
    eprint = "2311.12158",
    archivePrefix = "arXiv",
    primaryClass = "hep-ph",
    doi = "10.1007/JHEP02(2024)020",
    journal = "JHEP",
    volume = "02",
    pages = "020",
    year = "2024"
}

@article{Calibbi:2020jvd,
    author = "Calibbi, Lorenzo and Redigolo, Diego and Ziegler, Robert and Zupan, Jure",
    title = "{Looking forward to lepton-flavor-violating ALPs}",
    eprint = "2006.04795",
    archivePrefix = "arXiv",
    primaryClass = "hep-ph",
    reportNumber = "P3H-20-024, TTP20-025",
    doi = "10.1007/JHEP09(2021)173",
    journal = "JHEP",
    volume = "09",
    pages = "173",
    year = "2021"
}

@article{Calibbi:2022izs,
    author = "Calibbi, Lorenzo and Huang, Zijie and Qin, Shaoyang and Yang, Yiming and Yin, Xiaoyue",
    title = "{Testing axion couplings to leptons in Z decays at future e+e- colliders}",
    eprint = "2212.02818",
    archivePrefix = "arXiv",
    primaryClass = "hep-ph",
    doi = "10.1103/PhysRevD.108.015002",
    journal = "Phys. Rev. D",
    volume = "108",
    number = "1",
    pages = "015002",
    year = "2023"
}

@article{Calibbi:2024rcm,
    author = "Calibbi, Lorenzo and Li, Tong and Mukherjee, Lopamudra and Yang, Yiming",
    title = "{Probing ALP lepton flavor violation at {\ensuremath{\mu}}TRISTAN}",
    eprint = "2406.13234",
    archivePrefix = "arXiv",
    primaryClass = "hep-ph",
    doi = "10.1103/PhysRevD.110.115009",
    journal = "Phys. Rev. D",
    volume = "110",
    number = "11",
    pages = "115009",
    year = "2024"
}

@article{LHCHiggsCrossSectionWorkingGroup:2016ypw,
    author = "de Florian, D. and others",
    collaboration = "LHC Higgs Cross Section Working Group",
    title = "{Handbook of LHC Higgs Cross Sections: 4. Deciphering the Nature of the Higgs Sector}",
    eprint = "1610.07922",
    archivePrefix = "arXiv",
    primaryClass = "hep-ph",
    reportNumber = "CERN-2017-002-M, CERN-2017-002",
    doi = "10.23731/CYRM-2017-002",
    journal = "CERN Yellow Rep. Monogr.",
    volume = "2",
    pages = "1--869",
    year = "2017"
}

@article{Grzadkowski:2010es,
    author = "Grzadkowski, B. and Iskrzynski, M. and Misiak, M. and Rosiek, J.",
    title = "{Dimension-Six Terms in the Standard Model Lagrangian}",
    eprint = "1008.4884",
    archivePrefix = "arXiv",
    primaryClass = "hep-ph",
    reportNumber = "IFT-9-2010, TTP10-35",
    doi = "10.1007/JHEP10(2010)085",
    journal = "JHEP",
    volume = "10",
    pages = "085",
    year = "2010"
}

@article{Weinberg:1979sa,
    author = "Weinberg, Steven",
    title = "{Baryon and Lepton Nonconserving Processes}",
    reportNumber = "HUTP-79-A050",
    doi = "10.1103/PhysRevLett.43.1566",
    journal = "Phys. Rev. Lett.",
    volume = "43",
    pages = "1566--1570",
    year = "1979"
}

@article{Biekotter:2025fll,
    author = {Biek{\"o}tter, Anke and Mimasu, Ken},
    title = "{Axions and Axion-like particles: collider searches}",
    eprint = "2508.19358",
    archivePrefix = "arXiv",
    primaryClass = "hep-ph",
    doi = "10.1016/B978-0-443-26598-3.00049-3",
    journal = "Encycl. Part. Phys.",
    volume = "3",
    pages = "673--701",
    year = "2026"
}

@article{Alda:2025uwo,
    author = "Alda, Jorge and Fuentes Zamoro, Marta and Merlo, Luca and Ponce D{\'\i}az, Xavier and Rigolin, Stefano",
    title = "{Comprehensive ALP Searches in Meson Decays}",
    eprint = "2507.19578",
    archivePrefix = "arXiv",
    primaryClass = "hep-ph",
    reportNumber = "IFT-UAM/CSIC-25-55",
    month = "7",
    year = "2025"
}

@article{Wilczek:1977pj,
    author = "Wilczek, Frank",
    title = "{Problem of Strong  $P$  and  $T$  Invariance in the Presence of Instantons}",
    reportNumber = "Print-77-0939 (COLUMBIA)",
    doi = "10.1103/PhysRevLett.40.279",
    journal = "Phys. Rev. Lett.",
    volume = "40",
    pages = "279--282",
    year = "1978"
}

@article{Witten:1984dg,
    author = "Witten, Edward",
    title = "{Some Properties of O(32) Superstrings}",
    reportNumber = "Print-84-0838 (PRINCETON)",
    doi = "10.1016/0370-2693(84)90422-2",
    journal = "Phys. Lett. B",
    volume = "149",
    pages = "351--356",
    year = "1984"
}

@article{Choi:2006qj,
    author = "Choi, Kang-Sin and Kim, Ian-Woo and Kim, Jihn E.",
    title = "{String compactification, QCD axion and axion-photon-photon coupling}",
    eprint = "hep-ph/0612107",
    archivePrefix = "arXiv",
    reportNumber = "SNUTP-06-012",
    doi = "10.1088/1126-6708/2007/03/116",
    journal = "JHEP",
    volume = "03",
    pages = "116",
    year = "2007"
}

@article{Svrcek:2006yi,
    author = "Svrcek, Peter and Witten, Edward",
    title = "{Axions In String Theory}",
    eprint = "hep-th/0605206",
    archivePrefix = "arXiv",
    reportNumber = "SLAC-PUB-11894",
    doi = "10.1088/1126-6708/2006/06/051",
    journal = "JHEP",
    volume = "06",
    pages = "051",
    year = "2006"
}

@article{Arvanitaki:2009fg,
    author = "Arvanitaki, Asimina and Dimopoulos, Savas and Dubovsky, Sergei and Kaloper, Nemanja and March-Russell, John",
    title = "{String Axiverse}",
    eprint = "0905.4720",
    archivePrefix = "arXiv",
    primaryClass = "hep-th",
    doi = "10.1103/PhysRevD.81.123530",
    journal = "Phys. Rev. D",
    volume = "81",
    pages = "123530",
    year = "2010"
}

@article{Cicoli:2012sz,
    author = "Cicoli, Michele and Goodsell, Mark and Ringwald, Andreas",
    title = "{The type IIB string axiverse and its low-energy phenomenology}",
    eprint = "1206.0819",
    archivePrefix = "arXiv",
    primaryClass = "hep-th",
    reportNumber = "DESY-12-058, CERN-PH-TH-2012-153",
    doi = "10.1007/JHEP10(2012)146",
    journal = "JHEP",
    volume = "10",
    pages = "146",
    year = "2012"
}

@article{Bellazzini:2017neg,
    author = "Bellazzini, Brando and Mariotti, Alberto and Redigolo, Diego and Sala, Filippo and Serra, Javi",
    title = "{$R$-axion at colliders}",
    eprint = "1702.02152",
    archivePrefix = "arXiv",
    primaryClass = "hep-ph",
    reportNumber = "CERN-TH-2017-032, Saclay-t17/014",
    doi = "10.1103/PhysRevLett.119.141804",
    journal = "Phys. Rev. Lett.",
    volume = "119",
    number = "14",
    pages = "141804",
    year = "2017"
}

@article{Gelmini:1984pe,
    author = "Gelmini, G. and Schramm, David N. and Valle, J. W. F.",
    title = "{Majorons: A Simultaneous Solution to the Large and Small Scale Dark Matter Problems}",
    reportNumber = "CERN-TH-3865",
    doi = "10.1016/0370-2693(84)91703-9",
    journal = "Phys. Lett. B",
    volume = "146",
    pages = "311--317",
    year = "1984"
}

@article{Berezinsky:1993fm,
    author = "Berezinsky, V. and Valle, J. W. F.",
    title = "{The KeV majoron as a dark matter particle}",
    eprint = "hep-ph/9309214",
    archivePrefix = "arXiv",
    reportNumber = "FTUV-93-35, LNGS-93-79",
    doi = "10.1016/0370-2693(93)90140-D",
    journal = "Phys. Lett. B",
    volume = "318",
    pages = "360--366",
    year = "1993"
}

@article{Lattanzi:2007ux,
    author = "Lattanzi, M. and Valle, J. W. F.",
    title = "{Decaying warm dark matter and neutrino masses}",
    eprint = "0705.2406",
    archivePrefix = "arXiv",
    primaryClass = "astro-ph",
    reportNumber = "IFIC-06-39",
    doi = "10.1103/PhysRevLett.99.121301",
    journal = "Phys. Rev. Lett.",
    volume = "99",
    pages = "121301",
    year = "2007"
}

@article{Bazzocchi:2008fh,
    author = "Bazzocchi, Federica and Lattanzi, Massimiliano and Riemer-S{\o}rensen, Signe and Valle, Jose W. F.",
    title = "{X-ray photons from late-decaying majoron dark matter}",
    eprint = "0805.2372",
    archivePrefix = "arXiv",
    primaryClass = "astro-ph",
    reportNumber = "IFIC-08-25",
    doi = "10.1088/1475-7516/2008/08/013",
    journal = "JCAP",
    volume = "08",
    pages = "013",
    year = "2008"
}

@article{Lattanzi:2013uza,
    author = "Lattanzi, Massimiliano and Riemer-Sorensen, Signe and Tortola, Mariam and Valle, Jose W. F.",
    title = "{Updated CMB and x- and $\gamma$-ray constraints on Majoron dark matter}",
    eprint = "1303.4685",
    archivePrefix = "arXiv",
    primaryClass = "astro-ph.HE",
    reportNumber = "IFIC-13-15, IFIC-13-XXX",
    doi = "10.1103/PhysRevD.88.063528",
    journal = "Phys. Rev. D",
    volume = "88",
    number = "6",
    pages = "063528",
    year = "2013"
}

@article{Queiroz:2014yna,
    author = "Queiroz, Farinaldo S. and Sinha, Kuver",
    title = "{The Poker Face of the Majoron Dark Matter Model: LUX to keV Line}",
    eprint = "1404.1400",
    archivePrefix = "arXiv",
    primaryClass = "hep-ph",
    reportNumber = "CETUP2013-025",
    doi = "10.1016/j.physletb.2014.06.016",
    journal = "Phys. Lett. B",
    volume = "735",
    pages = "69--74",
    year = "2014"
}

@article{Ferreira:2018vjj,
    author = "Ferreira, Ricardo Z. and Notari, Alessio",
    title = "{Observable Windows for the QCD Axion Through the Number of Relativistic Species}",
    eprint = "1801.06090",
    archivePrefix = "arXiv",
    primaryClass = "hep-ph",
    doi = "10.1103/PhysRevLett.120.191301",
    journal = "Phys. Rev. Lett.",
    volume = "120",
    number = "19",
    pages = "191301",
    year = "2018"
}

@article{DEramo:2018vss,
    author = "D'Eramo, Francesco and Ferreira, Ricardo Z. and Notari, Alessio and Bernal, Jos{\'e} Luis",
    title = "{Hot Axions and the $H_0$ tension}",
    eprint = "1808.07430",
    archivePrefix = "arXiv",
    primaryClass = "hep-ph",
    doi = "10.1088/1475-7516/2018/11/014",
    journal = "JCAP",
    volume = "11",
    pages = "014",
    year = "2018"
}

@article{Escudero:2019gvw,
    author = "Escudero, Miguel and Witte, Samuel J.",
    title = "{A CMB search for the neutrino mass mechanism and its relation to the Hubble tension}",
    eprint = "1909.04044",
    archivePrefix = "arXiv",
    primaryClass = "astro-ph.CO",
    reportNumber = "KCL-2019-71",
    doi = "10.1140/epjc/s10052-020-7854-5",
    journal = "Eur. Phys. J. C",
    volume = "80",
    number = "4",
    pages = "294",
    year = "2020"
}

@article{Arias-Aragon:2020qtn,
    author = "Arias-Arag{\'o}n, Fernando and D'eramo, Francesco and Ferreira, Ricardo Z. and Merlo, Luca and Notari, Alessio",
    title = "{Cosmic Imprints of XENON1T Axions}",
    eprint = "2007.06579",
    archivePrefix = "arXiv",
    primaryClass = "hep-ph",
    reportNumber = "FTUAM-20-12, IFT-UAM/CSIC-20-104",
    doi = "10.1088/1475-7516/2020/11/025",
    journal = "JCAP",
    volume = "11",
    pages = "025",
    year = "2020"
}

@article{Arias-Aragon:2020qip,
    author = "Arias-Aragon, Fernando and Fernandez-Martinez, Enrique and Gonzalez-Lopez, Manuel and Merlo, Luca",
    title = "{Neutrino Masses and Hubble Tension via a Majoron in MFV}",
    eprint = "2009.01848",
    archivePrefix = "arXiv",
    primaryClass = "hep-ph",
    doi = "10.1140/epjc/s10052-020-08825-8",
    journal = "Eur. Phys. J. C",
    volume = "81",
    number = "1",
    pages = "28",
    year = "2021"
}

@article{Arias-Aragon:2020shv,
    author = "Arias-Arag{\'o}n, Fernando and D'Eramo, Francesco and Ferreira, Ricardo Z. and Merlo, Luca and Notari, Alessio",
    title = "{Production of Thermal Axions across the ElectroWeak Phase Transition}",
    eprint = "2012.04736",
    archivePrefix = "arXiv",
    primaryClass = "hep-ph",
    doi = "10.1088/1475-7516/2021/03/090",
    journal = "JCAP",
    volume = "03",
    pages = "090",
    year = "2021"
}

@article{Ferreira:2020bpb,
    author = "Ferreira, Ricardo Z. and Notari, Alessio and Rompineve, Fabrizio",
    title = "{Dine-Fischler-Srednicki-Zhitnitsky axion in the CMB}",
    eprint = "2012.06566",
    archivePrefix = "arXiv",
    primaryClass = "hep-ph",
    doi = "10.1103/PhysRevD.103.063524",
    journal = "Phys. Rev. D",
    volume = "103",
    number = "6",
    pages = "063524",
    year = "2021"
}

@article{Escudero:2021rfi,
    author = "Escudero, Miguel and Witte, Samuel J.",
    title = "{The hubble tension as a hint of leptogenesis and neutrino mass generation}",
    eprint = "2103.03249",
    archivePrefix = "arXiv",
    primaryClass = "hep-ph",
    reportNumber = "TUM-HEP 1318/21",
    doi = "10.1140/epjc/s10052-021-09276-5",
    journal = "Eur. Phys. J. C",
    volume = "81",
    number = "6",
    pages = "515",
    year = "2021"
}

@article{Araki:2021xdk,
    author = "Araki, Takeshi and Asai, Kento and Honda, Kei and Kasuya, Ryuta and Sato, Joe and Shimomura, Takashi and Yang, Masaki J. S.",
    title = "{Resolving the Hubble tension in a U(1)$_{L_\mu-L_\tau}$ model with the Majoron}",
    eprint = "2103.07167",
    archivePrefix = "arXiv",
    primaryClass = "hep-ph",
    reportNumber = "STUPP-20-242, UME-PP-18",
    doi = "10.1093/ptep/ptab108",
    journal = "PTEP",
    volume = "2021",
    number = "10",
    pages = "103B05",
    year = "2021"
}

@article{DEramo:2021psx,
    author = "D'Eramo, Francesco and Hajkarim, Fazlollah and Yun, Seokhoon",
    title = "{Thermal Axion Production at Low Temperatures: A Smooth Treatment of the QCD Phase Transition}",
    eprint = "2108.04259",
    archivePrefix = "arXiv",
    primaryClass = "hep-ph",
    doi = "10.1103/PhysRevLett.128.152001",
    journal = "Phys. Rev. Lett.",
    volume = "128",
    number = "15",
    pages = "152001",
    year = "2022"
}

@article{DEramo:2021lgb,
    author = "D'Eramo, Francesco and Hajkarim, Fazlollah and Yun, Seokhoon",
    title = "{Thermal QCD Axions across Thresholds}",
    eprint = "2108.05371",
    archivePrefix = "arXiv",
    primaryClass = "hep-ph",
    doi = "10.1007/JHEP10(2021)224",
    journal = "JHEP",
    volume = "10",
    pages = "224",
    year = "2021"
}

@article{DEramo:2022nvb,
    author = "D'Eramo, Francesco and Di Valentino, Eleonora and Giar{\`e}, William and Hajkarim, Fazlollah and Melchiorri, Alessandro and Mena, Olga and Renzi, Fabrizio and Yun, Seokhoon",
    title = "{Cosmological bound on the QCD axion mass, redux}",
    eprint = "2205.07849",
    archivePrefix = "arXiv",
    primaryClass = "astro-ph.CO",
    doi = "10.1088/1475-7516/2022/09/022",
    journal = "JCAP",
    volume = "09",
    pages = "022",
    year = "2022"
}

@article{Davidson:1981zd,
    author = "Davidson, Aharon and Wali, Kameshwar C.",
    title = "{MINIMAL FLAVOR UNIFICATION VIA MULTIGENERATIONAL PECCEI-QUINN SYMMETRY}",
    reportNumber = "WIS-81/40-Ph, SU-4217-206",
    doi = "10.1103/PhysRevLett.48.11",
    journal = "Phys. Rev. Lett.",
    volume = "48",
    pages = "11",
    year = "1982"
}

@article{Wilczek:1982rv,
    author = "Wilczek, Frank",
    title = "{Axions and Family Symmetry Breaking}",
    doi = "10.1103/PhysRevLett.49.1549",
    journal = "Phys. Rev. Lett.",
    volume = "49",
    pages = "1549--1552",
    year = "1982"
}

@article{Ema:2025bww,
    author = "Ema, Yohei and Fox, Patrick J. and Hostert, Matheus and Menzo, Tony and Pospelov, Maxim and Ray, Anupam and Zupan, Jure",
    title = "{Long-lived axionlike particles from tau decays}",
    eprint = "2507.15271",
    archivePrefix = "arXiv",
    primaryClass = "hep-ph",
    reportNumber = "CERN-TH-2025-123, FERMILAB-PUB-25-0408-T, N3AS-25-011",
    doi = "10.1103/51gx-m32t",
    journal = "Phys. Rev. D",
    volume = "112",
    number = "11",
    pages = "115028",
    year = "2025"
}

@article{Albertus:2026fbe,
    author = "Albertus, Conrado and others",
    title = "{WISPedia -- the WISPs Encyclopedia:~Cosmic WISPers 2026 {\textendash} V1.0}",
    eprint = "2602.09089",
    archivePrefix = "arXiv",
    primaryClass = "hep-ph",
    reportNumber = "IPPP/26/14, IFT-UAM/CSIC-26-9",
    month = "2",
    year = "2026"
}

@article{Arza:2026rsl,
    author = "Arza, Ariel and others",
    title = "{The COSMIC WISPers White Paper: The physics case for Weakly Interacting Slim Particles}",
    eprint = "2603.03433",
    archivePrefix = "arXiv",
    primaryClass = "hep-ph",
    reportNumber = "BARI-TH/784-26, CERN-TH-2026-016, hal-05395643, IPPP/26/13, IFT-UAM/CSIC-26-13, KCL-PH-TH/2026-04, KEK-Cosmo-0411, KEK-TH-2804, LAPTH-008/26, MPP-2026-21, RESCEU-5/26, SLAC-PUB-260219, ST/T006994/1, ST/Y004531/1",
    month = "3",
    year = "2026"
}

@article{Ema:2016ops,
    author = "Ema, Yohei and Hamaguchi, Koichi and Moroi, Takeo and Nakayama, Kazunori",
    title = "{Flaxion: a minimal extension to solve puzzles in the standard model}",
    eprint = "1612.05492",
    archivePrefix = "arXiv",
    primaryClass = "hep-ph",
    reportNumber = "UT-16-36, IPMU16-0189",
    doi = "10.1007/JHEP01(2017)096",
    journal = "JHEP",
    volume = "01",
    pages = "096",
    year = "2017"
}

@article{Calibbi:2016hwq,
    author = "Calibbi, Lorenzo and Goertz, Florian and Redigolo, Diego and Ziegler, Robert and Zupan, Jure",
    title = "{Minimal axion model from flavor}",
    eprint = "1612.08040",
    archivePrefix = "arXiv",
    primaryClass = "hep-ph",
    reportNumber = "TTP16-058, CERN-TH-2016-261",
    doi = "10.1103/PhysRevD.95.095009",
    journal = "Phys. Rev. D",
    volume = "95",
    number = "9",
    pages = "095009",
    year = "2017"
}

@article{Arias-Aragon:2017eww,
    author = "Arias-Aragon, F. and Merlo, L.",
    title = "{The Minimal Flavour Violating Axion}",
    eprint = "1709.07039",
    archivePrefix = "arXiv",
    primaryClass = "hep-ph",
    doi = "10.1007/JHEP10(2017)168",
    journal = "JHEP",
    volume = "10",
    pages = "168",
    year = "2017",
    note = "[Erratum: JHEP 11, 152 (2019)]"
}

@article{Arias-Aragon:2022ats,
    author = "Arias-Arag{\'o}n, Fernando and Fern{\'a}ndez-Mart{\'\i}nez, Enrique and Gonz{\'a}lez-L{\'o}pez, Manuel and Merlo, Luca",
    title = "{Dynamical Minimal Flavour Violating inverse seesaw}",
    eprint = "2204.04672",
    archivePrefix = "arXiv",
    primaryClass = "hep-ph",
    reportNumber = "IFT-UAM/CSIC-22-34",
    doi = "10.1007/JHEP09(2022)210",
    journal = "JHEP",
    volume = "09",
    pages = "210",
    year = "2022"
}

@article{DiLuzio:2023ndz,
    author = "Di Luzio, Luca and Guerrera, Alfredo Walter Mario and D{\'\i}az, Xavier Ponce and Rigolin, Stefano",
    title = "{On the IR/UV flavour connection in non-universal axion models}",
    eprint = "2304.04643",
    archivePrefix = "arXiv",
    primaryClass = "hep-ph",
    doi = "10.1007/JHEP06(2023)046",
    journal = "JHEP",
    volume = "06",
    pages = "046",
    year = "2023"
}

@article{Greljo:2024evt,
    author = "Greljo, Admir and Smolkovi{\v{c}}, Aleks and Valenti, Alessandro",
    title = "{Froggatt-Nielsen ALP}",
    eprint = "2407.02998",
    archivePrefix = "arXiv",
    primaryClass = "hep-ph",
    doi = "10.1007/JHEP09(2024)174",
    journal = "JHEP",
    volume = "09",
    pages = "174",
    year = "2024"
}

@article{Chikashige:1980qk,
    author = "Chikashige, Y. and Mohapatra, Rabindra N. and Peccei, R. D.",
    title = "{Spontaneously Broken Lepton Number and Cosmological Constraints on the Neutrino Mass Spectrum}",
    reportNumber = "MPI-PAE/PTh 40/80",
    doi = "10.1103/PhysRevLett.45.1926",
    journal = "Phys. Rev. Lett.",
    volume = "45",
    pages = "1926",
    year = "1980"
}

@article{Chikashige:1980ui,
    author = "Chikashige, Y. and Mohapatra, Rabindra N. and Peccei, R. D.",
    title = "{Are There Real Goldstone Bosons Associated with Broken Lepton Number?}",
    reportNumber = "MPI-PAE-PTH-36-80",
    doi = "10.1016/0370-2693(81)90011-3",
    journal = "Phys. Lett. B",
    volume = "98",
    pages = "265--268",
    year = "1981"
}

@article{Gelmini:1980re,
    author = "Gelmini, G. B. and Roncadelli, M.",
    title = "{Left-Handed Neutrino Mass Scale and Spontaneously Broken Lepton Number}",
    reportNumber = "MPI-PAE-PTH-50-80",
    doi = "10.1016/0370-2693(81)90559-1",
    journal = "Phys. Lett. B",
    volume = "99",
    pages = "411--415",
    year = "1981"
}

@article{deGiorgi:2023tvn,
    author = "de Giorgi, Arturo and Merlo, Luca and Ponce D{\'\i}az, Xavier and Rigolin, Stefano",
    title = "{The minimal massive Majoron Seesaw Model}",
    eprint = "2312.13417",
    archivePrefix = "arXiv",
    primaryClass = "hep-ph",
    doi = "10.1007/JHEP03(2024)094",
    journal = "JHEP",
    volume = "03",
    pages = "094",
    year = "2024"
}

@article{Choi:1986zw,
    author = "Choi, Kiwoon and Kang, Kyungsik and Kim, Jihn E.",
    title = "{Effects of $\eta^\prime$ in Low-energy Axion Physics}",
    reportNumber = "BROWN-HET-593, SNUHE 86/05",
    doi = "10.1016/0370-2693(86)91273-6",
    journal = "Phys. Lett. B",
    volume = "181",
    pages = "145--149",
    year = "1986"
}

@article{Salvio:2013iaa,
    author = "Salvio, Alberto and Strumia, Alessandro and Xue, Wei",
    title = "{Thermal axion production}",
    eprint = "1310.6982",
    archivePrefix = "arXiv",
    primaryClass = "hep-ph",
    reportNumber = "FTUAM-13-29, IFT-UAM-CSIC-13-113",
    doi = "10.1088/1475-7516/2014/01/011",
    journal = "JCAP",
    volume = "01",
    pages = "011",
    year = "2014"
}

@article{Brivio:2017ije,
    author = "Brivio, I. and Gavela, M. B. and Merlo, L. and Mimasu, K. and No, J. M. and del Rey, R. and Sanz, V.",
    title = "{ALPs Effective Field Theory and Collider Signatures}",
    eprint = "1701.05379",
    archivePrefix = "arXiv",
    primaryClass = "hep-ph",
    reportNumber = "IFT-UAM-CSIC-16-141, KCL-PH-TH-2016-72, FTUAM-16-49, CP3-17-04",
    doi = "10.1140/epjc/s10052-017-5111-3",
    journal = "Eur. Phys. J. C",
    volume = "77",
    number = "8",
    pages = "572",
    year = "2017"
}

@article{Alonso-Alvarez:2018irt,
    author = "Alonso-{\'A}lvarez, G. and Gavela, M. B. and Quilez, P.",
    title = "{Axion couplings to electroweak gauge bosons}",
    eprint = "1811.05466",
    archivePrefix = "arXiv",
    primaryClass = "hep-ph",
    reportNumber = "IFT-UAM/CSIC-18-110, FTUAM-18-25",
    doi = "10.1140/epjc/s10052-019-6732-5",
    journal = "Eur. Phys. J. C",
    volume = "79",
    number = "3",
    pages = "223",
    year = "2019"
}

@article{Gavela:2019wzg,
    author = "Gavela, M. B. and Houtz, R. and Quilez, P. and Del Rey, R. and Sumensari, O.",
    title = "{Flavor constraints on electroweak ALP couplings}",
    eprint = "1901.02031",
    archivePrefix = "arXiv",
    primaryClass = "hep-ph",
    doi = "10.1140/epjc/s10052-019-6889-y",
    journal = "Eur. Phys. J. C",
    volume = "79",
    number = "5",
    pages = "369",
    year = "2019"
}

@article{Chala:2020wvs,
    author = "Chala, Mikael and Guedes, Guilherme and Ramos, Maria and Santiago, Jose",
    title = "{Running in the ALPs}",
    eprint = "2012.09017",
    archivePrefix = "arXiv",
    primaryClass = "hep-ph",
    doi = "10.1140/epjc/s10052-021-08968-2",
    journal = "Eur. Phys. J. C",
    volume = "81",
    number = "2",
    pages = "181",
    year = "2021"
}

@article{Bonilla:2021ufe,
    author = "Bonilla, J. and Brivio, I. and Gavela, M. B. and Sanz, V.",
    title = "{One-loop corrections to ALP couplings}",
    eprint = "2107.11392",
    archivePrefix = "arXiv",
    primaryClass = "hep-ph",
    reportNumber = "IFT-UAM/CSIC-21-82",
    doi = "10.1007/JHEP11(2021)168",
    journal = "JHEP",
    volume = "11",
    pages = "168",
    year = "2021"
}

@article{Arias-Aragon:2022byr,
    author = "Arias-Arag{\'o}n, Fernando and Smith, Christopher",
    title = "{Leptoquarks, axions and the unification of B, L, and Peccei-Quinn symmetries}",
    eprint = "2206.09810",
    archivePrefix = "arXiv",
    primaryClass = "hep-ph",
    doi = "10.1103/PhysRevD.106.055034",
    journal = "Phys. Rev. D",
    volume = "106",
    number = "5",
    pages = "055034",
    year = "2022"
}

@article{Arias-Aragon:2022iwl,
    author = "Arias-Arag{\'o}n, Fernando and Quevillon, J{\'e}r{\'e}mie and Smith, Christopher",
    title = "{Axion-like ALPs}",
    eprint = "2211.04489",
    archivePrefix = "arXiv",
    primaryClass = "hep-ph",
    reportNumber = "CERN-TH-2022-182",
    doi = "10.1007/JHEP03(2023)134",
    journal = "JHEP",
    volume = "03",
    pages = "134",
    year = "2023"
}

@article{Jaeckel:2012yz,
    author = "Jaeckel, Joerg and Jankowiak, Martin and Spannowsky, Michael",
    title = "{LHC probes the hidden sector}",
    eprint = "1212.3620",
    archivePrefix = "arXiv",
    primaryClass = "hep-ph",
    reportNumber = "IPPP-12-94, DCPT-12-188",
    doi = "10.1016/j.dark.2013.06.001",
    journal = "Phys. Dark Univ.",
    volume = "2",
    pages = "111--117",
    year = "2013"
}

@article{Mimasu:2014nea,
    author = "Mimasu, Ken and Sanz, Ver{\'o}nica",
    title = "{ALPs at Colliders}",
    eprint = "1409.4792",
    archivePrefix = "arXiv",
    primaryClass = "hep-ph",
    doi = "10.1007/JHEP06(2015)173",
    journal = "JHEP",
    volume = "06",
    pages = "173",
    year = "2015"
}

@article{Jaeckel:2015jla,
    author = "Jaeckel, Joerg and Spannowsky, Michael",
    title = "{Probing MeV to 90 GeV axion-like particles with LEP and LHC}",
    eprint = "1509.00476",
    archivePrefix = "arXiv",
    primaryClass = "hep-ph",
    doi = "10.1016/j.physletb.2015.12.037",
    journal = "Phys. Lett. B",
    volume = "753",
    pages = "482--487",
    year = "2016"
}

@article{Alves:2016koo,
    author = "Alves, Alexandre and Dias, Alex G. and Sinha, Kuver",
    title = "{Diphotons at the $Z$-pole in Models of the 750 GeV Resonance Decaying to Axion-Like Particles}",
    eprint = "1606.06375",
    archivePrefix = "arXiv",
    primaryClass = "hep-ph",
    doi = "10.1007/JHEP08(2016)060",
    journal = "JHEP",
    volume = "08",
    pages = "060",
    year = "2016"
}

@article{Knapen:2016moh,
    author = "Knapen, Simon and Lin, Tongyan and Lou, Hou Keong and Melia, Tom",
    title = "{Searching for Axionlike Particles with Ultraperipheral Heavy-Ion Collisions}",
    eprint = "1607.06083",
    archivePrefix = "arXiv",
    primaryClass = "hep-ph",
    doi = "10.1103/PhysRevLett.118.171801",
    journal = "Phys. Rev. Lett.",
    volume = "118",
    number = "17",
    pages = "171801",
    year = "2017"
}

@article{Bauer:2017nlg,
    author = "Bauer, Martin and Neubert, Matthias and Thamm, Andrea",
    title = "{LHC as an Axion Factory: Probing an Axion Explanation for $(g-2)_\mu$ with Exotic Higgs Decays}",
    eprint = "1704.08207",
    archivePrefix = "arXiv",
    primaryClass = "hep-ph",
    doi = "10.1103/PhysRevLett.119.031802",
    journal = "Phys. Rev. Lett.",
    volume = "119",
    number = "3",
    pages = "031802",
    year = "2017"
}

@article{Mariotti:2017vtv,
    author = "Mariotti, Alberto and Redigolo, Diego and Sala, Filippo and Tobioka, Kohsaku",
    title = "{New LHC bound on low-mass diphoton resonances}",
    eprint = "1710.01743",
    archivePrefix = "arXiv",
    primaryClass = "hep-ph",
    reportNumber = "DESY-17-148",
    doi = "10.1016/j.physletb.2018.06.039",
    journal = "Phys. Lett. B",
    volume = "783",
    pages = "13--18",
    year = "2018"
}

@article{Bauer:2017ris,
    author = "Bauer, Martin and Neubert, Matthias and Thamm, Andrea",
    title = "{Collider Probes of Axion-Like Particles}",
    eprint = "1708.00443",
    archivePrefix = "arXiv",
    primaryClass = "hep-ph",
    reportNumber = "MITP-17-047",
    doi = "10.1007/JHEP12(2017)044",
    journal = "JHEP",
    volume = "12",
    pages = "044",
    year = "2017"
}

@article{Baldenegro:2018hng,
    author = "Baldenegro, Cristian and Fichet, Sylvain and von Gersdorff, Gero and Royon, Christophe",
    title = "{Searching for axion-like particles with proton tagging at the LHC}",
    eprint = "1803.10835",
    archivePrefix = "arXiv",
    primaryClass = "hep-ph",
    doi = "10.1007/JHEP06(2018)131",
    journal = "JHEP",
    volume = "06",
    pages = "131",
    year = "2018"
}

@article{Craig:2018kne,
    author = "Craig, Nathaniel and Hook, Anson and Kasko, Skyler",
    title = "{The Photophobic ALP}",
    eprint = "1805.06538",
    archivePrefix = "arXiv",
    primaryClass = "hep-ph",
    doi = "10.1007/JHEP09(2018)028",
    journal = "JHEP",
    volume = "09",
    pages = "028",
    year = "2018"
}

@article{Bauer:2018uxu,
    author = "Bauer, Martin and Heiles, Mathias and Neubert, Matthias and Thamm, Andrea",
    title = "{Axion-Like Particles at Future Colliders}",
    eprint = "1808.10323",
    archivePrefix = "arXiv",
    primaryClass = "hep-ph",
    reportNumber = "CERN-TH-2018-199, MITP/18-075",
    doi = "10.1140/epjc/s10052-019-6587-9",
    journal = "Eur. Phys. J. C",
    volume = "79",
    number = "1",
    pages = "74",
    year = "2019"
}

@article{Gavela:2019cmq,
    author = "Gavela, M. B. and No, J. M. and Sanz, V. and de Troc{\'o}niz, J. F.",
    title = "{Nonresonant Searches for Axionlike Particles at the LHC}",
    eprint = "1905.12953",
    archivePrefix = "arXiv",
    primaryClass = "hep-ph",
    doi = "10.1103/PhysRevLett.124.051802",
    journal = "Phys. Rev. Lett.",
    volume = "124",
    number = "5",
    pages = "051802",
    year = "2020"
}

@article{Haghighat:2020nuh,
    author = "Haghighat, Gholamhossein and Haji Raissi, Daruosh and Mohammadi Najafabadi, Mojtaba",
    title = "{New collider searches for axionlike particles coupling to gluons}",
    eprint = "2006.05302",
    archivePrefix = "arXiv",
    primaryClass = "hep-ph",
    doi = "10.1103/PhysRevD.102.115010",
    journal = "Phys. Rev. D",
    volume = "102",
    number = "11",
    pages = "115010",
    year = "2020"
}

@article{Wang:2021uyb,
    author = "Wang, Daohan and Wu, Lei and Yang, Jin Min and Zhang, Mengchao",
    title = "{Photon-jet events as a probe of axionlike particles at the LHC}",
    eprint = "2102.01532",
    archivePrefix = "arXiv",
    primaryClass = "hep-ph",
    doi = "10.1103/PhysRevD.104.095016",
    journal = "Phys. Rev. D",
    volume = "104",
    number = "9",
    pages = "095016",
    year = "2021"
}

@article{deGiorgi:2022oks,
    author = "de Giorgi, Arturo and Merlo, Luca and Tastet, Jean-Loup",
    title = "{Probing HNL-ALP couplings at colliders}",
    eprint = "2212.11290",
    archivePrefix = "arXiv",
    primaryClass = "hep-ph",
    reportNumber = "IFT-UAM-CSIC-22-152",
    doi = "10.1002/prop.202300027",
    journal = "Fortsch. Phys.",
    volume = "71",
    number = "4-5",
    pages = "2300027",
    year = "2023"
}

@article{Bonilla:2022pxu,
    author = "Bonilla, J. and Brivio, I. and Machado-Rodr{\'\i}guez, J. and de Troc{\'o}niz, J. F.",
    title = "{Nonresonant searches for axion-like particles in vector boson scattering processes at the LHC}",
    eprint = "2202.03450",
    archivePrefix = "arXiv",
    primaryClass = "hep-ph",
    reportNumber = "IFT-UAM/CSIC-22-7, VBSCAN-PUB-01-22",
    doi = "10.1007/JHEP06(2022)113",
    journal = "JHEP",
    volume = "06",
    pages = "113",
    year = "2022"
}

@article{Ghebretinsaea:2022djg,
    author = "Ghebretinsaea, Filmon Andom and Wang, Zeren Simon and Wang, Kechen",
    title = "{Probing axion-like particles coupling to gluons at the LHC}",
    eprint = "2203.01734",
    archivePrefix = "arXiv",
    primaryClass = "hep-ph",
    doi = "10.1007/JHEP07(2022)070",
    journal = "JHEP",
    volume = "07",
    pages = "070",
    year = "2022"
}

@article{Vileta:2022jou,
    author = "Vileta, Victor Enguita and Gavela, Belen and Houtz, Rachel and Quilez, Pablo",
    title = "{Discrete Goldstone bosons}",
    eprint = "2205.09131",
    archivePrefix = "arXiv",
    primaryClass = "hep-ph",
    reportNumber = "DESY-22-082, IFT-UAM/CSIC-20-144, FTUAM-20-21, IPPP/22/31",
    doi = "10.1103/PhysRevD.107.035009",
    journal = "Phys. Rev. D",
    volume = "107",
    number = "3",
    pages = "035009",
    year = "2023"
}

@article{Marcos:2024yfm,
    author = "Marcos, Marta Burgos and de Giorgi, Arturo and Merlo, Luca and Tastet, Jean-Loup",
    title = "{ALPs and HNLs at LHC and Muon Colliders: Uncovering New Couplings and Signals}",
    eprint = "2407.14970",
    archivePrefix = "arXiv",
    primaryClass = "hep-ph",
    reportNumber = "IFT-UAM/CSIC-24-103",
    doi = "10.21468/SciPostPhys.18.3.084",
    journal = "SciPost Phys.",
    volume = "18",
    pages = "084",
    year = "2025"
}

@article{Arias-Aragon:2024gpm,
    author = "Arias-Arag{\'o}n, Fernando and Darm{\'e}, Luc and di Cortona, Giovanni Grilli and Nardi, Enrico",
    title = "{Atoms as Electron Accelerators for Measuring the Cross Section of e+e-{\textrightarrow}Hadrons}",
    eprint = "2407.15941",
    archivePrefix = "arXiv",
    primaryClass = "hep-ph",
    doi = "10.1103/PhysRevLett.134.061802",
    journal = "Phys. Rev. Lett.",
    volume = "134",
    number = "6",
    pages = "061802",
    year = "2025"
}

@article{Izaguirre:2016dfi,
    author = "Izaguirre, Eder and Lin, Tongyan and Shuve, Brian",
    title = "{Searching for Axionlike Particles in Flavor-Changing Neutral Current Processes}",
    eprint = "1611.09355",
    archivePrefix = "arXiv",
    primaryClass = "hep-ph",
    reportNumber = "SLAC-PUB-16876",
    doi = "10.1103/PhysRevLett.118.111802",
    journal = "Phys. Rev. Lett.",
    volume = "118",
    number = "11",
    pages = "111802",
    year = "2017"
}

@article{Merlo:2019anv,
    author = "Merlo, L. and Pobbe, F. and Rigolin, S. and Sumensari, O.",
    title = "{Revisiting the production of ALPs at B-factories}",
    eprint = "1905.03259",
    archivePrefix = "arXiv",
    primaryClass = "hep-ph",
    reportNumber = "IFT-UAM/CISC-19-56, FTUAM-19-8",
    doi = "10.1007/JHEP06(2019)091",
    journal = "JHEP",
    volume = "06",
    pages = "091",
    year = "2019"
}

@article{Aloni:2019ruo,
    author = "Aloni, Daniel and Fanelli, Cristiano and Soreq, Yotam and Williams, Mike",
    title = "{Photoproduction of Axionlike Particles}",
    eprint = "1903.03586",
    archivePrefix = "arXiv",
    primaryClass = "hep-ph",
    reportNumber = "CERN-TH-2019-023",
    doi = "10.1103/PhysRevLett.123.071801",
    journal = "Phys. Rev. Lett.",
    volume = "123",
    number = "7",
    pages = "071801",
    year = "2019"
}

@article{Bauer:2019gfk,
    author = "Bauer, Martin and Neubert, Matthias and Renner, Sophie and Schnubel, Marvin and Thamm, Andrea",
    title = "{Axionlike Particles, Lepton-Flavor Violation, and a New Explanation of $a_\mu$ and $a_e$}",
    eprint = "1908.00008",
    archivePrefix = "arXiv",
    primaryClass = "hep-ph",
    reportNumber = "CERN-TH-2019-124, IPPP/19/64, MITP/19-053",
    doi = "10.1103/PhysRevLett.124.211803",
    journal = "Phys. Rev. Lett.",
    volume = "124",
    number = "21",
    pages = "211803",
    year = "2020"
}

@article{Bauer:2020jbp,
    author = "Bauer, Martin and Neubert, Matthias and Renner, Sophie and Schnubel, Marvin and Thamm, Andrea",
    title = "{The Low-Energy Effective Theory of Axions and ALPs}",
    eprint = "2012.12272",
    archivePrefix = "arXiv",
    primaryClass = "hep-ph",
    reportNumber = "IPPP/20/69, MITP/20-070 SISSA 30/2020/FISI, ZH-TH-47/20",
    doi = "10.1007/JHEP04(2021)063",
    journal = "JHEP",
    volume = "04",
    pages = "063",
    year = "2021"
}

@article{Bauer:2021mvw,
    author = "Bauer, Martin and Neubert, Matthias and Renner, Sophie and Schnubel, Marvin and Thamm, Andrea",
    title = "{Flavor probes of axion-like particles}",
    eprint = "2110.10698",
    archivePrefix = "arXiv",
    primaryClass = "hep-ph",
    reportNumber = "MITP/21-025, CERN-TH-2021-148, IPPP/21/37",
    doi = "10.1007/JHEP09(2022)056",
    journal = "JHEP",
    volume = "09",
    pages = "056",
    year = "2022"
}

@article{Guerrera:2021yss,
    author = "Guerrera, Alfredo Walter Mario and Rigolin, Stefano",
    title = "{Revisiting $K \rightarrow \pi a$ decays}",
    eprint = "2106.05910",
    archivePrefix = "arXiv",
    primaryClass = "hep-ph",
    doi = "10.1140/epjc/s10052-022-10146-x",
    journal = "Eur. Phys. J. C",
    volume = "82",
    number = "3",
    pages = "192",
    year = "2022"
}

@article{Gallo:2021ame,
    author = "Alda, Jorge and Guerrera, Alfredo Walter Mario and Pe{\~n}aranda, Siannah and Rigolin, Stefano",
    title = "{Leptonic meson decays into invisible ALP}",
    eprint = "2111.02536",
    archivePrefix = "arXiv",
    primaryClass = "hep-ph",
    doi = "10.1016/j.nuclphysb.2022.115791",
    journal = "Nucl. Phys. B",
    volume = "979",
    pages = "115791",
    year = "2022"
}

@article{Bonilla:2022qgm,
    author = "Bonilla, J. and de Giorgi, A. and Gavela, B. and Merlo, L. and Ramos, M.",
    title = "{The cost of an ALP solution to the neutral B-anomalies}",
    eprint = "2209.11247",
    archivePrefix = "arXiv",
    primaryClass = "hep-ph",
    doi = "10.1007/JHEP02(2023)138",
    journal = "JHEP",
    volume = "02",
    pages = "138",
    year = "2023"
}

@article{Bonilla:2022vtn,
    author = "Bonilla, Jesus and de Giorgi, Arturo and Ramos, Maria",
    title = "{Neutral $B$-anomalies from an $\mathit{on\text{-}shell}$ scalar exchange}",
    eprint = "2211.05135",
    archivePrefix = "arXiv",
    primaryClass = "hep-ph",
    month = "11",
    year = "2022"
}

@article{deGiorgi:2022vup,
    author = "de Giorgi, Arturo and Piazza, Gioacchino",
    title = "{A Lesson from R{\ensuremath{\tau}}{\ensuremath{\tau}}K(*) and R{\ensuremath{\nu}}{\ensuremath{\nu}}K(*) at Belle~II}",
    eprint = "2211.05595",
    archivePrefix = "arXiv",
    primaryClass = "hep-ph",
    doi = "10.1002/prop.202300200",
    journal = "Fortsch. Phys.",
    volume = "72",
    number = "1",
    pages = "2300200",
    year = "2024"
}

@article{Guerrera:2022ykl,
    author = "Guerrera, Alfredo Walter Mario and Rigolin, Stefano",
    title = "{ALP Production in Weak Mesonic Decays}",
    eprint = "2211.08343",
    archivePrefix = "arXiv",
    primaryClass = "hep-ph",
    doi = "10.1002/prop.202200192",
    journal = "Fortsch. Phys.",
    volume = "71",
    number = "2-3",
    pages = "2200192",
    year = "2023"
}

@article{Bonilla:2023dtf,
    author = "Bonilla, J. and Gavela, B. and Machado-Rodr{\'\i}guez, J.",
    title = "{Limits on ALP-neutrino couplings from loop-level processes}",
    eprint = "2309.15910",
    archivePrefix = "arXiv",
    primaryClass = "hep-ph",
    reportNumber = "IFT-UAM/CSIC-23-121",
    doi = "10.1103/PhysRevD.109.055023",
    journal = "Phys. Rev. D",
    volume = "109",
    number = "5",
    pages = "055023",
    year = "2024"
}

@article{Arias-Aragon:2023ehh,
    author = "Arias-Arag{\'o}n, Fernando and Brdar, Vedran and Quevillon, J{\'e}r{\'e}mie",
    title = "{New Directions for Axionlike Particle Searches Combining Nuclear Reactors and Haloscopes}",
    eprint = "2310.03631",
    archivePrefix = "arXiv",
    primaryClass = "hep-ph",
    reportNumber = "CERN-TH-2023-178",
    doi = "10.1103/PhysRevLett.132.211802",
    journal = "Phys. Rev. Lett.",
    volume = "132",
    number = "21",
    pages = "211802",
    year = "2024"
}

@article{DiLuzio:2024jip,
    author = "Di Luzio, Luca and Guerrera, Alfredo Walter Mario and Ponce D{\'\i}az, Xavier and Rigolin, Stefano",
    title = "{Axion-like particles in radiative quarkonia decays}",
    eprint = "2402.12454",
    archivePrefix = "arXiv",
    primaryClass = "hep-ph",
    doi = "10.1007/JHEP06(2024)217",
    journal = "JHEP",
    volume = "06",
    pages = "217",
    year = "2024"
}

@article{deGiorgi:2024str,
    author = "de Giorgi, Arturo and Fuentes Zamoro, Marta and Merlo, Luca",
    title = "{Visible GeV ALP from TeV Vector-Like Leptons}",
    eprint = "2402.14059",
    archivePrefix = "arXiv",
    primaryClass = "hep-ph",
    reportNumber = "IFT-UAM/CSIC-23-17",
    doi = "10.1002/prop.202400165",
    journal = "Fortsch. Phys.",
    volume = "73",
    number = "3",
    pages = "2400165",
    year = "2025"
}

@article{Alda:2024cxn,
    author = "Alda, Jorge and Levati, Gabriele and Paradisi, Paride and Rigolin, Stefano and Selimovic, Nudzeim",
    title = "{Collider and astrophysical signatures of light scalars with enhanced {\ensuremath{\tau}} couplings}",
    eprint = "2407.18296",
    archivePrefix = "arXiv",
    primaryClass = "hep-ph",
    doi = "10.1007/JHEP06(2025)008",
    journal = "JHEP",
    volume = "06",
    pages = "008",
    year = "2025"
}

@article{Alda:2024xxa,
    author = "Alda, Jorge and Broggini, Carlo and Di Carlo, Giuseppe and Di Luzio, Luca and Piatti, Denise and Rigolin, Stefano and Toni, Claudio",
    title = "{Weak nuclear decays deep-underground as a probe of axion dark matter}",
    eprint = "2412.20932",
    archivePrefix = "arXiv",
    primaryClass = "hep-ph",
    doi = "10.1103/PhysRevD.111.035022",
    journal = "Phys. Rev. D",
    volume = "111",
    number = "3",
    pages = "035022",
    year = "2025"
}

@article{Arias-Aragon:2024qji,
    author = "Arias-Arag{\'o}n, Fernando and Darm{\'e}, Luc and di Cortona, Giovanni Grilli and Nardi, Enrico",
    title = "{Production of Dark Sector Particles via Resonant Positron Annihilation on Atomic Electrons}",
    eprint = "2403.15387",
    archivePrefix = "arXiv",
    primaryClass = "hep-ph",
    doi = "10.1103/PhysRevLett.132.261801",
    journal = "Phys. Rev. Lett.",
    volume = "132",
    number = "26",
    pages = "261801",
    year = "2024",
    note = "[Erratum: Phys.Rev.Lett. 133, 219901 (2024)]"
}

@article{Cheng:2021kjg,
    author = "Cheng, Hsin-Chia and Li, Lingfeng and Salvioni, Ennio",
    title = "{A theory of dark pions}",
    eprint = "2110.10691",
    archivePrefix = "arXiv",
    primaryClass = "hep-ph",
    reportNumber = "CERN-TH-2021-150",
    doi = "10.1007/JHEP01(2022)122",
    journal = "JHEP",
    volume = "01",
    pages = "122",
    year = "2022"
}

@article{Bisht:2024hbs,
    author = "Bisht, Deepanshu and Chakraborty, Sabyasachi and Samanta, Atanu",
    title = "{A comprehensive study of ALPs from B-decays}",
    eprint = "2412.09678",
    archivePrefix = "arXiv",
    primaryClass = "hep-ph",
    doi = "10.1007/JHEP07(2025)092",
    journal = "JHEP",
    volume = "07",
    pages = "092",
    year = "2025"
}

@article{MartinCamalich:2025srw,
    author = "Martin Camalich, Jorge and Ziegler, Robert",
    title = "{Flavor Phenomenology of Light Dark Sectors}",
    eprint = "2503.17323",
    archivePrefix = "arXiv",
    primaryClass = "hep-ph",
    doi = "10.1146/annurev-nucl-121423-100931",
    journal = "Ann. Rev. Nucl. Part. Sci.",
    volume = "75",
    number = "1",
    pages = "223--246",
    year = "2025"
}

@article{Bertholet:2021hjl,
    author = "Bertholet, Emilie and Chakraborty, Sabyasachi and Loladze, Vazha and Okui, Takemichi and Soffer, Abner and Tobioka, Kohsaku",
    title = "{Heavy QCD axion at Belle II: Displaced and prompt signals}",
    eprint = "2108.10331",
    archivePrefix = "arXiv",
    primaryClass = "hep-ph",
    reportNumber = "KEK-TH-2343",
    doi = "10.1103/PhysRevD.105.L071701",
    journal = "Phys. Rev. D",
    volume = "105",
    number = "7",
    pages = "L071701",
    year = "2022"
}

@article{Arias-Aragon:2024gdz,
    author = "Arias-Arag{\'o}n, Fernando and Giannotti, Maurizio and di Cortona, Giovanni Grilli and Mescia, Federico",
    title = "{Axion-induced pair production: A new strategy for axion detection}",
    eprint = "2411.19327",
    archivePrefix = "arXiv",
    primaryClass = "hep-ph",
    doi = "10.1103/PhysRevD.111.043021",
    journal = "Phys. Rev. D",
    volume = "111",
    number = "4",
    pages = "043021",
    year = "2025"
}

@article{Zhitnitsky:1980tq,
    author = "Zhitnitsky, A. R.",
    title = "{On Possible Suppression of the Axion Hadron Interactions}",
    note = "[Yad. Fiz. {\bf 31}, 497 (1980)]",
    journal = "Sov. J. Nucl. Phys.",
    volume = "31",
    pages = "260",
    year = "1980"
}

@article{Dine:1981rt,
    author = "Dine, Michael and Fischler, Willy and Srednicki, Mark",
    title = "{A Simple Solution to the Strong CP Problem with a Harmless Axion}",
    reportNumber = "Print-81-0320 (IAS,PRINCETON)",
    doi = "10.1016/0370-2693(81)90590-6",
    journal = "Phys. Lett. B",
    volume = "104",
    pages = "199--202",
    year = "1981"
}

@article{Kim:1979if,
    author = "Kim, Jihn E.",
    title = "{Weak Interaction Singlet and Strong CP Invariance}",
    reportNumber = "UPR-0120T",
    doi = "10.1103/PhysRevLett.43.103",
    journal = "Phys. Rev. Lett.",
    volume = "43",
    pages = "103",
    year = "1979"
}

@article{Shifman:1979if,
    author = "Shifman, Mikhail A. and Vainshtein, A. I. and Zakharov, Valentin I.",
    title = "{Can Confinement Ensure Natural CP Invariance of Strong Interactions?}",
    reportNumber = "ITEP-64-1979",
    doi = "10.1016/0550-3213(80)90209-6",
    journal = "Nucl. Phys. B",
    volume = "166",
    pages = "493--506",
    year = "1980"
}

@article{Alda:2025nsz,
    author = "Alda, Jorge and Fuentes Zamoro, Marta and Merlo, Luca and Ponce D{\'\i}az, Xavier and Rigolin, Stefano",
    title = "{ALPaca: The ALP Automatic Computing Algorithm}",
    eprint = "2508.08354",
    archivePrefix = "arXiv",
    primaryClass = "hep-ph",
    reportNumber = "IFT-UAM/CSIC-25-82",
    month = "8",
    year = "2025"
}

@article{DiLuzio:2020wdo,
    author = "Di Luzio, Luca and Giannotti, Maurizio and Nardi, Enrico and Visinelli, Luca",
    title = "{The landscape of QCD axion models}",
    eprint = "2003.01100",
    archivePrefix = "arXiv",
    primaryClass = "hep-ph",
    reportNumber = "DESY 20-036, DESY-20-036",
    doi = "10.1016/j.physrep.2020.06.002",
    journal = "Phys. Rept.",
    volume = "870",
    pages = "1--117",
    year = "2020"
}

@article{Merlo:2017sun,
    author = "Merlo, L. and Pobbe, F. and Rigolin, S.",
    title = "{The Minimal Axion Minimal Linear $\sigma$ Model}",
    eprint = "1710.10500",
    archivePrefix = "arXiv",
    primaryClass = "hep-ph",
    reportNumber = "FTUAM-17-16, IFT-UAM-CSIC-17-082",
    doi = "10.1140/epjc/s10052-018-5892-z",
    journal = "Eur. Phys. J. C",
    volume = "78",
    number = "5",
    pages = "415",
    year = "2018",
    note = "[Erratum: Eur.Phys.J.C 79, 963 (2019)]"
}

@article{Brivio:2017sdm,
    author = "Brivio, I. and Gavela, M. B. and Pascoli, S. and del Rey, R. and Saa, S.",
    title = "{The axion and the Goldstone Higgs}",
    eprint = "1710.07715",
    archivePrefix = "arXiv",
    primaryClass = "hep-ph",
    reportNumber = "IFT-UAM/CSIC-17-103; FTUAM-17-24, IFT-UAM-CSIC-17-103, FTUAM-17-24",
    doi = "10.1016/j.cjph.2019.06.018",
    journal = "Chin. J. Phys.",
    volume = "61",
    pages = "55--71",
    year = "2019"
}

@article{Alonso-Gonzalez:2020wst,
    author = "Alonso-Gonzalez, J. and Lizana, J. M. and Martinez-Fernandez, V. and Merlo, L. and Pokorski, S.",
    title = "{Probing effective field theory approach in the CP violating minimal linear $\sigma $ model}",
    eprint = "2012.03990",
    archivePrefix = "arXiv",
    primaryClass = "hep-ph",
    doi = "10.1140/epjc/s10052-021-09326-y",
    journal = "Eur. Phys. J. C",
    volume = "81",
    number = "6",
    pages = "538",
    year = "2021"
}

@article{DiLuzio:2016sbl,
    author = "Di Luzio, Luca and Mescia, Federico and Nardi, Enrico",
    title = "{Redefining the Axion Window}",
    eprint = "1610.07593",
    archivePrefix = "arXiv",
    primaryClass = "hep-ph",
    reportNumber = "IPPP-16-99",
    doi = "10.1103/PhysRevLett.118.031801",
    journal = "Phys. Rev. Lett.",
    volume = "118",
    number = "3",
    pages = "031801",
    year = "2017"
}

@article{DiLuzio:2017pfr,
    author = "Di Luzio, Luca and Mescia, Federico and Nardi, Enrico",
    title = "{Window for preferred axion models}",
    eprint = "1705.05370",
    archivePrefix = "arXiv",
    primaryClass = "hep-ph",
    reportNumber = "IPPP-17-41",
    doi = "10.1103/PhysRevD.96.075003",
    journal = "Phys. Rev. D",
    volume = "96",
    number = "7",
    pages = "075003",
    year = "2017"
}

@article{Gaillard:2018xgk,
    author = "Gaillard, M. K. and Gavela, M. B. and Houtz, R. and Quilez, P. and Del Rey, R.",
    title = "{Color unified dynamical axion}",
    eprint = "1805.06465",
    archivePrefix = "arXiv",
    primaryClass = "hep-ph",
    reportNumber = "IFT-UAM/CSIC-18-050, FTUAM-18-13, IFT-UAM-CSIC-18-050",
    doi = "10.1140/epjc/s10052-018-6396-6",
    journal = "Eur. Phys. J. C",
    volume = "78",
    number = "11",
    pages = "972",
    year = "2018"
}

@article{Hook:2019qoh,
    author = "Hook, Anson and Kumar, Soubhik and Liu, Zhen and Sundrum, Raman",
    title = "{High Quality QCD Axion and the LHC}",
    eprint = "1911.12364",
    archivePrefix = "arXiv",
    primaryClass = "hep-ph",
    reportNumber = "UMD-PP-019-07",
    doi = "10.1103/PhysRevLett.124.221801",
    journal = "Phys. Rev. Lett.",
    volume = "124",
    number = "22",
    pages = "221801",
    year = "2020"
}

@article{DiLuzio:2021pxd,
    author = "Di Luzio, Luca and Gavela, Belen and Quilez, Pablo and Ringwald, Andreas",
    title = "{An even lighter QCD axion}",
    eprint = "2102.00012",
    archivePrefix = "arXiv",
    primaryClass = "hep-ph",
    reportNumber = "DESY-21-010, DESY 21-010, IFT-UAM/CSIC-20-143, FTUAM-20-21",
    doi = "10.1007/JHEP05(2021)184",
    journal = "JHEP",
    volume = "05",
    pages = "184",
    year = "2021"
}

@article{Gavela:2023tzu,
    author = "Gavela, Bel{\'e}n and Qu{\'\i}lez, Pablo and Ramos, Maria",
    title = "{The QCD axion sum rule}",
    eprint = "2305.15465",
    archivePrefix = "arXiv",
    primaryClass = "hep-ph",
    reportNumber = "IFT-UAM/CSIC-23-58",
    doi = "10.1007/JHEP04(2024)056",
    journal = "JHEP",
    volume = "04",
    pages = "056",
    year = "2024"
}

@article{Cox:2023dou,
    author = "Cox, Peter and Gherghetta, Tony and Paul, Arpon",
    title = "{A common origin for the QCD axion and sterile neutrinos from $SU(5)$ strong dynamics}",
    eprint = "2310.08557",
    archivePrefix = "arXiv",
    primaryClass = "hep-ph",
    reportNumber = "UMN-TH-4301/23",
    doi = "10.1007/JHEP12(2023)180",
    journal = "JHEP",
    volume = "12",
    pages = "180",
    year = "2023"
}

@article{deGiorgi:2024elx,
    author = "de Giorgi, Arturo and Ramos, Maria",
    title = "{Extra-dimensional axion patterns}",
    eprint = "2412.00179",
    archivePrefix = "arXiv",
    primaryClass = "hep-ph",
    reportNumber = "IPPP/24/75, CERN-TH-2024-205",
    doi = "10.1103/PhysRevD.111.075006",
    journal = "Phys. Rev. D",
    volume = "111",
    number = "7",
    pages = "075006",
    year = "2025"
}

@article{Alonso-Gonzalez:2018vpc,
    author = "Alonso-Gonz{\'a}lez, Javier and Merlo, Luca and Pobbe, Federico and Rigolin, Stefano and Sumensari, Olcyr",
    title = "{Testable axion-like particles in the minimal linear {\ensuremath{\sigma}} model}",
    eprint = "1807.08643",
    archivePrefix = "arXiv",
    primaryClass = "hep-ph",
    doi = "10.1016/j.nuclphysb.2019.114839",
    journal = "Nucl. Phys. B",
    volume = "950",
    pages = "114839",
    year = "2020"
}

@article{DiLuzio:2021gos,
    author = "Di Luzio, Luca and Gavela, Belen and Quilez, Pablo and Ringwald, Andreas",
    title = "{Dark matter from an even lighter QCD axion: trapped misalignment}",
    eprint = "2102.01082",
    archivePrefix = "arXiv",
    primaryClass = "hep-ph",
    reportNumber = "DESY 21-011, DESY-21-011, IFT-UAM/CSIC-20-144, FTUAM-20-21",
    doi = "10.1088/1475-7516/2021/10/001",
    journal = "JCAP",
    volume = "10",
    pages = "001",
    year = "2021"
}

@article{Carmona:2021seb,
    author = "Carmona, Adrian and Scherb, Christiane and Schwaller, Pedro",
    title = "{Charming ALPs}",
    eprint = "2101.07803",
    archivePrefix = "arXiv",
    primaryClass = "hep-ph",
    reportNumber = "MITP-21-003",
    doi = "10.1007/JHEP08(2021)121",
    journal = "JHEP",
    volume = "08",
    pages = "121",
    year = "2021"
}

@article{Greljo:2025suh,
    author = "Greljo, Admir and Ponce D{\'\i}az, Xavier and Thomsen, Anders Eller",
    title = "{Insights on the cosmic origin of matter from proton stability}",
    eprint = "2505.18259",
    archivePrefix = "arXiv",
    primaryClass = "hep-ph",
    doi = "10.1088/1475-7516/2025/11/043",
    journal = "JCAP",
    volume = "11",
    pages = "043",
    year = "2025"
}

@article{Liang:2024vnd,
    author = "Liang, Qiuyue and Ponce D{\'\i}az, Xavier and Yanagida, Tsutomu T.",
    title = "{Axion Detection Experiments Can Probe Majoron Models}",
    eprint = "2406.19083",
    archivePrefix = "arXiv",
    primaryClass = "hep-ph",
    doi = "10.1103/PhysRevLett.134.151803",
    journal = "Phys. Rev. Lett.",
    volume = "134",
    number = "15",
    pages = "151803",
    year = "2025"
}

@article{ParticleDataGroup:2024cfk,
    author = "Navas, S. and others",
    collaboration = "Particle Data Group",
    title = "{Review of particle physics}",
    doi = "10.1103/PhysRevD.110.030001",
    journal = "Phys. Rev. D",
    volume = "110",
    number = "3",
    pages = "030001",
    year = "2024"
}

@inproceedings{Anisimov:2006hv,
    author = "Anisimov, Alexey",
    title = "{Majorana Dark Matter}",
    booktitle = "{6th International Workshop on the Identification of Dark Matter}",
    eprint = "hep-ph/0612024",
    publisher = "World Scientific",
    address = "Singapore",
    archivePrefix = "arXiv",
    doi = "10.1142/9789812770288_0058",
    pages = "439--449",
    month = "12",
    year = "2006"
}

@article{Graesser:2007yj,
    author = "Graesser, Michael L.",
    title = "{Broadening the Higgs boson with right-handed neutrinos and a higher dimension operator at the electroweak scale}",
    eprint = "0704.0438",
    archivePrefix = "arXiv",
    primaryClass = "hep-ph",
    reportNumber = "RUNHETC-04-2007",
    doi = "10.1103/PhysRevD.76.075006",
    journal = "Phys. Rev. D",
    volume = "76",
    pages = "075006",
    year = "2007"
}

@article{Graesser:2007pc,
    author = "Graesser, Michael L.",
    title = "{Experimental Constraints on Higgs Boson Decays to TeV-scale Right-Handed Neutrinos}",
    eprint = "0705.2190",
    archivePrefix = "arXiv",
    primaryClass = "hep-ph",
    reportNumber = "RUNHETC-04-2007",
    month = "5",
    year = "2007"
}

@article{delAguila:2008ir,
    author = "del Aguila, Francisco and Bar-Shalom, Shaouly and Soni, Amarjit and Wudka, Jose",
    title = "{Heavy Majorana Neutrinos in the Effective Lagrangian Description: Application to Hadron Colliders}",
    eprint = "0806.0876",
    archivePrefix = "arXiv",
    primaryClass = "hep-ph",
    reportNumber = "UG-FT-230-08, CAFPE-100-08, UCI-TR-2008-21, BNL-HET-08-14",
    doi = "10.1016/j.physletb.2008.11.031",
    journal = "Phys. Lett. B",
    volume = "670",
    pages = "399--402",
    year = "2009"
}

@article{Aparici:2009fh,
    author = "Aparici, Alberto and Kim, Kyungwook and Santamaria, Arcadi and Wudka, Jose",
    title = "{Right-handed neutrino magnetic moments}",
    eprint = "0904.3244",
    archivePrefix = "arXiv",
    primaryClass = "hep-ph",
    reportNumber = "FTUV-09-0421, IFIC-09-15, UCRHEP-T466",
    doi = "10.1103/PhysRevD.80.013010",
    journal = "Phys. Rev. D",
    volume = "80",
    pages = "013010",
    year = "2009"
}

@article{Liao:2016qyd,
    author = "Liao, Yi and Ma, Xiao-Dong",
    title = "{Operators up to Dimension Seven in Standard Model Effective Field Theory Extended with Sterile Neutrinos}",
    eprint = "1612.04527",
    archivePrefix = "arXiv",
    primaryClass = "hep-ph",
    doi = "10.1103/PhysRevD.96.015012",
    journal = "Phys. Rev. D",
    volume = "96",
    number = "1",
    pages = "015012",
    year = "2017"
}

@article{Li:2022ipc,
    author = "Li, Xu and Zhang, Di and Zhou, Shun",
    title = "{One-loop matching of the type-II seesaw model onto the Standard Model effective field theory}",
    eprint = "2201.05082",
    archivePrefix = "arXiv",
    primaryClass = "hep-ph",
    doi = "10.1007/JHEP04(2022)038",
    journal = "JHEP",
    volume = "04",
    pages = "038",
    year = "2022"
}

@article{Bhattacharya:2015vja,
    author = "Bhattacharya, Subhaditya and Wudka, Jos{\'e}",
    title = "{Dimension-seven operators in the standard model with right handed neutrinos}",
    eprint = "1505.05264",
    archivePrefix = "arXiv",
    primaryClass = "hep-ph",
    doi = "10.1103/PhysRevD.94.055022",
    journal = "Phys. Rev. D",
    volume = "94",
    number = "5",
    pages = "055022",
    year = "2016",
    note = "[Erratum: Phys.Rev.D 95, 039904 (2017)]"
}

@article{Li:2021tsq,
    author = "Li, Hao-Lin and Ren, Zhe and Xiao, Ming-Lei and Yu, Jiang-Hao and Zheng, Yu-Hui",
    title = "{Operator bases in effective field theories with sterile neutrinos: d {\ensuremath{\leq}} 9}",
    eprint = "2105.09329",
    archivePrefix = "arXiv",
    primaryClass = "hep-ph",
    doi = "10.1007/JHEP11(2021)003",
    journal = "JHEP",
    volume = "11",
    pages = "003",
    year = "2021"
}

@article{Balkin:2018tma,
    author = "Balkin, Reuven and Ruhdorfer, Maximilian and Salvioni, Ennio and Weiler, Andreas",
    title = "{Dark matter shifts away from direct detection}",
    eprint = "1809.09106",
    archivePrefix = "arXiv",
    primaryClass = "hep-ph",
    reportNumber = "TUM-HEP-1162-18",
    doi = "10.1088/1475-7516/2018/11/050",
    journal = "JCAP",
    volume = "11",
    pages = "050",
    year = "2018"
}

@article{Broncano:2002rw,
    author = "Broncano, A. and Gavela, M. B. and Jenkins, Elizabeth Ellen",
    title = "{The Effective Lagrangian for the seesaw model of neutrino mass and leptogenesis}",
    eprint = "hep-ph/0210271",
    archivePrefix = "arXiv",
    reportNumber = "FTUAM-02-26, IFT-UAM-CSIC-02-46, UCSD-PTH-02-25",
    doi = "10.1016/S0370-2693(02)03130-1",
    journal = "Phys. Lett. B",
    volume = "552",
    pages = "177--184",
    year = "2003",
    note = "[Erratum: Phys.Lett.B 636, 332 (2006)]"
}

@article{Blennow:2023mqx,
    author = "Blennow, Mattias and Fern{\'a}ndez-Mart{\'\i}nez, Enrique and Hern{\'a}ndez-Garc{\'\i}a, Josu and L{\'o}pez-Pav{\'o}n, Jacobo and Marcano, Xabier and Naredo-Tuero, Daniel",
    title = "{Bounds on lepton non-unitarity and heavy neutrino mixing}",
    eprint = "2306.01040",
    archivePrefix = "arXiv",
    primaryClass = "hep-ph",
    reportNumber = "IFT-UAM/CSIC-23-60, FTUV-23-0531.7594, IFIC/23-19",
    doi = "10.1007/JHEP08(2023)030",
    journal = "JHEP",
    volume = "08",
    pages = "030",
    year = "2023"
}

@article{Arhrib:2011uy,
    author = "Arhrib, A. and Benbrik, R. and Chabab, M. and Moultaka, G. and Peyranere, M. C. and Rahili, L. and Ramadan, J.",
    title = "{The Higgs Potential in the Type II Seesaw Model}",
    eprint = "1105.1925",
    archivePrefix = "arXiv",
    primaryClass = "hep-ph",
    doi = "10.1103/PhysRevD.84.095005",
    journal = "Phys. Rev. D",
    volume = "84",
    pages = "095005",
    year = "2011"
}

@article{Moultaka:2020dmb,
    author = "Moultaka, Gilbert and Peyran{\`e}re, Michel C.",
    title = "{Vacuum stability conditions for Higgs potentials with $SU(2)_L$ triplets}",
    eprint = "2012.13947",
    archivePrefix = "arXiv",
    primaryClass = "hep-ph",
    doi = "10.1103/PhysRevD.103.115006",
    journal = "Phys. Rev. D",
    volume = "103",
    number = "11",
    pages = "115006",
    year = "2021"
}

@article{ATLAS:2023tkt,
    author = "Aad, Georges and others",
    collaboration = "ATLAS",
    title = "{Combination of searches for invisible decays of the Higgs boson using 139 fb$^{-1}$ of proton-proton collision data at $\sqrt{s}=13$ TeV collected with the ATLAS experiment}",
    eprint = "2301.10731",
    archivePrefix = "arXiv",
    primaryClass = "hep-ex",
    reportNumber = "CERN-EP-2022-289",
    doi = "10.1016/j.physletb.2023.137963",
    journal = "Phys. Lett. B",
    volume = "842",
    pages = "137963",
    year = "2023"
}

@article{KamLAND-Zen:2012uen,
    author = "Gando, A. and others",
    collaboration = "KamLAND-Zen",
    title = "{Limits on Majoron-emitting double-beta decays of Xe-136 in the KamLAND-Zen experiment}",
    eprint = "1205.6372",
    archivePrefix = "arXiv",
    primaryClass = "hep-ex",
    doi = "10.1103/PhysRevC.86.021601",
    journal = "Phys. Rev. C",
    volume = "86",
    pages = "021601",
    year = "2012"
}

@article{Sandner:2023ptm,
    author = "Sandner, Stefan and Escudero, Miguel and Witte, Samuel J.",
    title = "{Precision CMB constraints on eV-scale bosons coupled to neutrinos}",
    eprint = "2305.01692",
    archivePrefix = "arXiv",
    primaryClass = "hep-ph",
    reportNumber = "IFIC/23-13, FTUV-23-0413.0599, CERN-TH-2023-073",
    doi = "10.1140/epjc/s10052-023-11864-6",
    journal = "Eur. Phys. J. C",
    volume = "83",
    number = "8",
    pages = "709",
    year = "2023"
}

@article{TWIST:2014ymv,
    author = "Bayes, R. and others",
    collaboration = "TWIST",
    title = "{Search for two body muon decay signals}",
    eprint = "1409.0638",
    archivePrefix = "arXiv",
    primaryClass = "hep-ex",
    doi = "10.1103/PhysRevD.91.052020",
    journal = "Phys. Rev. D",
    volume = "91",
    number = "5",
    pages = "052020",
    year = "2015"
}

@article{Lessa:2007up,
    author = "Lessa, A. P. and Peres, O. L. G.",
    title = "{Revising limits on neutrino-Majoron couplings}",
    eprint = "hep-ph/0701068",
    archivePrefix = "arXiv",
    doi = "10.1103/PhysRevD.75.094001",
    journal = "Phys. Rev. D",
    volume = "75",
    pages = "094001",
    year = "2007"
}

@article{Jodidio:1986mz,
    author = "Jodidio, A. and others",
    title = "{Search for Right-Handed Currents in Muon Decay}",
    reportNumber = "LBL-21616",
    doi = "10.1103/PhysRevD.34.1967",
    journal = "Phys. Rev. D",
    volume = "34",
    pages = "1967",
    year = "1986",
    note = "[Erratum: Phys.Rev.D 37, 237 (1988)]"
}

@article{ATLAS:2022vkf,
    author = "Aad, Georges and others",
    collaboration = "ATLAS",
    title = "{A detailed map of Higgs boson interactions by the ATLAS experiment ten years after the discovery}",
    eprint = "2207.00092",
    archivePrefix = "arXiv",
    primaryClass = "hep-ex",
    reportNumber = "CERN-EP-2022-057",
    doi = "10.1038/s41586-022-04893-w",
    journal = "Nature",
    volume = "607",
    number = "7917",
    pages = "52--59",
    year = "2022",
    note = "[Erratum: Nature 612, E24 (2022)]"
}

@article{FernandezNavarro:2026cyu,
    author = "Fern{\'a}ndez Navarro, Mario and Zamoro, Marta F. and Pesut, Marko and Ponce D{\'\i}az, Xavier",
    title = "{Structure of multiaxion solutions to the strong CP problem}",
    eprint = "2605.06787",
    doi = "10.1103/hw7t-45p2",
    journal = "Phys. Rev. D",
    volume = "114",
    pages = "035019",
    archivePrefix = "arXiv",
    primaryClass = "hep-ph",
    month = "5",
    year = "2026"
}

@article{Cheng:2020rla,
    author = "Cheng, Yu and Chiang, Cheng-Wei and He, Xiao-Gang and Sun, Jin",
    title = "{Flavor-changing Majoron interactions with leptons}",
    eprint = "2012.15287",
    archivePrefix = "arXiv",
    primaryClass = "hep-ph",
    doi = "10.1103/PhysRevD.104.013001",
    journal = "Phys. Rev. D",
    volume = "104",
    number = "1",
    pages = "013001",
    year = "2021"
}

@article{LHCHiggsCrossSectionWorkingGroup:2012nn,
    author = "David, A. and Denner, A. and Duehrssen, M. and Grazzini, M. and Grojean, C. and Passarino, G. and Schumacher, M. and Spira, M. and Weiglein, G. and Zanetti, M.",
    collaboration = "LHC Higgs Cross Section Working Group",
    title = "{LHC HXSWG interim recommendations to explore the coupling structure of a Higgs-like particle}",
    eprint = "1209.0040",
    archivePrefix = "arXiv",
    primaryClass = "hep-ph",
    reportNumber = "CERN-PH-TH-2012-284, LHCHXSWG-2012-001",
    month = "9",
    year = "2012"
}

@article{Sun:2021jpw,
    author = "Sun, Jin and Cheng, Yu and He, Xiao-Gang",
    title = "{Structure Of Flavor Changing Goldstone Boson Interactions}",
    eprint = "2101.06055",
    archivePrefix = "arXiv",
    primaryClass = "hep-ph",
    doi = "10.1007/JHEP04(2021)141",
    journal = "JHEP",
    volume = "04",
    pages = "141",
    year = "2021"
}

@article{Zamoro:2026ily,
    author = "Zamoro, Marta F. and Lozano-Onrubia, {\'A}lvaro and Merlo, Luca and Rosende Herrero, Samuel",
    title = "{Crossing into the $m_a > f_a$ Region for Leptophilic ALPs}",
    eprint = "2605.00115",
    archivePrefix = "arXiv",
    primaryClass = "hep-ph",
    reportNumber = "IFT-UAM/CSIC-26-21",
    month = "4",
    year = "2026"
}

@misc{PonceDiaz:2021,
    author = "Ponce D{\'i}az, Xavier",
    title = "{$\Delta$EFT: An Effective Field Theory of the Type-2 Seesaw Mechanism}",
    howpublished = "{Talk given at the Internal HIDDeN Webinar (Universit{\"a}t Heidelberg and Max-Planck-Institut f{\"u}r Kernphysik, 2021), based on the author's MSc thesis (supervisors: T.~Plehn and W.~Rodejohann)}",
    year = "2021",
    url = "https://projects.ift.uam-csic.es/hidden-virtual-institute/files/PresentacionHiddenESR_PonceDiaz_Xavier.pdf",
}

@article{Akita:2026gzk,
    author = "Akita, Kensuke and Hamaguchi, Koichi and Kitagawa, Haruto and Yokoyama, Tatsuya",
    title = "{Minimal Majoron Dark Matter}",
    eprint = "2605.12946",
    archivePrefix = "arXiv",
    primaryClass = "hep-ph",
    month = "5",
    year = "2026"
}

@article{deGiorgi:2026jqn,
    author = "de Giorgi, Arturo and Naredo-Tuero, Daniel and Ponce D{\'\i}az, Xavier",
    title = "{The Majoron Cosmological Window: Dark Matter and Thermal Leptogenesis}",
    eprint = "2605.18944",
    archivePrefix = "arXiv",
    primaryClass = "hep-ph",
    reportNumber = "IPPP/26/39",
    month = "5",
    year = "2026"
}

@article{Batell:2026avi,
    author = "Batell, Brian and Dasgupta, Arnab and Dutta, Swapnil and Ghalsasi, Akshay",
    title = "{Majoron dark matter, high-scale seesaw framework, and leptogenesis}",
    eprint = "2606.02706",
    doi = "10.1103/qjrh-r2tv",
    journal = "Phys. Rev. D",
    volume = "114",
    pages = "055018",
    archivePrefix = "arXiv",
    primaryClass = "hep-ph",
    reportNumber = "PITT-PACC-2606",
    month = "6",
    year = "2026"
}

@article{Debnath:2026ebg,
    author = "Debnath, Hridoy and Fileviez Perez, Pavel and Miller, Grace",
    title = "{The QCD Axion and Neutrino Masses}",
    eprint = "2608.05150",
    archivePrefix = "arXiv",
    primaryClass = "hep-ph",
    month = "8",
    year = "2026"
}

@article{Palavric:2026vej,
    author = "Palavri{\'c}, Ajdin and Ponce D{\'\i}az, Xavier and Tiblom, Hector",
    title = "{The KSVZ Atlas: A Unified SMEFT-ALP Framework}",
    eprint = "2606.12298",
    archivePrefix = "arXiv",
    primaryClass = "hep-ph",
    month = "6",
    year = "2026"
}

@article{Debnath:2024vpf,
    author = "Debnath, Hridoy and Fileviez Perez, Pavel and Gonzalez-Quesada, Kevin",
    title = "{Majorana neutrinos and dark matter from anomaly cancellation}",
    eprint = "2403.17144",
    archivePrefix = "arXiv",
    primaryClass = "hep-ph",
    doi = "10.1103/PhysRevD.109.115030",
    journal = "Phys. Rev. D",
    volume = "109",
    number = "11",
    pages = "115030",
    year = "2024"
}

@article{Garcia-Cely:2017oco,
    author = "Garcia-Cely, Camilo and Heeck, Julian",
    title = "{Neutrino Lines from Majoron Dark Matter}",
    eprint = "1701.07209",
    archivePrefix = "arXiv",
    primaryClass = "hep-ph",
    doi = "10.1007/JHEP05(2017)102",
    journal = "JHEP",
    volume = "05",
    pages = "102",
    year = "2017"
}

@article{Heeck:2019guh,
    author = "Heeck, Julian and Patel, Hiren H.",
    title = "{Majoron at two loops}",
    eprint = "1909.02029",
    archivePrefix = "arXiv",
    primaryClass = "hep-ph",
    doi = "10.1103/PhysRevD.100.095015",
    journal = "Phys. Rev. D",
    volume = "100",
    number = "9",
    pages = "095015",
    year = "2019"
}

@article{Akhmedov:1992hi,
    author = "Akhmedov, Evgeny Kh. and Berezhiani, Z. G. and Mohapatra, R. N. and Senjanovic, G.",
    title = "{Planck scale effects on the majoron}",
    eprint = "hep-ph/9209285",
    archivePrefix = "arXiv",
    primaryClass = "hep-ph",
    doi = "10.1016/0370-2693(93)90887-N",
    journal = "Phys. Lett. B",
    volume = "299",
    pages = "90--93",
    year = "1993"
}

@article{Rothstein:1992rh,
    author = "Rothstein, Ira Z. and Babu, K. S. and Seckel, D.",
    title = "{Planck scale symmetry breaking and majoron physics}",
    eprint = "hep-ph/9301213",
    archivePrefix = "arXiv",
    primaryClass = "hep-ph",
    doi = "10.1016/0550-3213(93)90368-Y",
    journal = "Nucl. Phys. B",
    volume = "403",
    pages = "725--748",
    year = "1993"
}

@article{Bhattacharya:2025hme,
    author = "Bhattacharya, Subhaditya and Jahedi, Sahabub and Manna, Soumen Kumar and Sil, Arunansu",
    title = "{Probing ALP-portal fermionic dark matter at the $e^+e^-$ colliders}",
    eprint = "2505.00478",
    archivePrefix = "arXiv",
    primaryClass = "hep-ph",
    doi = "10.1103/n7bh-t2kd",
    journal = "Phys. Rev. D",
    volume = "113",
    pages = "115015",
    year = "2026"
}

\end{document}